\documentclass[authoryear]{elsarticle} 

\usepackage{fullpage}
\journal{Ocean Engineering}

\usepackage{enumitem}

\usepackage{upgreek}
\usepackage{mathtools}
\usepackage{rotating}
\usepackage{amsmath}
\usepackage{amssymb}
\usepackage{bbm}
\usepackage{subcaption}
\usepackage{graphicx}
\usepackage{url}
\usepackage[linesnumbered]{algorithm2e}
\usepackage{placeins}
\usepackage{xcolor}
\usepackage{multirow}
\usepackage{mathrsfs}

\definecolor{Ph}{rgb}{1,0,0}

\def\bSig\mathbf{\Sigma}

\newcommand {\pbi}{\begin{itemize}}
\newcommand {\pei}{\end{itemize}}

\newcommand {\pbc}{\begin{center}}
\newcommand {\pec}{\end{center}}
\newcommand {\pbe}{\begin{eqnarray*}}
\newcommand {\pee}{\end{eqnarray*}}
\newcommand {\pben}{\begin{eqnarray}}
\newcommand {\peen}{\end{eqnarray}}
\let\hat=\widehat

\newcommand {\maxover}[1]{\underset{#1}{\mathrm{max} \text{ }}}

\newcommand {\meanover}[1]{\underset{#1}{\mathrm{mean} \text{ }}}

\newcommand {\ed}[1]{\textcolor{black}{#1}}
\usepackage {tikz}
\usetikzlibrary{arrows}
\usepackage{xspace} 

\providecommand{\RS}{\texttt{rsds}\xspace}
\providecommand{\WS}{\texttt{sfcWind}\xspace}
\providecommand{\WM}{\texttt{sfcWindmax}\xspace}
\providecommand{\TA}{\texttt{tas}\xspace}

\providecommand{\AC}{\texttt{ACCESS-CM2}\xspace}
\providecommand{\CA}{\texttt{CAMS-CSM1-0}\xspace}
\providecommand{\CE}{\texttt{CESM2}\xspace}
\providecommand{\EC}{\texttt{EC-Earth3}\xspace}
\providecommand{\MR}{\texttt{MRI-ESM2-0}\xspace}
\providecommand{\No}{\texttt{NorESM2-0-LL}\xspace}
\providecommand{\UK}{\texttt{UKESM1-0-LL}\xspace}

\providecommand{\SL}{\texttt{SSP126}\xspace}
\providecommand{\SM}{\texttt{SSP245}\xspace}
\providecommand{\SH}{\texttt{SSP585}\xspace}

\providecommand{\NA}{{North Atlantic}\xspace}
\providecommand{\CS}{{Celtic Sea}\xspace}

\begin{document}

\begin{frontmatter}
	
	\title{Changes over time in the 100-year return value of climate model variables}
	
	\author[shelluk]{Callum Leach}
	\address[shelluk]{Shell Information Technology International Ltd., London SE1 7NA, United Kingdom.}
	
	\author[mos]{Kevin Ewans}
	\address[mos]{MetOcean Research Ltd, New Plymouth 4310, New Zealand.}	
	
	\author[lancs]{Philip Jonathan}
	\address[lancs]{School of Mathematical Sciences, Lancaster University LA1 4YF, United Kingdom.}
		
	\begin{abstract}
	\ed{We assess evidence for changes in tail characteristics of wind, solar irradiance and temperature variables output from CMIP6 global climate models (GCMs) due to climate forcing. We estimate global and climate zone annual maximum and annual means for period (2015, 2100) from daily output of seven GCMs for daily wind speed, maximum wind speed, solar irradiance and near-surface temperature. We calculate corresponding annualised data for individual locations within neighbourhoods of the North Atlantic and Celtic Sea region. We consider output for three climate scenarios and multiple climate ensembles. We estimate non-stationary extreme value models for annual extremes, and non-homogeneous Gaussian regressions for annual means, using Bayesian inference. We use estimated statistical models to quantify the distribution of (i) the change in 100-year return value for annual extremes, and (2) the change in annual mean, over the period (2025, 2125). To summarise results, we estimate linear mixed effects models for observed variation of (i) and (ii). Evidence for changes in the 100-year return value for annual maxima of solar irradiance and temperature is much stronger than for wind variables over time and with climate scenario.} 
	\end{abstract}
	
\end{frontmatter}

\section{Introduction} \label{Sct:Int}
%
The Summary for Policymakers from the 6$^{\text{th}}$ assessment report of the Intergovernmental Panel on Climate Change (\citealt{IPCC-SPM}) states unequivocally that human activity has warmed the Earth, resulting in changes to the atmosphere and oceans in particular. It is likely that global average precipitation has increased, that near-surface ocean salinity patterns have changed, that global mean sea level has \ed{increased}, that mid-latitude storm tracks have shifted poleward, that sea ice has retreated in the Arctic (but not in the Antarctic), and that the surface of the Greenland Ice Sheet has melted, all due to human activity. The rate of occurrence of temperature extremes has increased, as has that of Category 3-5 tropical cyclones, as a result of human activity.

Numerous numerical studies (e.g. \citealt{YngRbl19}, \citealt{MccEA20}, \citealt{MccEA22}, \citealt{EwnJnt23a}) reporting likely changes in wind and wave climate, have been summarised in \cite{EwnJnt23a}. Most historical studies are based on global or regional analysis of output from Coupled Model Intercomparison Project (CMIP) General Circulation Models (GCMs). Output is available from multiple institutions (including e.g. the UK Metoffice, the Norwegian Climate Centre, the Chinese Meteorological Administration and the Japanese Meteorological Research Institute) for a large set of climate variables. For the 6$^{\text{th}}$ phase of CMIP (CMIP6), daily values of climate variables are provided for the period 2015-2100, under different assumptions combining future Shared Socioeconomic Pathways (SSPs) and Representative Concentration Pathways (RCPs); see Section~\ref{Sct:Dat} for further details.

GCMs behave somewhat differently to each other, due to different modelling assumptions made (e.g. \citealt{IPCC-WG1-2013}, Chapter 8). Moreover, just as wave hindcast and forecast model output is generally calibrated (e.g. \citealt{SngTuo24}) to provide better in-situ agreement with measurements, it is to be expected that GCM output should be calibrated similarly for better agreement with measurements (e.g. \citealt{BllEA12}, \citealt{LmsEA23}, \citealt{LavEA24}, \citealt{MccEA24}). There is a difference however. For calibration of a hindcast, both hindcast and measurement data are typically available for a relatively large proportion of the period of the hindcast. For calibration of short-term forecasts similarly, historical forecasts and corresponding measurements are both available over a relatively long period, with which to perform a reasonable calibration. For GCM output, by definition, the proportion of the total time interval of the GCM output for which measured data are available is relatively small, and focussed on the early years of GCM output by definition; specifically, for CMIP6 GCMs, there is no means by which we can be confident that a calibration developed for the years 2015-2024 would be appropriate in 2100. \ed{Analysis of GCM output provides indications of likely long-term climate effects which cannot be achieved for example from calibration of typical hindcast output for short-term purposes.} Nevertheless, we note the work of \cite{TttEA22} regarding reduction of GCM variability using calibration, and recently produced guidance (\citealt{IOGP24}) from the International Association of Oil and Gas Producers on potential climate change effects on metocean design and operating criteria. 

Some authors (e.g. \citealt{SvrEA24}, \citealt{WdlEA24}) seek to couple the output of CMIP6 models with climate risk assessment models, to obtain predictions of future impacts (e.g. cost of damage to infrastructure). \citealt{SvrEA24} in particular emphasise the importance of incorporating climate model uncertainty in predictions. From the perspective of wind energy resources, \cite{MiaEA23} find that wind speeds generally decrease under CMIP6 scenarios in the Northern Hemisphere; again, the variability in estimates due to climate model uncertainty is emphasised. \cite{IbrEA23} find that climate-related changes in electricity generating capacity from wind and wave occur in at most 15\% of coastal locations studied under CMIP6 SSP126 and SSP585 scenarios.


\ed{Engineering design for a physical environment subject to climate change is challenging, because estimating future anthropogenic climate change forcing and the planet's response to it is subject to large uncertainties (e.g. \citealt{EwnJnt23a}). Nevertheless it is important to understand the practical implications of the latest models forecasting the Earth's future environment, and its extremes, and use that knowledge as wisely as possible. The literature addressing the uncertainties in future climate directly from GCM output, summarised above and in \cite{EwnJnt23a}, is limited and does not often undertake statistical analysis in the most careful manner; this is a gap we hope the current work will contribute to closing, quantifying uncertainties across multiple GCMs, climate variables, climate scenarios and ensembles. Environmental extremes are often characterised in terms of a return value, an extreme quantile of the distribution of the annual maximum of an environmental variable (e.g. \citealt{JntEA20}). The 100-year return value, with an annual exceedance rate of 1/100, is in widespread use. In the current work, we consider extremes of climate variables of obvious interest, including average and maximum daily wind speed (for offshore design generally, and wind and wave energy applications), solar irradiance (for solar energy applications), and near-surface atmospheric temperature. We emphasise however that the current work is academic research examining key outputs from CMIP6 global coupled models, to form a view of what state-of-the-art climate science is telling us about climate change effects in variables of potential interest to the offshore engineering community. Discussion of practical design steps, already provided in documents such as \cite{IOGP24}, is far beyond the scope of this article.}

\subsection*{Objectives and outline}
The objective of the current work is to examine key output from CMIP6 global coupled models at site-specific, regional and global scales, to form a view of what state-of-the-art science is telling us about climate change effects. We focus on assessing the change in extreme quantiles of annual distributions of climate variables, specifically the change in the 100-year return value for a climate variable over the next 100 years (from 2025 to 2125), for surface downwelling short wave radiation, near-surface wind speed, daily maximum near-surface wind speed and near-surface air temperature. For the global and climate zone analyses considered, we think it \ed{is} interesting to compare inferences for extremes with those for changes in the spatial means (e.g. global and climate zone averages) also.

The layout of the paper is as follows. The GCM output considered in the current study is introduced and summarised in Section~\ref{Sct:Dat}, for both global and climate zone analysis (Section~\ref{Sct:Dat:GlbClmZon}) and North Atlantic-Celtic Sea point location analysis (Section~\ref{Sct:Dat:NACS}). Section~\ref{Sct:Mth} outlines the statistical methodology adopted to estimate the change in the 100-year return value for each climate variable over the next 100 years (i.e. the period 2025-2125), and changes in global and climate zone means over the same period. Extreme value analysis of annual maxima data is achieved by fitting non-stationary generalised extreme value (GEV) models, described in Section~\ref{Sct:Mth:GEVR}. Trends in spatial means over time are quantified using non-homogeneous Gaussian regression (NHGR, Section~\ref{Sct:Mth:NHGR}). Results of applying the models from Section~\ref{Sct:Mth} to the data from Section~\ref{Sct:Dat} are presented in Section~\ref{Sct:Rsl}. Section~\ref{Sct:Rsl:Glb} provides a summary of the analysis for global annual extreme and mean data, Section~\ref{Sct:Rsl:ClmZon} for climate zone annual extreme and mean data, and Section~\ref{Sct:Rsl:NACS} for annual maxima from specific locations in the North Atlantic and Celtic Sea neighbourhoods. Discussions and conclusions are given in Section~\ref{Sct:DscCnc}. Supporting illustrations are provided in the accompanying online Supplementary Material (SM), and referenced e.g. by Figure~SM3. All data used in this work (global and climate zone annual maximum, minimum and mean data, and all North Sea and Celtic Sea annual maximum data) is accessible at \cite{Lch24}. The software used for the analysis is provided at \cite{LchJnt24}.

\FloatBarrier
\section{Global coupled model output}  \label{Sct:Dat}
%
Output for 7 GCMs from CMIP6 was accessed via the UK Centre for Environmental Analysis (CEDA) archive during the Spring of 2024. For each of these GCMs, gridded output for the whole globe is generally available daily for the time period 2015-2100. We choose to examine annualised data for a total of 4 environmental variables of engineering interest: surface downwelling shortwave radiation (\RS, Wm$^{-2}$), near-surface wind speed (\WS, ms$^{-1}$), daily maximum near-surface wind speed (\WM, ms$^{-1}$) and near-surface air temperature (\TA, K). In addition, we examine three climate scenarios (or climate experiments): SSP126, SSP245 and SSP585. Each of these scenarios combines a SSP with a RCP. Experiment SSP126 follows SSP1, a storyline with low climate change mitigation and adaptation challenges, and RCP2.6 which leads to (additional) radiative forcing of 2.6 Wm$^{-2}$ by the year 2100. Analogously, experiment SSP245 (SSP585) follows SSP2 (SSP3), a storyline with intermediate (high) climate change mitigation and adaptation challenges, and RCP4.5 (RDC8.5) which leads to (additional) radiative forcing of 4.5 (8.5) Wm$^{-2}$ by the year 2100. 

Daily data are accessed directly from the CEDA archive. For \RS, \WS and \TA, output corresponds to daily averages over higher-frequency underlying model time-steps over the past 24 hours. For \WM, the data corresponds to the maximum over higher-frequency model time-steps over the past 24 hours. Annual maxima are then extracted for each calendar year, and annual arithmetic means calculated, for all 4 climate variables; annual minima are also extracted per calendar year for \TA. 

Then, for each combination of GCM, climate scenario and variable, \ed{we examine output for five climate model ensemble members (rX) where available; these correspond to a common initialisation (iX), physics (pX) and forcing (fX) per GCM,} and are listed in Table~\ref{Tbl:GcmDat}. For each combination of GCM, climate scenario, variable and ensemble member, we therefore have access to four time-series each of 86 values of annual maximum, mean and \TA minimum for the period 2015-2100, for each spatial grid location on the surface of the Earth.

\ed{The choice of data to analyse was entirely pragmatic. We accessed output from all GCMs available on the CEDA archive at the time of data harvesting in early 2024, which was sufficiently comprehensive in terms of climate variables, ensembles and scenarios.}
\newcommand\Tstrut{\rule{0pt}{2.6ex}} 
\begin{table}[!ht]
	\resizebox{\columnwidth}{!}{%
		\begin{tabular}{|c|l|l|l|l|}
			\hline
			\multicolumn{1}{|c|}{GCM\Tstrut} &
			\multicolumn{1}{|c|}{Variable\Tstrut} &
			\multicolumn{1}{|c|}{SSP126\Tstrut} &
			\multicolumn{1}{|c|}{SSP245\Tstrut} &
			\multicolumn{1}{|c|}{SSP585\Tstrut} \\ \hline  

			\multirow{4}{*}{
				\begin{tabular}{@{}c@{}}ACCESS-CM2 \\ (AC)\end{tabular}} &
			rsds\Tstrut &
			r1i1p1f1, r4i1p1f1, r5i1p1f1 &
			r1i1p1f1, r4i1p1f1, r5i1p1f1 &
			r1i1p1f1, r4i1p1f1, r5i1p1f1 \\	\cline{2-5} 	
			&
			sfcWind\Tstrut &
			r1i1p1f1, r2i1p1f1, r3i1p1f1, r4i1p1f1, r5i1p1f1 &
			r1i1p1f1, r2i1p1f1, r3i1p1f1, r4i1p1f1, r5i1p1f1 &
			r1i1p1f1, r2i1p1f1, r3i1p1f1, r4i1p1f1, r5i1p1f1 \\ \cline{2-5} 
			&
			sfcWindmax\Tstrut &
			r4i1p1f1, r5i1p1f1 &
			r4i1p1f1, r5i1p1f1 &
			r4i1p1f1, r5i1p1f1 \\ \cline{2-5} 
			&
			tas\Tstrut &
			r1i1p1f1, r2i1p1f1, r3i1p1f1, r4i1p1f1, r5i1p1f1 &
			r1i1p1f1, r2i1p1f1, r3i1p1f1, r4i1p1f1, r5i1p1f1 &
			r1i1p1f1, r2i1p1f1, r3i1p1f1, r4i1p1f1, r5i1p1f1 \\ \hline

			\begin{tabular}{@{}c@{}}CAMS-CSM1-0\Tstrut \\ (CA)\end{tabular} &
			tas &
			r2i1p1f1 &
			r2i1p1f1 &
			r2i1p1f1 \\ \hline
			
			\multirow{3}{*}{\begin{tabular}{@{}c@{}}CESM2 \\ (CE)\end{tabular}} &
			sfcWind\Tstrut &
			r4i1p1f1,   r10i1p1f1,r11i1p1f1 &
			r4i1p1f1,   r10i1p1f1,r11i1p1f1 &
			r4i1p1f1,   r10i1p1f1,r11i1p1f1 \\ \cline{2-5} 
			&
			tas\Tstrut &
			r4i1p1f1,   r10i1p1f1,r11i1p1f1 &
			r4i1p1f1,   r10i1p1f1,r11i1p1f1 &
			r4i1p1f1,   r10i1p1f1,r11i1p1f1 \\ \cline{2-5} 
			&
			rsds\Tstrut &
			r4i1p1f1,   r10i1p1f1,r11i1p1f1 &
			r4i1p1f1,   r10i1p1f1,r11i1p1f1 &
			r4i1p1f1,   r10i1p1f1,r11i1p1f1 \\ \hline
			
			\multirow{4}{*}{\begin{tabular}{@{}c@{}}EC-Earth3 \\ (EC)\end{tabular}} &
			rsds\Tstrut &
			r1i1p1f1,r4i1p1f1 &
			r1i1p1f1,r4i1p1f1 &
			r1i1p1f1,r4i1p1f1 \\ \cline{2-5}
			&		
			sfcWind\Tstrut &
			r1i1p1f1,r4i1p1f1 &
			r1i1p1f1,r4i1p1f1 &
			r1i1p1f1,r4i1p1f1 \\ \cline{2-5} 
			&
			sfcWindmax\Tstrut &
			r1i1p1f1,r4i1p1f1 &
			r1i1p1f1,r4i1p1f1 &
			r1i1p1f1,r4i1p1f1 \\ \cline{2-5} 
			&
			tas\Tstrut &
			r1i1p1f1,r4i1p1f1,   r9i1p1f1, r11i1p1f1 &
			r1i1p1f1,r4i1p1f1,   r9i1p1f1, r11i1p1f1 &
			r1i1p1f1,r4i1p1f1,   r9i1p1f1, r11i1p1f1 \\ \hline 
			
			\multirow{4}{*}{\begin{tabular}{@{}c@{}}MRI-ESM2-0 \\ (MR)\end{tabular}} &
			rsds\Tstrut &
			r1i1p1f1, r2i1p1f1,   r3i1p1f1, r4i1p1f1, r5i1p1f1 &
			r1i1p1f1, r2i1p1f1,   r3i1p1f1, r4i1p1f1, r5i1p1f1 &
			r1i1p1f1, r2i1p1f1,   r3i1p1f1, r4i1p1f1, r5i1p1f1 \\ \cline{2-5}			
			&
			sfcWind\Tstrut &
			r1i1p1f1, r2i1p1f1,   r3i1p1f1, r4i1p1f1, r5i1p1f1 &
			r1i1p1f1, r2i1p1f1,   r3i1p1f1, r4i1p1f1, r5i1p1f1 &
			r1i1p1f1, r2i1p1f1,   r3i1p1f1, r4i1p1f1, r5i1p1f1 \\ \cline{2-5} 
			&
			sfcWindmax\Tstrut &
			r1i1p1f1, r2i1p1f1,   r3i1p1f1, r4i1p1f1, r5i1p1f1 &
			r1i1p1f1, r2i1p1f1,   r3i1p1f1, r4i1p1f1, r5i1p1f1 &
			r1i1p1f1, r2i1p1f1,   r3i1p1f1, r4i1p1f1, r5i1p1f1 \\ \cline{2-5} 
			&
			tas\Tstrut &
			r1i1p1f1, r2i1p1f1,   r3i1p1f1, r4i1p1f1, r5i1p1f1 &
			r1i1p1f1, r2i1p1f1,   r3i1p1f1, r4i1p1f1, r5i1p1f1 &
			r1i1p1f1, r2i1p1f1,   r3i1p1f1, r4i1p1f1, r5i1p1f1 \\ \hline 
			
			\multirow{3}{*}{\begin{tabular}{@{}c@{}}NorESM2-MM \\ (No)\end{tabular}} &
			rsds\Tstrut &
			r1i1p1f1 &
			r1i1p1f1, r2i1p1f1 &
			r1i1p1f1 \\ \cline{2-5} 
			&
			sfcWind\Tstrut &
			r1i1p1f1 &
			r1i1p1f1, r2i1p1f1 &
			r1i1p1f1 \\ \cline{2-5} 
			&
			tas\Tstrut &
			r1i1p1f1 &
			r1i1p1f1, r2i1p1f1 &
			r1i1p1f1 \\ \hline
			
			\multirow{4}{*}{\begin{tabular}{@{}c@{}}UKESM1-0-LL \\ (UK)\end{tabular}} &
			rsds\Tstrut &
			r1i1p1f2, r2i1p1f2, r3i1p1f2, r4i1p1f2, r8i1p1f2 &
			r1i1p1f2, r2i1p1f2, r3i1p1f2, r4i1p1f2, r8i1p1f2 &
			r1i1p1f2, r2i1p1f2, r3i1p1f2, r4i1p1f2, r8i1p1f2 \\ \cline{2-5}			
			&
			sfcWind\Tstrut &
			r1i1p1f2, r2i1p1f2, r3i1p1f2, r4i1p1f2, r8i1p1f2 &
			r1i1p1f2, r2i1p1f2, r3i1p1f2, r4i1p1f2, r8i1p1f2 &
			r1i1p1f2, r2i1p1f2, r3i1p1f2, r4i1p1f2, r8i1p1f2 \\ \cline{2-5} 
			&
			sfcWindmax\Tstrut &
			r1i1p1f2, r2i1p1f2, r3i1p1f2, r4i1p1f2, r8i1p1f2 &
			r1i1p1f2, r2i1p1f2, r3i1p1f2, r4i1p1f2, r8i1p1f2 &
			r1i1p1f2, r2i1p1f2, r3i1p1f2, r4i1p1f2, r8i1p1f2 \\ \cline{2-5} 
			&
			tas\Tstrut &
			r1i1p1f2, r2i1p1f2, r3i1p1f2, r4i1p1f2, r8i1p1f2 &
			r1i1p1f2, r2i1p1f2, r3i1p1f2, r4i1p1f2, r8i1p1f2 &
			r1i1p1f2, r2i1p1f2, r3i1p1f2, r4i1p1f2, r8i1p1f2 \\ \hline
		\end{tabular}%
	}
	\caption{Summary of global coupled model (GCM) output considered. A total of 7 GCMs are used, listed in alphabetical order together with two-letter acronym (column 1). For each GCM, up to four climate variables are used, depending on their availability (column 2), for each of three climate scenarios (row 1, columns 3-5). A total of up to 5 ensemble members are considered for each combination of climate variable and scenario, again depending on availability (columns 3-5).}
	\label{Tbl:GcmDat}
\end{table}

\FloatBarrier
\subsection{Compilation of spatial summaries: global and climate zone data}  \label{Sct:Dat:GlbClmZon}
%
We consider analysis of spatial maxima, minima and means of corresponding location-specific annual maximum, minimum and mean data, for spatial domains corresponding to the whole globe and its partition into 5 climatic zones. We also present analysis of annual maximum data for specific locations in the \NA and the \CS neighbourhood; see Section~\ref{Sct:Dat:NACS}. \ed{Analysis of spatial extremes and of spatial means across multiple models and scenarios provides a more comprehensive assessment of future GCM-predicted climate variability that typically considered for offshore design.}

Global annual maximum data are calculated over the complete set of global grid locations $\mathcal{G}$ available for each GCM. Thus, given annual maximum output $\{y(i,j)\}$ for year $i=1, 2, ..., 86$ at grid location $j \in \mathcal{G}$, the global (spatial) maximum time series $\{y_{G,\max}(i)\}$, $i=1, 2, ..., 86$ is extracted from the location-specific maximum using
\begin{eqnarray}
	y_{G,\max}(i) = \maxover{j \in \mathcal{G}}\{y(i,j)\} .
	\label{Eqt:Clc-GlbMxm}
\end{eqnarray}
In a similar fashion, we also calculate the annual maximum for each of the Antarctic, Temperate South, Tropical, Temperate North and Arctic climate zones, defined by the latitudinal intervals $(-90^\circ, -66.5^\circ)$, $(-66.5^\circ, -23.5^\circ)$, $(-23.5^\circ, 23.5^\circ)$, $(23.5^\circ, 66.5^\circ)$ and $(66.5^\circ, 90^\circ)$ respectively. This calculation is made using Equation~\ref{Eqt:Clc-GlbMxm} (with $\mathcal{G}$ replaced by $\mathcal{G}_k$), for the appropriate choice of set $\mathcal{G}_k$ of climate zone locations, $k=1, 2, ..., 5$. For variable \TA, it is also interesting to consider global and climate zone annual minimum time-series, which are calculated in the analogous fashion from location-specific annual minimum data.

\begin{figure}[!ht]
	\centering
	\includegraphics[width=0.6\textwidth]{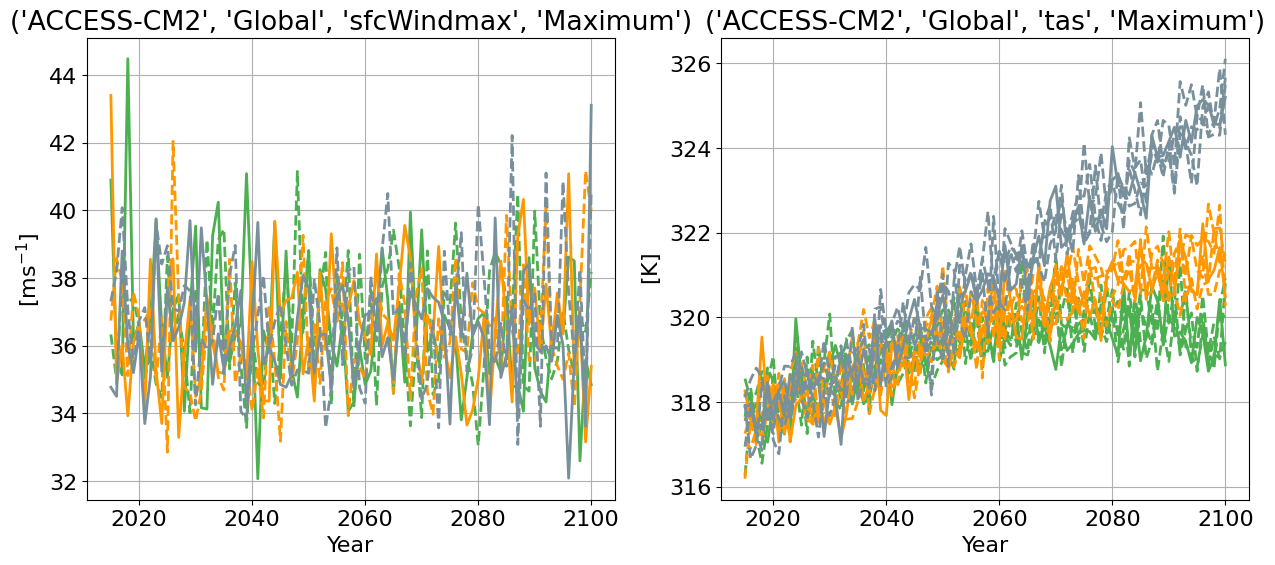}
	\caption{Illustrations of global annual maxima time-series for \WM (left) and \TA (right) from \AC, for different climate scenarios (\SL, green; \SM, orange; \SH, grey). Line styles distinguish different ensemble members.}
	\label{Fgr-Glb-TimSrs-WMAC-TAAC}
\end{figure}

Illustrations of global annual maximum time-series for 5 ensemble members of \WM and \TA from \AC are given in the left and right hand panels of Figure~\ref{Fgr-Glb-TimSrs-WMAC-TAAC}. Colours distinguish climate scenarios: \SL (green), \SM (orange) and \SH (grey); and line styles distinguish individual ensemble member output. 
It is apparent that the effect of climate scenario is more pronounced for \TA than for \WM. There is some evidence also under scenario \SL, that global maximum \TA asymptotes to a constant value with time (to around 320K for the \AC GCM illustrated in the right hand panel of Figure~\ref{Fgr-Glb-TimSrs-WMAC-TAAC}), and that perhaps \TA under scenario \SM asymptotes also (to around 322K); however similar trends are rarely seen for other climate variable, or for \TA under \SH for the period (2015,2100). Supporting illustrations for all other combinations of climate variables and GCMs are provided in Figure~SM1 (of the SM); this figure also provides illustrations of generally reducing trends in global annual maximum \RS, and increasing trends in global \ed{annual} minimum of \TA. Again, trends with time or the effect of climate scenario are difficult to discern for \WS and \WM relative to \TA (in particular) and \RS.

We also calculate climate zone and global annual mean time-series from the corresponding location-specific annual mean data. To simplify the calculation per climate zone, we assume it reasonable to approximate the climate zone mean using the arithmetic mean over locations within the climate zone. With $\{y(i,j)\}$, $i=1, 2, ..., 86$, $j \in \mathcal{G}_k$, $k=1, 2, ..., 5$ now representing location-specific annual mean time-series per climate zone, the climate zone annual mean time-series is calculated using
\begin{eqnarray}
	y_{G_k, \text{mean}}(i) = \meanover{j \in \mathcal{G}_k}\{y(i,j)\} .
	\label{Eqt:Clc-ClmZonMen}
\end{eqnarray}
We then estimate the global mean as a weighted average of climate zone means, with weight equal to the surface area $\mathcal{A}_k$, $k=1, 2, ..., 5$ of the climate zone, using
\begin{eqnarray}
	y_{G, \text{mean}}(i) = \sum_k \mathcal{A}_k \ y_{G_k, \text{mean}}(i) \quad / \quad \sum_{k'} \mathcal{A}_{k'} .
	\label{Eqt:Clc-GlbMen}
\end{eqnarray}
Each climate zone surface area $\mathcal{A}_k$ is estimated assuming a spherical Earth. The resulting time-series of global annual mean data for the four climate variables are shown in Figure~SM2. Obvious time trends and scenario effects are visible for \TA; corresponding effects for \RS are clearer for global annual means than for global annual maxima (in Figure~SM1). Regardless of GCM, \RS exhibits a positive slope in time under scenario \SL, with more negative slope for \SM and especially for \SH; this results in a negative gradient for \RS under \SH for three GCMs. \WS and \WM again show much weaker trends in time and with scenario; no scenario effect is discrenable, but both \AC and \UK GCMs (which share some modelling components) suggest a minor increase in \WS and \WM of approximately 0.1ms$^{-1}$ over the 86 years. In contrast, \EC suggests that \WS and \WM reduce by the same amount in general over the same period.

\begin{figure}[!ht]
	\centering
	\includegraphics[width=1\textwidth]{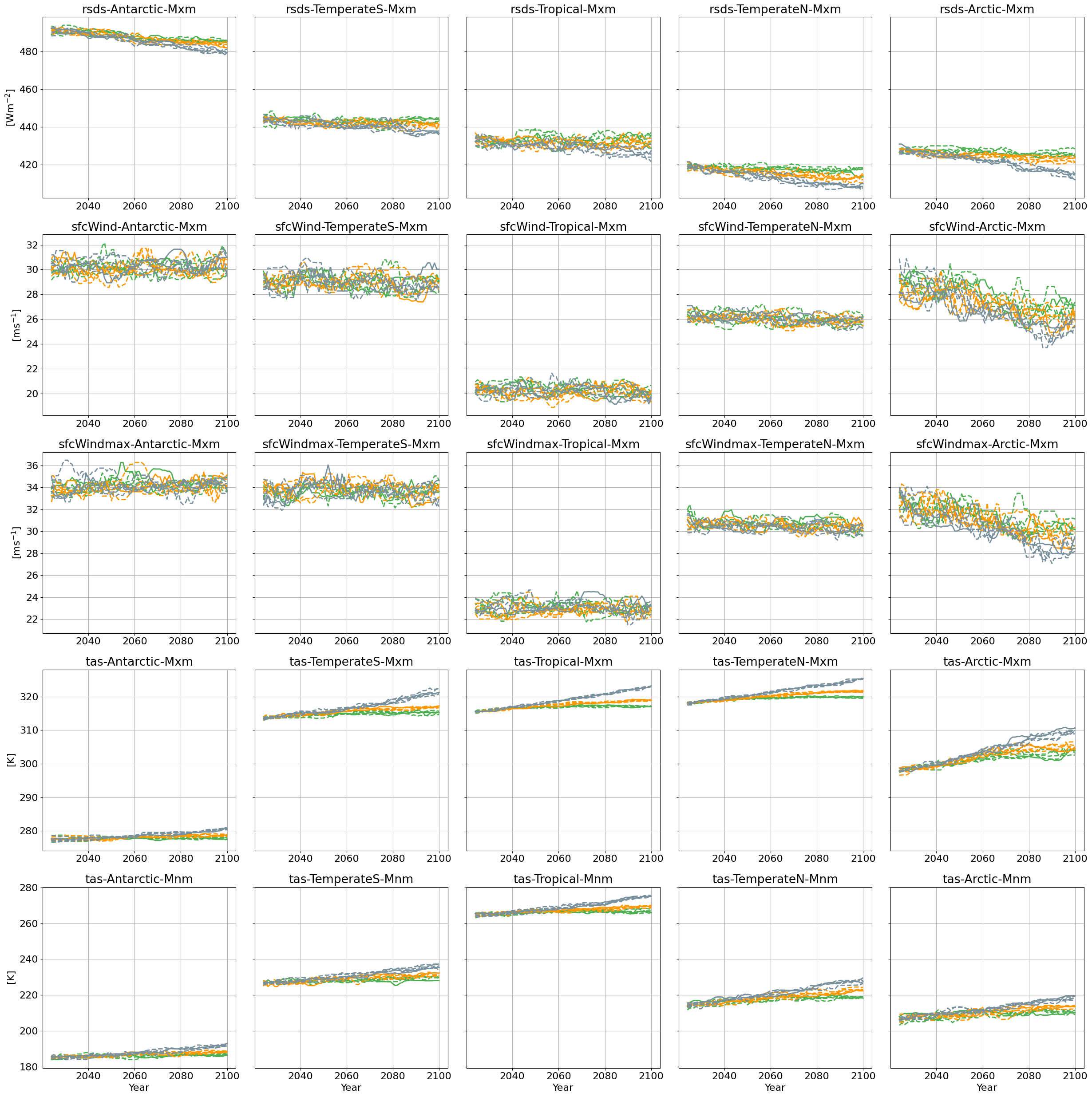}
	\caption{Time-series plots of smoothed annual maximum \RS, \WS, \WM and \TA (rows 1-4) and annual minimum \TA (row 5), for each of 5 climate zones (columns), for \UK. Colour indicates climate scenario (\SL, green; \SM, orange; \SH, grey) with different line style for each ensemble member. Smoothing performed using moving 10 year median filter. \ed{``Mxm'' and ``Mnm'' in titles refer to maximum and minimum respectively.}}
	\label{Fgr-ClmZon-TS-UK-MxmMnm-Smt}
\end{figure}

Figure~\ref{Fgr-ClmZon-TS-UK-MxmMnm-Smt} illustrates time-series of annual maximum and minima data for the climate variables (rows) across climate zones (columns) for the \UK GCM. To emphasise trends here, data have been smoothed using a moving $\pm$ 10-year median filter. Rows 1-4 represent annual maxima for \RS, \WS, \WM and \TA, and row 5 annual minima of \TA. Columns represent the Antarctic, Temperate South, Tropical, Temperate North and Arctic climate zones. Within each panel, colour distinguishes scenarios, and line style ensemble members, as described in the figure caption.
Obvious differences in level are apparent across climate zones for all climate variables. Temporal trends are clearer for \TA then for \RS than for \WS and \WM, although there are stronger time trends for \WS and \WM in the Arctic zone. Differences due to climate scenario are clearer for \TA and then for \RS than for \WS and \WM. \RS shows a generally decreasing level across climate zones from the Antarctic. As would be expected due to reduced Coriolis effect, \WS and \WM are smaller in the Tropical zone. For \TA, similar trends with climate zone are seen for both annual maxima and annual minima. The corresponding data (without smoothing) are illustrated in Figures~SM3-SM9 for each of the seven GCMs in turn, for comparison, showing generally consistent features. 

Note however that \MR (Figure~SM7) offers dubious values ($\gg$ 100ms$^{-1}$) for \WM in the Temperate North, considered unreasonable. For this reason, \MR data for \WM is not admitted into the analysis. Note also the large growth in \WS and \WM under scenario \SL for \EC in the Tropical zone (Figure~SM6); these values ($\approx$ 35ms$^{-1}$) are not unusual compared to values in other zones, and for this reason they are retained for the analysis. Note also a suspect value in \texttt{r1i1p1r1} for \No \TA minimum for scenario \SM in the Antarctic (and hence Global), see Figures~SM\ed{8} and SM\ed{1} for illustration: we have ignored this run for analysis of annual minima, but included for analysis of means and maxima (since the offending point is of negligible influence).

Regarding the trends observed in \RS across climate zones, physically we know that the Antarctic typically has clearer skies, especially during the summer months, whereas the Arctic often has more persistent cloud cover. Further, during the respective summer seasons, the sun angle in the Antarctic can be slightly higher on average due to the Earth's tilt. Moreover, the Antarctic ozone hole can lead to higher UV radiation levels reaching the surface during the spring and early summer. Downward trends with time per climate zone reflect increase concentrations of aerosols. For \TA, lower values in the Antarctic compared to the Arctic occur due to the moderating influence of oceans in the latter.

In summary, climate scenario effects on wind speed variables are less pronounced than on \RS and \TA, with the exception of the Arctic zone; physically, perhaps this is related to the occurrence of polar lows and cyclonic systems there, but it not clear why these effects would also influence the Temperate North, at least to some extent. 

\ed{Figures~SM10-SM16 show corresponding climate zone annual mean time-series for the seven GCMs in turn, for further comparison. Again, there is good general consistency across GCMs. However, trends across climate zones for annual means are quite different to those for annual maxima; e.g. compare \RS across climate zones in Figures~SM9 and SM16. This suggests that the shape of the distribution of climate zone annual \RS in particular changes with climate zone.}

\begin{figure}[!ht]
	\centering
	\includegraphics[width=1\textwidth]{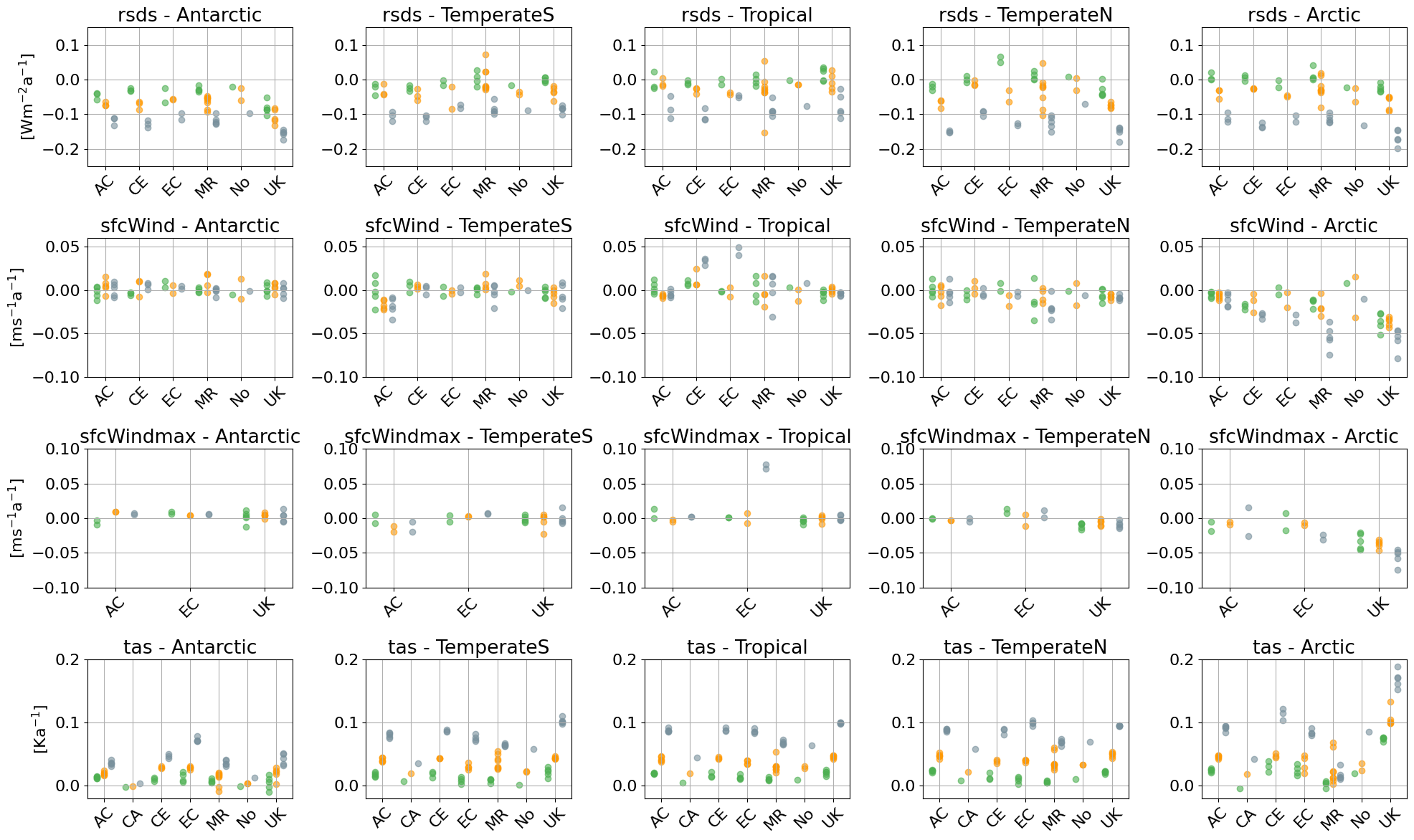}
	\caption{Slopes of linear regression lines fitted directly to the annual maximum data for each climate variable, given climate zone, GCM, climate scenario and ensemble member. Rows indicate different climate variables. Columns indicate climate zone. \ed{Discs indicate slopes for (multiple) ensemble members: disc colour indicates climate scenario (\SL, green; \SM, orange; \SH, grey), and disc opacity distinguishes individual ensemble members given GCM.} Units of slopes are rates of change of value per annum.}
	\label{Fgr-ClmZon-Mxm-LnrRgrTrn}
\end{figure}

In an attempt to summarise temporal trends and scenario trends for all GCMs and climate zones concisely, Figure~\ref{Fgr-ClmZon-Mxm-LnrRgrTrn} reports the linear trend slope found in the annual maximum data calculated for each combination of climate variable, ensemble member, climate zone and GCM, from a linear regression analysis. We omit uncertainty bars on slope estimates for clarity. Note that the figure summarises the change in value of a climate variable within a climate zone, rather than the general level of a climate variable in that zone (for which illustrations like those in Figure~\ref{Fgr-ClmZon-TS-UK-MxmMnm-Smt} and Figures~SM3-9 are more useful). Note also that the fitting of a linear model is performed only to provide a statistic to summarise temporal trends. More suitable approaches to model fitting for annual extremes and annual means are discussed in Section~\ref{Sct:Mth}. 
Again, clearest trends with scenario are observed for \TA and then for \RS, particularly in the Arctic zone. Here, for \TA under \SH, some GCMs indicate a rate of increase of temperature of $>0.1$K per annum; the \UK GCM appears to produce particularly high annual rates of temperature increase. The Arctic zone also provides the clearest trends with scenario for \WS and \WM, with a small decrease in wind speeds (of $\approx$ 0.05ms$^{-1}$a$^{-1}$) with increased climate forcing. Outside the Arctic, scenario trends in \WS and \WM are negligible. As noted previously, \MR GCM output is suspect for \WM in the Temperate North zone and not admitted into the analysis. Corresponding linear slope estimates for climate zone minima of \TA and climate zone means for all climate variables are summarised in Figures~SM17 and SM18. Trends in \WS and \WM are negligible in magnitude, but again it is interesting to observe e.g. consistent downward (upwards) trends with increasing climate forcing in the Temperate North (Arctic) across GCM.

\FloatBarrier
\subsection{North Atlantic and Celtic Sea data}  \label{Sct:Dat:NACS}
\begin{figure}[!ht]
	\centering
	\includegraphics[width=0.7\textwidth]{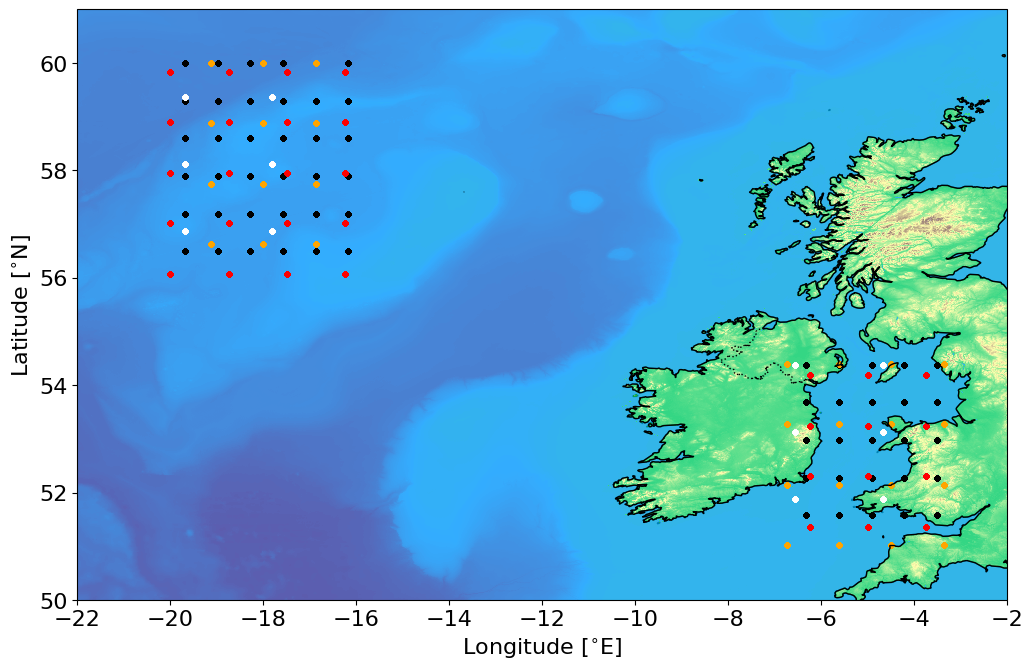}
	\caption{Discs showing grid locations of CMIP6 data accessed in the North Atlantic and the neighbourhood of the Celtic Sea. Disc colours represent climate model grids: \AC and \UK (white), \CA and \MR (orange), \CE and \No (red), \EC (black). Background bathymetry-topography data accessed from \cite{GEBCO}; blue bathymetry palette on [-5000,0]m, and green-yellow topography palette on [0,1500]m.}
	\label{Fgr-NACS-Grd}
\end{figure}

%
From an engineering perspective, it is interesting to examine the impact of climate change on specific locations, as well as more globally. For this reason, we also undertake statistical analysis of location-specific annual maxima of \RS, \WS, \WM and \TA for two neighbourhoods of locations in the North Atlantic and in the vicinity of the Celtic Sea, illustrated in Figure~\ref{Fgr-NACS-Grd}. The North Atlantic neighbourhood corresponds to one of the stormiest regions on the planet, whereas the Celtic Sea neighbourhood encompasses sheltered locations in shallow water and on land. We might expect therefore that the characteristics of extreme events would be materially different in the two neighbourhoods, despite their geographic proximity.
In the figure, note that GCMs use different spatial grids in general, but that some GCMs share a common grid; for the 7 GCMs considered here, in fact only 4 different spatial grids are employed, with GCM pairs \AC and \UK, \CA and \MR, and \CE and \No using common grids.

Illustrations of time-series of annual data for centre locations of the North Atlantic and Celtic Sea neighbourhoods considered, for the four climate variables from each of the seven GCMs, \ed{are} given in Figures~SM19-20. For both neighbourhoods, there are no obvious temporal or scenario trends visible of variables other than \TA. Even for \TA, there is no consistency in trends across GCM in the North Atlantic; indeed, \MR and \No suggest decreasing trends in time under scenarios \SL and \SM. In the \CS however, trends for \TA across GCMs appear more similar.

\FloatBarrier
\section{Methodology}  \label{Sct:Mth}
%
We consider two statistical models in the current work. The first uses generalised extreme value regression (GEVR) to estimate models for (temporal and spatio-temporal) block maxima (BM) of climate variables, and is outlined in Section~\ref{Sct:Mth:GEVR}. The second uses non-homogeneous Gaussian regression (NHGR) to estimate models for (temporal and spatio-temporal) means of climate variables, outlined in Section~\ref{Sct:Mth:NHGR}. For both GEVR and NHGR, temporal blocks correspond to years, and spatial blocks to the whole globe or to one of five climate zones, as set out in Section~\ref{Sct:Dat:GlbClmZon}. For both models, following \cite{EwnJnt23a}, we assume that all model parameters vary linearly over the period of observation, unless stated otherwise. That is, for any model parameter $\eta$, we assume that
\begin{eqnarray}
	\eta_t = \eta(t) = \eta^0 + \frac{t-2015}{P-1} \eta^1, \text{   for   } t=2015, 2016, ...
	\label{Eqt:CvrTim}
\end{eqnarray}
in year $t$, where $\eta^0$ is the parameter value at the start year (i.e. $t=2015$), and $\eta^1$ is the change in $\eta$ over the period (2015, 2100) of length $P=86$ years to be estimated. Further, we assume access to a sample of annual observations $\{x_i\}_{i=1}^P$ of events from $X_{t}$ for analysis, for different choices of CMIP6 GCM output.

\FloatBarrier
\subsection{Non-stationary generalised extreme value regression}  \label{Sct:Mth:GEVR}
%
Asymptotic extreme value theory (e.g. \citealt{Cls01}, \citealt{JntEwn13}) shows that block maxima of random draws from a max-stable distribution follow the GEV distribution for sufficiently large block length. We therefore assume that the sample $\{x_i\}_{i=1}^P$ of (temporal or spatio-temporal) BM $X_{t}$ ($t=1, 2, 3, ..., P$) follows the GEV distribution with non-stationary location parameter $\mu_t \in \mathbb{R}$, scale $\sigma_t>0$, shape $\xi_t \in \mathbb{R}$ and log density function
\begin{align}
	\log f_{\text{GEVR}}(x|\mu_t, \sigma_t, \xi_t) =
	\begin{cases} 
		- \left[ 1 + (\xi_t/\sigma_t)\left(x-\mu_t\right)\right]^{-1/\xi_t}  &\xi_t \neq 0\\
		\exp(-(x-\mu_t)/\sigma_t) &\xi_t = 0 .
	\end{cases}
	\label{Eqt:GEVR:Dns}
\end{align}
Model parameters $\eta_t \in \{\mu_t, \sigma_t, \xi_t\}$ vary with $t$ as described in Equation~\ref{Eqt:CvrTim}. The $T$-year return value $Q_t$ for year $t$ is estimated as the $p=1-1/T$ quantile of the corresponding GEVR distribution function $F_{GEVR}(x;t)$, so that
\begin{align}
	Q_t =
	\begin{cases} 
		(\sigma_t/\xi_t) \left[ \left(-\log p\right)^{-{\xi}_t} -1\right] + {\mu}_t  &\xi_t \neq 0\\
		\mu_t-\sigma_t \log\left(-\log p\right) &\xi_t = 0 .
	\end{cases}
	\label{Eqt:GEVR:RV}
\end{align}
Note that we only consider the case $T=100$ years in the current work for definiteness.

Parameter estimation is performed using Bayesian inference as described in \ref{Sct:App:BysInf}, yielding a sample of $n_I$ joint posterior estimates $\{\hat{\mu}^0_{(k)},\hat{\mu}^1_{(k)},\hat{\sigma}^0_{(k)},\hat{\sigma}^1_{(k)},\hat{\xi}^0_{(k)},\hat{\xi}^1_{(k)}\}_{k=1}^{n_I}$ where $n_I=10,000$. These can be used to estimate the empirical distribution of quantities of interest, such as return values $Q_{2025}$, $Q_{2125}$, and the difference 
\begin{eqnarray}
	\Delta^Q = Q_{2125}-Q_{2025}
	\label{Eqt:DeltaQ}
\end{eqnarray}
in return value over the 100 years from 2025 to 2125. Discussion and illustration of estimated posterior cumulative distribution functions of parameters, $Q_{2025}$, $Q_{2125}$ and $\Delta^Q$ is provided in Section~\ref{Sct:Rsl} below, in panels (a) of Figures~SM21-24, and in Figure~SM25. Note that extreme value analysis for annual minimum data for \TA (over some spatial domain) is performed via characterisation of the right hand tail of the distribution of negated data.

\FloatBarrier
\subsection{Non-homogeneous Gaussian regression}  \label{Sct:Mth:NHGR}
%
\ed{We expect mean values of climate variables to vary slowly and smoothly with time, and that a Gaussian distribution might be appropriate to characterise fluctuations in mean GCM output relative to an underlying smooth trend. However, we also might anticipate that the extent of these fluctuations may itself vary in time. Given these expectations, we} assume that the sample $\{x_i\}_{i=1}^P$ of (temporal or spatio-temporal) means $X_{t}$ ($t=1, 2, 3, ..., P$) follows a non-homogeneous Gaussian distribution with non-stationary mean parameter $\alpha_t \in \mathbb{R}$, scale $\beta_t>0$ and log density function
\begin{align}
	\log f_{\text{NHGR}}(x|\alpha_t, \beta_t) =
		- \log\left(2 \pi \beta_t^2\right)/2
		- \left(x-\alpha_t\right)^2/(2\beta_t^2) .
	\label{Eqt:NHGR:Dns}
\end{align}
Model parameters $\eta_t \in \{\alpha_t, \beta_t\}$ again vary with $t$ as described in Equation~\ref{Eqt:CvrTim}. Using the Bayesian inference scheme outlined for GEVR, we are able to estimate a chain of posterior estimates $\{\hat{\alpha}^0_{(k)},\hat{\alpha}^1_{(k)},\hat{\beta}^0_{(k)},\hat{\beta}^1_{(k)}\}_{k=1}^{n_I}$ for parameters with which to estimate empirical distributions of any quantities of interest, such as the change in mean $(100/86)\alpha_1$ over any 100-year period, and specifically for the period (2025, 2125). The predicted value of $X_{t}$ in year $t$ is given by 
\begin{align}
	M_t \sim N(\alpha_t, \beta_t^2) .
	\label{Eqt:NHGR:PrdVal}
\end{align}
Discussion and illustration of estimated posterior cumulative distribution functions of model parameters and the change 
\begin{eqnarray}
	\Delta^M = M_{2125}-M_{2025}
	\label{Eqt:DeltaM}
\end{eqnarray}
in mean over the period (2025, 2125) is provided in Section~\ref{Sct:Rsl} below and panels (b) of Figures~SM21-24.\\

A synoptic analysis (using linear mixed effects models) and discussion of findings for variation of both $\Delta^Q$ and $\Delta^M$ with climate variable, climate scenario, climate model and its consistent ensembles is provided in Section~\ref{Sct:DscCnc} and Section~SM6.

\FloatBarrier
\section{Results}  \label{Sct:Rsl}
%
In this section, we apply the GEVR methodology from Section~\ref{Sct:Mth:GEVR} to estimate changes in 100-year return values over the period (2025,2125) for samples of spatio-temporal extremes (e.g. global or climate zone annual maxima or annual minima) for climate variables corresponding to different choices of GCM, climate scenario and ensemble member. We also apply the NHGR methodology from Section~\ref{Sct:Mth:NHGR} to estimate changes in annual global and climate zone means over the same period. Inferences for global and climate zone annual means and extremes are interesting to characterise risk over large spatial scales. However, metocean design is usually performed for specific locations. For this reason we also consider GEVR inferences for annual maxima from individual locations in the neighbourhoods of the North Atlantic and Celtic Sea. As outlined in Section~\ref{Sct:Mth} and \ref{Sct:App:BysInf}, all GEVR and NHGR inferences are made using Markov chain Monte Carlo (MCMC). The section is structured as follows. In Section~\ref{Sct:Rsl:Glb}, we consider inferences for global maxima, means and minima. In Section~\ref{Sct:Rsl:ClmZon}, inferences for maxima, means and minima partitioned by climate zone are discussed. Finally, in Section~\ref{Sct:Rsl:NACS}, results for annual maxima at multiple individual locations in the North Atlantic and Celtic Sea neighbourhood are compared.

\FloatBarrier
\subsection{Global}  \label{Sct:Rsl:Glb}
\begin{figure}[!ht]
	\centering
	\includegraphics[width=1\textwidth]{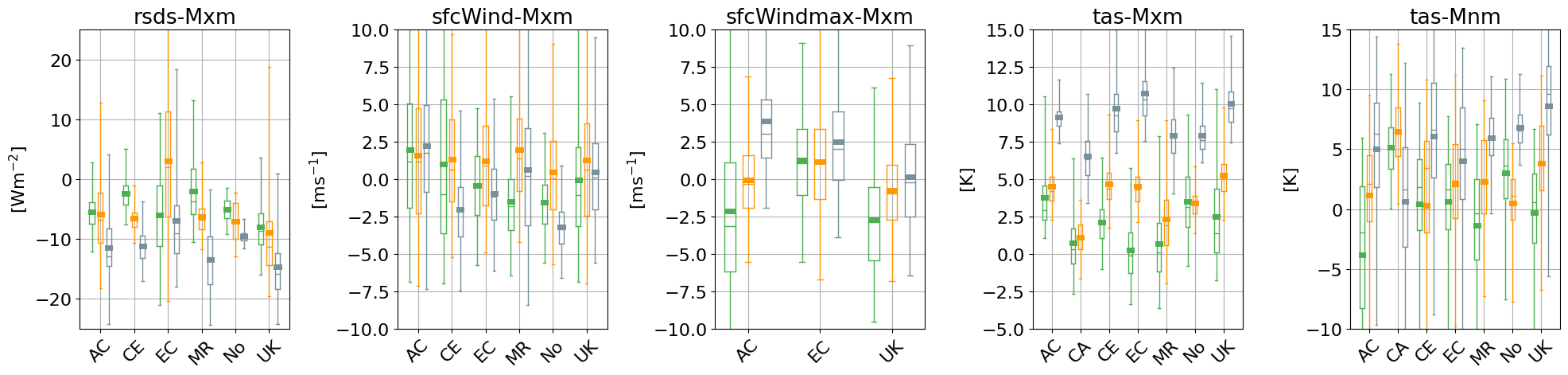}
	\caption{Box-whisker plots for the posterior distribution of the change $\Delta^Q$ in 100-year return value $Q_t$ of global annual maximum and minima over the next 100-years, for each of \AC, \CE, \EC, \MR, \No and \UK GCMs; the first two characters of the GCM name are used for concise labelling. Columns 1-4 correspond to change $\Delta^Q$ for global annual maxima of \RS, \WS, \WM and \TA. Column 5 corresponds to change $\Delta^Q$ for global annual minima of \TA; ``Mxm'' and ``Mnm'' in titles refer to maximum and minimum respectively. For each GCM, colour distinguishes climate scenarios (\SL, green; \SM, orange; \SH, grey). For each box-whisker, the box represents the central (25\%, 75\%) interval for the posterior distribution, and the whisker the central (2.5\%, 97.5\%) interval. The posterior median is shown as a thin horizontal line (of the same width as the box), and the posterior mean as a thicker horizontal line. Vertical extents of panels have been restricted so that all central (25\%, 75\%) intervals are relatively clear. Relevant samples from posterior distributions for each of multiple ensembles are aggregated to estimate box-whisker structures.}
	\label{Fgr-Glb-BW-MxmMnm}
\end{figure}

%
Here we consider GEVR analysis of annual global maximum and minimum data, and NHGR analysis of annual global mean data, prepared from the underlying location data using the approach described in Section~\ref{Sct:Dat:GlbClmZon}. We summarise results in terms of box-whisker plots for the posterior distribution of the change in 100-year return value $\Delta^Q$, aggregated over all relevant climate ensemble members, for each climate variable over the period (2025, 2125), for different climate scenarios. Results are illustrated in Figure~\ref{Fgr-Glb-BW-MxmMnm}.
For \RS, the GCMs generally indicate a small decrease in the 100-year return value of the global annual maximum over next 100 years. Generally, there is also some indication that $\Delta^Q$ reduces with increased climate forcing. For \WS, all central (25\%, 75\%) box intervals for \SL and \SM scenarios include zero, suggesting no strong evidence for a change in return value. For the \SH scenario, the \CE and \No GCMs in particular provide stronger evidence for a small reduction in return value of around 3ms$^{-1}$ (but the central (2.5\%, 97.5\%) whisker intervals nevertheless include zero). For \WM, there is weak evidence for a trend of increasing change in $\Delta^Q$ with climate forcing, of the order of 5ms$^{-1}$. The least ambiguous inferences can be drawn for global annual maxima for \TA in the fourth panel, which generally show an increasing trend in $\Delta^Q$ for all GCMs, with values of around 9K under scenario \SH. Results for global annual minima for \TA in the last panel are noisier, but generally similar: the figure indicates that the change in the 100-year return value for annual minimum of \TA is generally positive with increasing climate forcing; that is, extremely low temperatures occur less often in 2125 than 2025. In summary, results for \TA are considerably clearer than those for other climate variables. For \WS and \WM, of most interest for offshore and coastal design, results are somewhat contradictory: there is some evidence that return values for \WS reduce under \SH, whereas there is more consistent evidence across GCMs that \WM increases with climate forcing. 

\begin{figure}[!ht]
	\centering
	\includegraphics[width=0.8\textwidth]{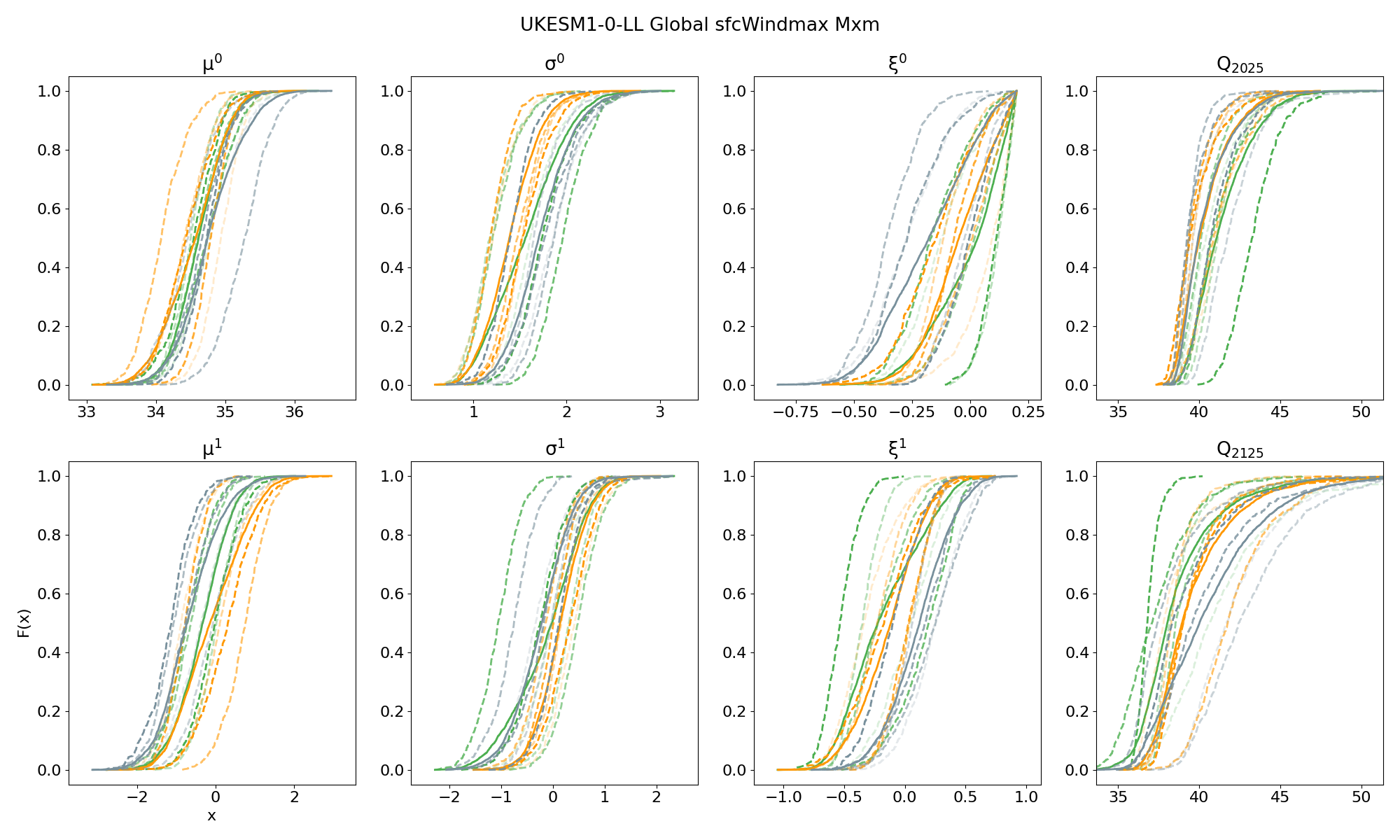}
	\caption{Estimated posterior cumulative distribution functions for GEVR model parameters and 100-year return values ($Q_{2025}$, $Q_{2125}$, for years 2025 and 2125), for the global annual maximum of \WM from \UK. Colours indicate climate scenario (\SL, green; \SM, orange; \SH, grey). Dashed line styles indicate inferences for different ensemble members; a solid line corresponds to the ensemble mean cumulative distribution function. Model parameters are the GEV location $\mu^0$, scale $\sigma^0$ and shape $\xi^0$ in 2015, and the changes $\mu^1$, $\sigma^1$ and $\xi^1$ in those parameters over the period (2015,2100).}
	\label{Fgr-Glb-Cdf-UK-WM-Mxm}
\end{figure}

\begin{figure}[!ht]
	\centering
	\includegraphics[width=0.8\textwidth]{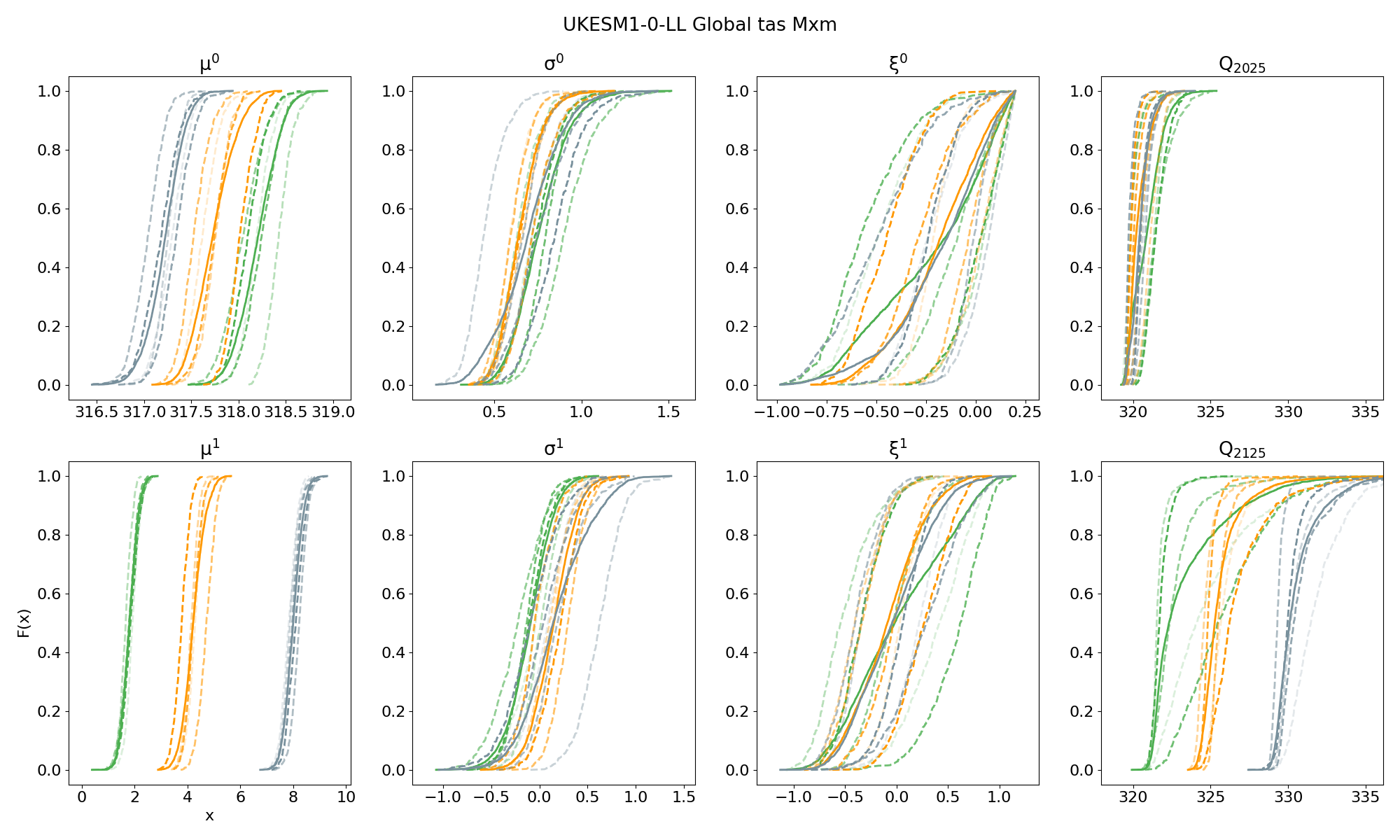}
	\caption{Estimated posterior cumulative distribution functions for GEVR model parameters and 100-year return values ($Q_{2025}$, $Q_{2125}$, for years 2025 and 2125), for the global annual maximum of \TA from \UK. For other details, see Figure~\ref{Fgr-Glb-Cdf-UK-WM-Mxm}.}
	\label{Fgr-Glb-Cdf-UK-TA-Mxm}
\end{figure}

To understand the nature of the change in the tail of the distribution of global annual maxima further, Figures~\ref{Fgr-Glb-Cdf-UK-WM-Mxm} and \ref{Fgr-Glb-Cdf-UK-TA-Mxm} show estimated posterior cumulative distribution functions for the parameters of the GEVR models (see Section~\ref{Sct:Mth:GEVR}) for \WM and \TA from the \UK GCM. 
In Figure~\ref{Fgr-Glb-Cdf-UK-WM-Mxm}, it is difficult to discern any systematic trends between climate scenarios; indeed the variability between inferences for different ensemble members for the same scenario appears to be of the same order as that for different scenarios. In Figure~\ref{Fgr-Glb-Cdf-UK-TA-Mxm} however, the effect of scenario on $\mu^1$ is most obvious; this indicates a change of around 2K in the 100-year return value for global annual \TA over the period (2015,2100) under scenario \SL, rising to around 4K under \SM and 8K under \SH. That is, changes in the general level of \TA lead to clear differences in $Q_{2125}$ shown in the bottom right panel. In contrast, there is considerably less evidence for clustering with respect to scenario for variables $\sigma$ and $\xi$, responsible for the shape of the distributional tail. That is, there is little evidence in the data that the shape of the distributional tail changes in time. We also note a change in $\mu^0$ (the value of GEV ``level'' in 2015, with posterior medians varying by around 1K from around 317.2k to 318.2K under the different scenarios) which provides a minor counter-effect for the change in $\mu^1$. 

\begin{figure}[!ht]
	\centering
	\includegraphics[width=1\textwidth]{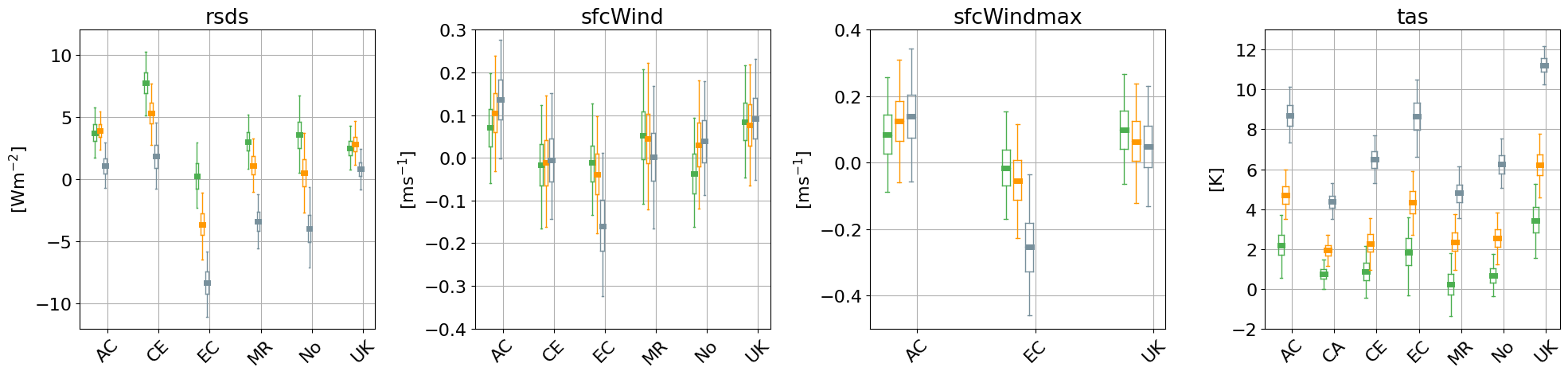}
	\caption{Box-whisker plots for the change $\Delta^M$ in global mean  over the next 100-years, for each of the four climate variables. For details, see Figure~\ref{Fgr-Glb-BW-MxmMnm}.}
	\label{Fgr-Glb-BW-Men}
\end{figure}

Figure~\ref{Fgr-Glb-BW-Men} summarises the change $\Delta^M$ in the global mean of each of the four climate variables over the next 100 years. Results reflect data characteristics mentioned in Section~\ref{Sct:Dat}. For \RS, there is agreement across GCMs that increased climate forcing will reduce mean value. Findings for \TA are consistent across GCM. Relative to the size of change indicated by \TA, changes in global mean \WS and \WM are very small ($\ll$1ms$^{-1}$ in magnitude), with some evidence for minor climate scenario effect. 

Supporting Figures~SM21-24 (panels (b)) provide estimated posterior cumulative distribution functions for parameters of NHGR models for all four climate variables from \UK GCM. \ed{(Note that diagnostic plots were produced for all analyses undertaken. The UKESM1-0-LL GCM is used as a convenient exemplar here.)} \ed{P}anels (a) of the same figures report the corresponding estimates for GEVR parameters. The most obvious feature of these figures for GEVR models of global annual maxima and minima, is again evidence for a scenario-related change in level $\mu^1$ (rather than for tail parameters $\sigma^1$ and $\xi^1$), for \RS and \TA. For NHGR models of global mean data, evidence for scenario-related change in level $\alpha^1$ is present for both \RS and \TA. For the latter, there is also evidence for change in standard deviation $\beta^1$. Smaller competing changes in $\alpha^0$ and $\beta^0$ are present, particular for \TA, suggesting that the modelling assumption of a linear trend in NHGR parameters with time may not always be reasonable for \TA.

\FloatBarrier
\subsection{Climate zones}  \label{Sct:Rsl:ClmZon}
\begin{figure}[!ht]
	\centering
	\begin{subfigure}{1\textwidth}
		\includegraphics[width=1\textwidth]{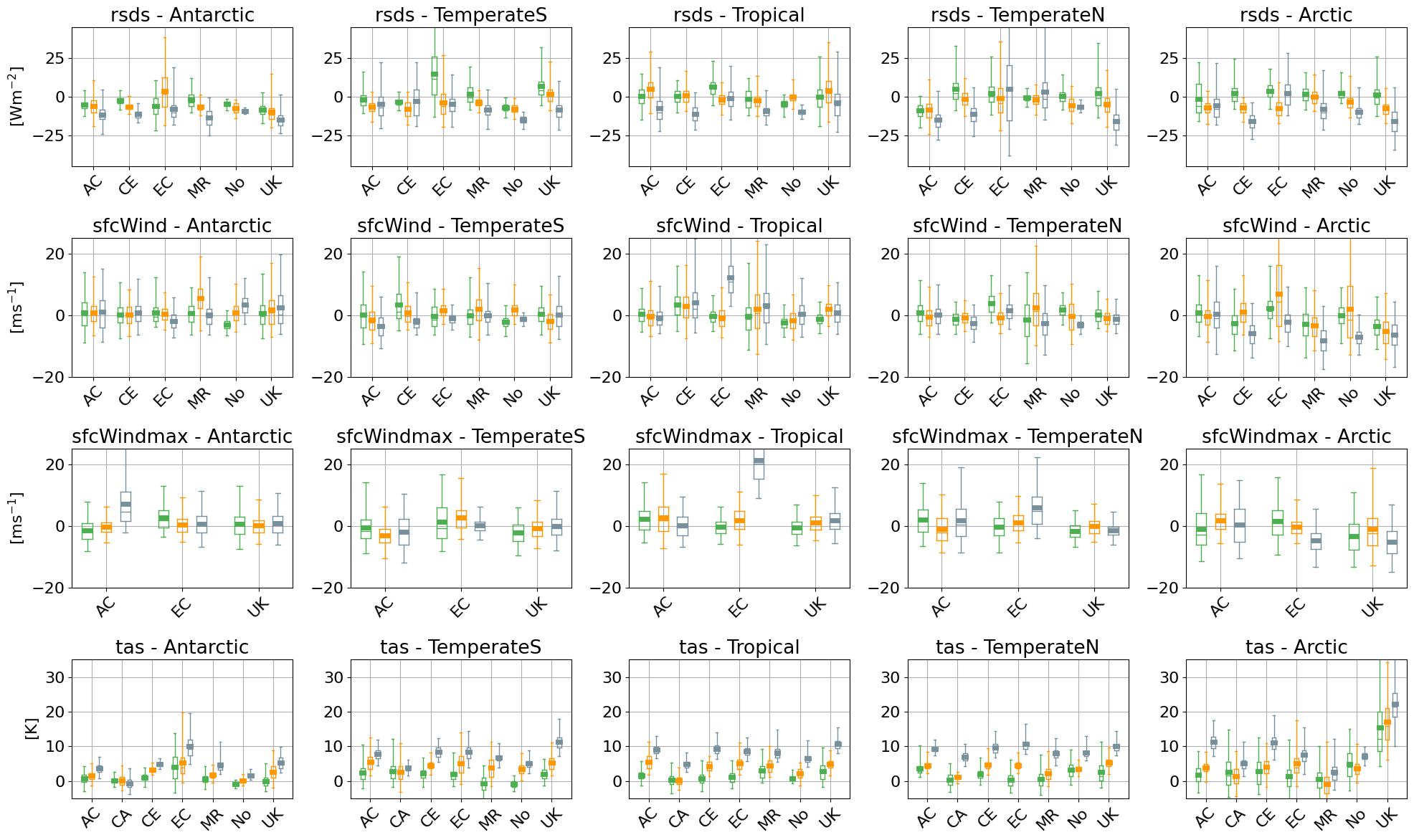}
		\caption{Maximum.}
	\end{subfigure}
	\begin{subfigure}{1\textwidth}
		\includegraphics[width=1\textwidth]{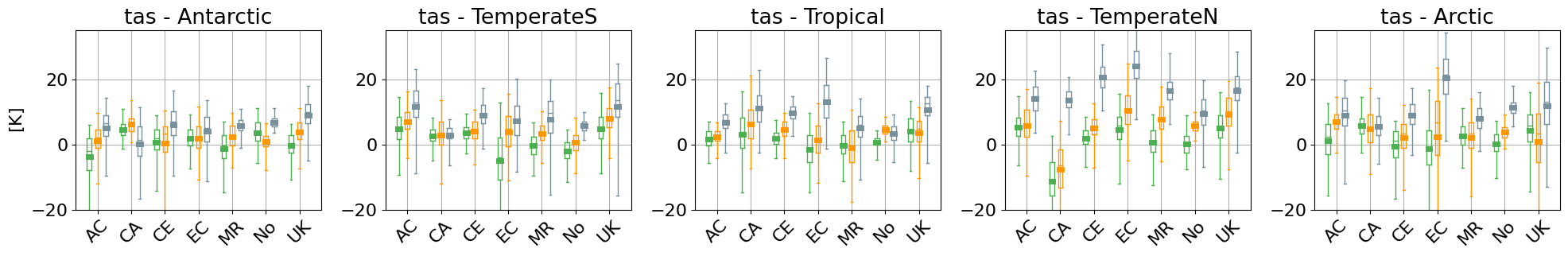}
		\caption{Minimum.}
	\end{subfigure}
	\caption{Box-whisker plots for the estimated posterior distribution of the change $\Delta^Q$ in (a) the 100-year return value of the climate zone annual maximum, and (b) the 100-year return value of the climate zone annual minimum for \TA. Columns provide results for the Antarctic, Temperate South, Tropical, Temperate North and Arctic climate zones. For other details, see Figure~\ref{Fgr-Glb-BW-MxmMnm}.  Results from analysis of one ensemble member for global minima of \No \TA under scenario \SM omitted; see notes in Section~\ref{Sct:Dat:GlbClmZon}.}
	\label{Fgr-ClmZon-BW-MnmMxm}
\end{figure}

\begin{figure}[!ht]
	\centering
	\includegraphics[width=1\textwidth]{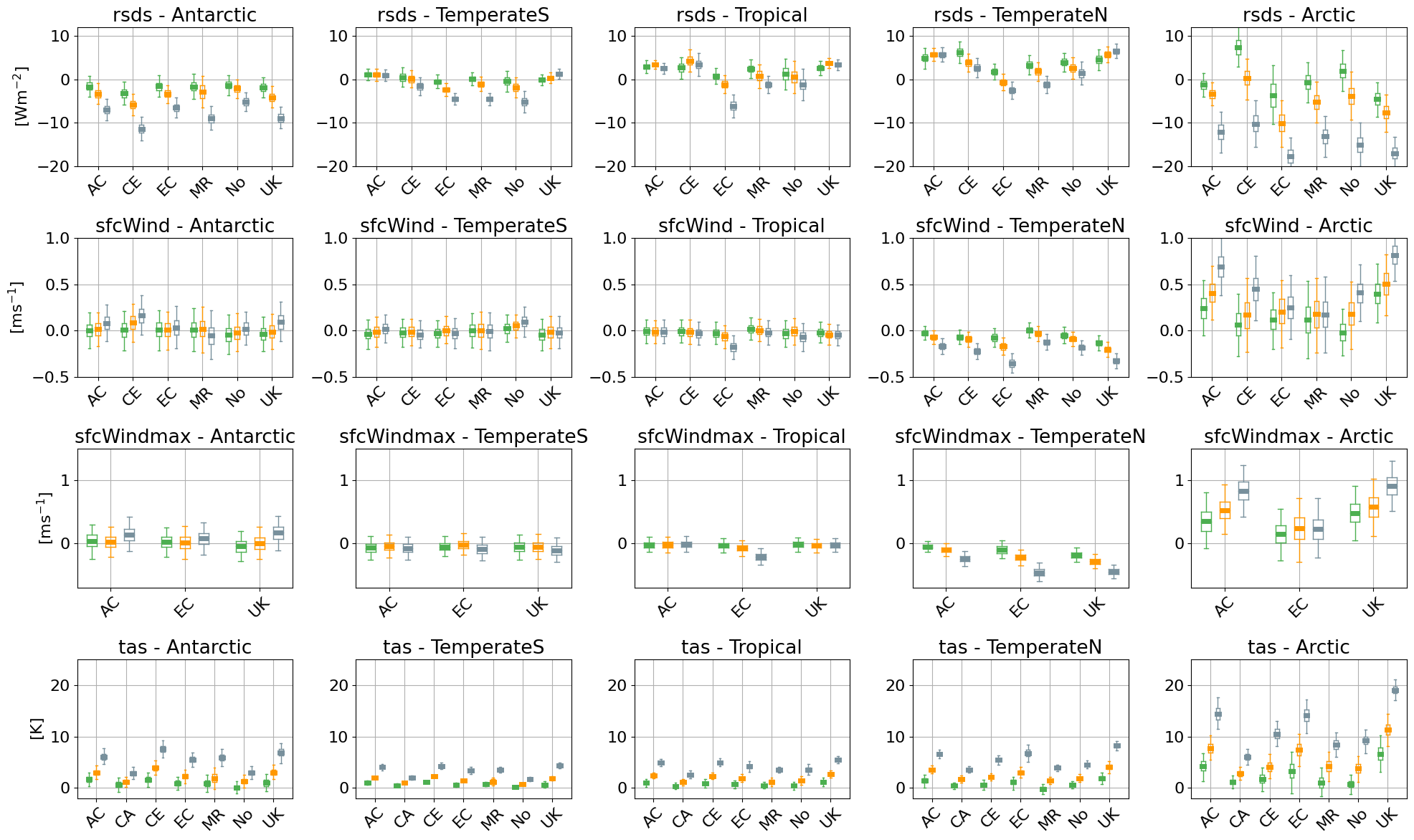}
	\caption{Box-whisker plots for the change $\Delta^M$ in the climate zone annual mean value over the next 100-years. Rows indicate different climate variables. Columns indicate climate zone. For other details, see Figure~\ref{Fgr-Glb-BW-MxmMnm}.}
	\label{Fgr-ClmZon-BW-Men}
\end{figure}

%
Physical intuition, together with the exploratory analysis in Section~\ref{Sct:Dat}, suggests that there may be considerable variation in changes $\Delta^Q$ and $\Delta^M$ across climate zones. For this reason, we extend the GEVR analysis of global annual extremes and NHGR analysis of global annual means to the corresponding climate zone annual extremes and means, for the Antarctic, Temperate South, Tropical, Temperate North and Arctic climate zones (see Section~\ref{Sct:Dat:GlbClmZon}). Results of the analysis are summarised in Figure~\ref{Fgr-ClmZon-BW-MnmMxm} in terms of the change $\Delta^Q$ in 100-year return value, and Figure~\ref{Fgr-ClmZon-BW-Men} in terms of the change $\Delta^M$ in mean for NHGR.

Results in Figure~\ref{Fgr-ClmZon-BW-MnmMxm} are generally similar to those for global annual extremes in Figure~\ref{Fgr-Glb-BW-MxmMnm}. The estimated posterior distributions for $\Delta^Q$ are wider that the corresponding global estimates, possibly due to the reduction in size of spatial domain over which annual maxima are taken per climate zone. For \WS and \WM, the central (25\%,75\%) box region typically includes zero for all combinations of GCM and climate scenario, indicating no obvious change in 100-year return values over the next 100 years for all climate zones, with the exception of the Arctic: here, there is weak evidence for more negative $\Delta^Q$ under \SH. For \RS, there is general evidence for reduction of $\Delta^Q$ with increased forcing. For annual maxima and minima of climate zone \TA, there is strong evidence for a scenario effect, with the \UK GCM suggesting considerably more Arctic heating than other models and other climate zones. There is also surprisingly consistent evidence across GCMs for large $\Delta^Q \approx $20K for \TA minima in the Temperate North under \SH.
The anomalous behaviour of some estimates in Figure~\ref{Fgr-ClmZon-BW-MnmMxm} (e.g. for \WM from \EC in the tropics under \SH) is to be expected given similar anomalous behaviour in the underlying climate zone annual maximum data (e.g. Figure~SM6 for \WM in the tropics). \ed{Figures SM26-29 provide supporting findings for $\Delta^Q$ per ensemble member, plotted seperately per climate scenario.}

Corresponding estimates for change $\Delta^M$ in the climate zone annual mean values are given in Figure~\ref{Fgr-ClmZon-BW-Men}. Here, it is noticeable that trends with scenario are clearer across GCM than for changes in 100-year return value; this is due in part at least to the relative efficiency of estimating sample means rather than sample tails. However, the trends also appear to be more consistent across GCM. The exaggerated extent of changes in the Arctic climate zone is also clear, presumably related to a reduction in the extent of surface ice. Visually at least, partitioning the globe into climate zones appears to improve the consistency of trends with scenario across GCMs for \RS. In the Antarctic, there is evidence for a reduction in annual mean \RS under all scenarios, with the amplitude of reduction increasing with climate forcing. The climate scenario effect is strongest in the Arctic. For \WS and \WM, changes in mean climate zone values are small with magnitudes $\ll$1ms$^{-1}$ everywhere except the Arctic. In the Arctic, a number of GCMs suggest that the annual mean for both \WS and \WM will increase with climate forcing. For \TA, we again see consistent trends across GCMs, with largest effects of climate forcing amounting the changes $\Delta^M$ of around 10K in the Arctic.

It is interesting that in general changes $\Delta^Q$ reflect changes $\Delta^M$. That is, in broad terms, change in 100-year return value over the next 100 years is largely driven by change in the general level of the climate variable, rather than change to the shape of the tail of the distribution of the climate variable; a similar observation was made in Section~\ref{Sct:Rsl:Glb} regarding the importance of $\mu^1$ in the analysis of global annual maxima. 

\FloatBarrier
\subsection{North Atlantic and Celtic Sea}  \label{Sct:Rsl:NACS}
\begin{figure}[!ht]
	\centering
	\includegraphics[width=1\textwidth]{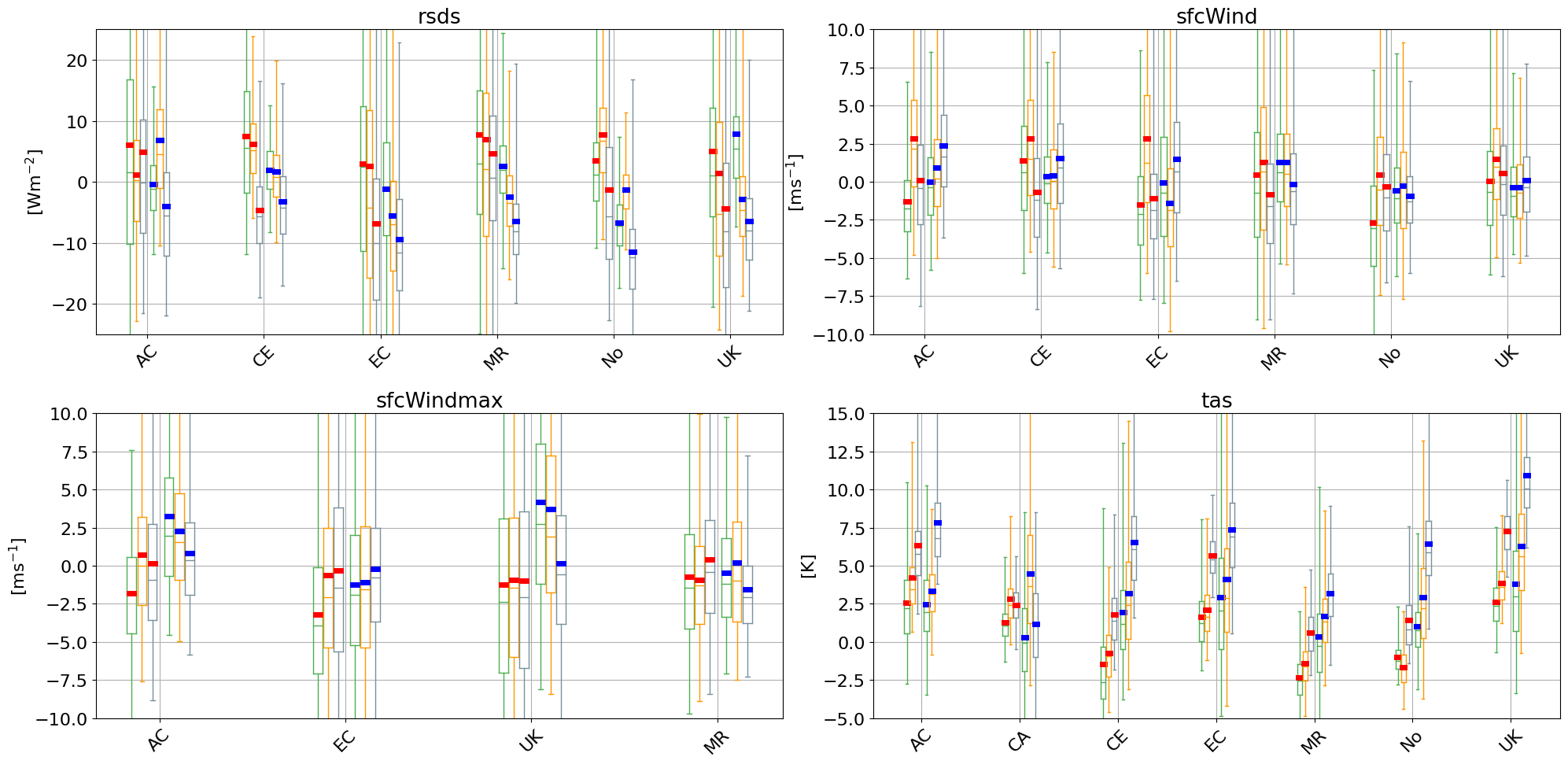}
	\caption{Box-whisker plots for the change $\Delta^Q$ in 100-year return values over the next 100-years for the North Atlantic and Celtic Sea neighbourhoods. Panels show results for different climate variables. Within each panel, inferences per climate model are given as a cluster of 6 box-whiskers: left-hand box-whiskers (with means in red) for the North Atlantic and right-hand (with means in blue) for the Celtic Sea. Box-whisker outline colour indicates climate scenario (\SL, green; \SM, orange; \SH, grey).}
	\label{Fgr-NACS-BW-Mxm}
\end{figure}

%
For annual maximum data from locations in the North Atlantic and Celtic Sea neighbourhoods, we estimate the posterior distribution of the change $\Delta^Q$ in 100-year return value over the next 100-years independently for each combination of location, GCM, climate variable, scenario and ensemble member. For conciseness of presentation, we then aggregate estimates for the posterior distribution of $\Delta^Q$ over location (within neighbourhood) and climate ensemble, resulting in distributional estimates for $\Delta^Q$ from each combination of climate variable, GCM, geographic neighbourhood and climate scenario. These are summarised in Figure~\ref{Fgr-NACS-BW-Mxm} in terms of box-whisker plots.
Inspection of the figure suggests that changes $\Delta^Q$ in \RS are somewhat more positive in the \NA compared with the \CS neighbourhood, although the central (25\%,75\%) box region often includes zero, indicating little evidence for a change in 100-year return value. There is also some evidence for reduction of \RS with increased forcing, particularly for the \CS neighbourhood; a reduction by as much as 10Wm$^{-2}$ is suggested under scenario \SH. There are no trends at all of interest for \WS and \WM, with almost all central (25\%,75\%) box regions including zero. There is evidence for increasing \TA with climate forcing in both the \NA and \CS neigbourhoods, and that changes in the \CS neighbourhood are likely to be somewhat larger than in the \NA. 

\FloatBarrier
\section{Discussion and conclusions}  \label{Sct:DscCnc}

There is a wealth of evidence that climate change is occurring (\citealt{IPCC-WG1-2013}, \citealt{IPCC-SPM}). It is pertinent therefore to assess what evidence is present in output of the latest climate models for changing characteristics of extremes of wind, surface solar irradiance and temperature. 

\ed{The objective of the current work was to examine key output from CMIP6 global coupled models at site-specific, regional and global scales, to form a view of what state-of-the-art science is telling us about climate change effects. To achieve this, we estimate} global and climate zone annual maximum and annual mean data for four climate variables (\WS, \WM, \RS and \TA) and annual minimum also for \TA, for the period (2015,2100) from output of seven CMIP6 GCMs. We also calculate the corresponding annual data for neighbourhoods of locations in the \NA and the \CS region. For each data source we access output corresponding to three climate scenarios and multiple climate ensemble runs. We then estimate non-stationary extreme value models (GEVR) for annual extremes, and perform non-homogeneous Gaussian regression (NHGR) for annual means. Using the estimated statistical models, we further estimate (i) the change $\Delta^Q$ in the 100-year return value for annual extremes, and (ii) the change $\Delta^M$ in the annual mean level over the period (2025, 2125). 

Inspection of results from Section~\ref{Sct:Rsl} reveals inconsistencies in estimates of $\Delta^Q$ and $\Delta^M$ across GCMs, but in general we can draw the following conclusions. Unsurprisingly, but importantly, inference for mean trends is more efficient than for extreme quantiles. That is, it is easier to identify trends in means than in extreme quantiles with confidence. For climate zone annual means, there is evidence for small changes in $\Delta^M$ for wind variables as a function of climate scenario, but the direction of these changes differs by GCM and climate zone. For extremes of wind variables, the estimated posterior distribution of $\Delta^Q$ is considerably broader and centred near zero; there is no consistent evidence for change in return value (except perhaps in the Arctic zone), and changes in posterior mean of around $\pm$ 5ms$^{-1}$ are typical across GCMs. In contrast, estimated changes for \RS and \TA are considerably more convincing. For \RS there is clear evidence for a reduction in mean of around 10Wm$^{-2}$ when comparing \SH with \SL in the Antarctic, and even larger reductions in Arctic. There is some evidence for reduction in $\Delta^Q$ for climate zone annual maxima with increased climate forcing. For \TA there is clear evidence for increase in global and climate zone annual means, maxima and minima with increased forcing. Inspection of estimated posterior distributions of extreme value model parameters reveals that changes in $\Delta^Q$ for \RS and \TA can be explained by a change in level $\mu^1$ rather than changes in the tail characteristics of the distribution of climate variable quantified by $\sigma^1$ and $\xi^1$ in the GEVR model. For single location analysis of annual maxima in the \NA and \CS region, uncertainties in inferences for $\Delta^Q$ are large. Hence it is difficult to be confident about the presence of climate change effects even for changes in the 100-year return values of wind variables and \RS; systematic climate effects in \TA are nevertheless present.

\subsection{Expected changes in return value, and probability of return value increase} \label{Sct:DscCnc:EmpDff}
It is important to consider the practical implications of the findings of this work, particularly for changes in the 100-year return value of the annual maximum over the next 100 years of wind variables. Despite the large variability of inferences for $\Delta^Q$ for wind variables, if we are prepared to assume that all GCMs are equally informative, we can estimate the expected change $\Delta^Q$ in return value for each of the climate variables empirically, simply by aggregating the samples of estimates for the return value differences from Section~\ref{Sct:Rsl}. We can also estimate the probability that an increase in return value will occur. Results per climate zone and scenario are given in Table~\ref{Tbl:ExpPrb:GATTTA}. \ed{Thus the expected change in return value is calculated as the arithmetic mean of the composite predictive distribution of $\Delta^Q$ from Bayesian inference aggregated over ensembles and GCMs, for a given combination of climate variable, scenario and climate zone. Similarly, the probability of increasing return value is estimated as the probability that $\Delta^Q>0$ from the same composite distribution.} \ed{We emphasise that viewing each GCM as an equally informative independent source of information about the climate is a useful simple working assumption rather than fact, since it is known that certain GCMs (e.g. \AC and \UK) exploit common modelling components.}
\begin{table}[!ht]
	\begin{center}
		\begin{tabular}{|c|c|c|c|c|r|r|r|}
			\hline
			\multicolumn{1}{|c|}{\multirow{2}{*}{Variable\Tstrut}} &
			\multicolumn{1}{|c|}{\multirow{2}{*}{Zone\Tstrut}} &
			\multicolumn{3}{|c|}{\(\mathbb{E}(\Delta^Q)\)\Tstrut} &
			\multicolumn{3}{|c|}{\(\mathbb{P}(\Delta^Q > 0)\)\Tstrut} \\ \cline{3-8}
			& & \(SSP126\) & \(SSP245\) & \(SSP585\) & \(SSP126\) & \(SSP245\) & \(SSP585\) \\ \hline
			\multirow{6}{*}{\begin{tabular}{c} \RS \\ $[\text{Wm}^{-2}]$ \end{tabular}} & GL\Tstrut & -4.89 & -6.06 & -12.15 & 0.15 & 0.13 & 0.04 \\    
			\cline{2-8}
			& AN\Tstrut & -4.93 & -6.02 & -12.54 & 0.14 & 0.12 & 0.04 \\
			& TS\Tstrut & 2.78 & -2.54 & -6.46 & 0.49 & 0.27 & 0.18 \\
			& TR\Tstrut & 0.24 & 2.37 & -6.12 & 0.51 & 0.49 & 0.21 \\
			& TN\Tstrut & 0.09 & -3.62 & -7.47 & 0.41 & 0.27 & 0.21 \\
			& AR\Tstrut & 1.46 & -4.51 & -9.18 & 0.52 & 0.2 & 0.18 \\
			\hline
			\multirow{6}{*}{\begin{tabular}{c} \WS \\ $[\text{ms}^{-1}]$ \end{tabular}} & GL\Tstrut & 0.21 & 1.41 & 0.23 & 0.42 & 0.58 & 0.47 \\        
			\cline{2-8}
			& AN\Tstrut & 0.45 & 1.85 & 1.21 & 0.47 & 0.59 & 0.51 \\
			& TS\Tstrut & 0.55 & 0.05 & -1.16 & 0.47 & 0.48 & 0.32 \\
			& TR\Tstrut & 0.1 & 0.98 & 2.05 & 0.44 & 0.53 & 0.54 \\
			& TN\Tstrut & 0.28 & -0.08 & -1.25 & 0.5 & 0.4 & 0.32 \\
			& AR\Tstrut & -1.4 & -0.97 & -4.64 & 0.36 & 0.35 & 0.19 \\
			\hline
			\multirow{6}{*}{\begin{tabular}{c} \WM \\ $[\text{ms}^{-1}]$ \end{tabular}} & GL\Tstrut & -1.71 & -0.16 & 1.52 & 0.33 & 0.44 & 0.63 \\      
			\cline{2-8}
			& AN\Tstrut & 0.68 & 0.15 & 2.19 & 0.52 & 0.47 & 0.6 \\
			& TS\Tstrut & -1.05 & -0.46 & -0.39 & 0.34 & 0.41 & 0.43 \\
			& TR\Tstrut & 0.14 & 1.67 & 5.83 & 0.45 & 0.62 & 0.69 \\
			& TN\Tstrut & -0.5 & -0.02 & 0.96 & 0.39 & 0.46 & 0.42 \\
			& AR\Tstrut & -1.68 & -0.21 & -3.75 & 0.36 & 0.43 & 0.23 \\
			\hline
			\multirow{6}{*}{\begin{tabular}{c} \TA \\ $[\text{K}]$ \end{tabular}} & GL\Tstrut & 2.0 & 4.1 & 9.32 & 0.75 & 0.97 & 1.0 \\
			\cline{2-8}
			& AN\Tstrut & 1.05 & 2.26 & 5.07 & 0.63 & 0.83 & 0.97 \\
			& TS\Tstrut & 1.52 & 4.24 & 8.2 & 0.75 & 0.93 & 1.0 \\
			& TR\Tstrut & 1.81 & 4.1 & 8.83 & 0.8 & 0.95 & 1.0 \\
			& TN\Tstrut & 1.9 & 4.05 & 9.2 & 0.73 & 0.96 & 1.0 \\
			& AR\Tstrut & 4.75 & 5.75 & 10.62 & 0.72 & 0.84 & 0.95 \\
			\hline
		\end{tabular}%
	\end{center}

	\caption{Estimated expected change $\mathbb{E}(\Delta^Q)$ and probability of increase $\mathbb{P}(\Delta^Q>0)$ of 100-year return values over the period (2025,2125) for annual maxima of four climate variables over 6 climate zones (GL: global; AN: Antarctic; TS: Temperate South; TR: Tropical; TN: Temperate North; AR: Arctic). Columns under ``$\mathbb{E}(\Delta^Q)$'' show expected changes per scenario, in the units of the variable; thus we estimate a reduction in 4.64ms$^{-1}$ in the return value for \WS under scenario \SH in the Arctic. Columns under ``$\mathbb{P}(\Delta^Q>0)$'' show corresponding probabilities of increasing return value. Estimates are calculated assuming equal weighting for each climate model. Note that the first (year 2015) observation for rsds in all \CE runs is spurious and has been omitted. Multiple values for \WM from \MR are also suspect (with some values $>100$ ms$^{-1}$; see Figure~SM7); for this reason \MR output is ignored for variable \WM only.}

	\label{Tbl:ExpPrb:GATTTA}

\end{table}
Empirical probabilities $\mathbb{P}(\Delta^Q>0)$ near zero or unity are indicative of agreement between climate models (and their constituent ensembles) regarding a change in $\Delta^Q = Q_{2125}-Q_{2025}$. For \WS and \WM, we therefore see that only in the Arctic under scenario \SH is there evidence of agreement between models regarding a change in return value: specifically, values of $\mathbb{P}(\Delta^Q>0) \notin (0.25.0.75)$ are observed only in the Arctic zone under scenario \SH. Here, mean reductions of $\Delta^Q \approx$ 4ms$^{-1}$ are estimated for \WS and \WM. For comparison, \RS shows significant reduction in the Antarctic and hence globally under \SH; there is also a consistent pattern of larger reductions in polar zones than elsewhere under both \SM and \SH. For \TA, $\mathbb{P}(\Delta^Q>0)$ is near unity under \SM and \SH for all climate zones.

The ocean engineer is typically interested not in summary statistics for large spatial regions such as climate zones, but rather in analysis for specific locations. To illustration the latter, Table~\ref{Tbl:ExpPrb:NACS} provides estimates for the expected change $\Delta^Q$, and the probability that $Q_{2125}$ exceeds $Q_{2025}$, for the central locations of the \NA and \CS regions in Figure~\ref{Fgr-NACS-Grd}.
\begin{table}[!ht]
	\begin{center}
		\begin{tabular}{|c|c|c|c|c|r|r|r|}
			\hline
			\multicolumn{1}{|c|}{\multirow{2}{*}{Variable\Tstrut}} &
			\multicolumn{1}{|c|}{\multirow{2}{*}{Zone\Tstrut}} &
			\multicolumn{3}{|c|}{\(\mathbb{E}(\Delta^Q)\)\Tstrut} &
			\multicolumn{3}{|c|}{\(\mathbb{P}(\Delta^Q > 0)\)\Tstrut} \\ \cline{3-8}
			& & \(126\) & \(245\) & \(585\) & \(126\) & \(245\) & \(585\) \\ \hline
			\multirow{2}{*}{\begin{tabular}{c} \RS \\ $[\text{Wm}^{-2}]$ \end{tabular}} & NA\Tstrut & 12.58 & 6.83 & -6.93 & 0.74 & 0.59 & 0.26 \\
			& CS\Tstrut & 0.76 & -1.15 & -7.77 & 0.53 & 0.27 & 0.16 \\
			\hline
			\multirow{2}{*}{\begin{tabular}{c} \WS \\ $[\text{ms}^{-1}]$ \end{tabular}} & NA\Tstrut & -0.73 & 1.68 & -0.12 & 0.24 & 0.57 & 0.52 \\
			& CS\Tstrut & 1.98 & 0.3 & 1.42 & 0.52 & 0.38 & 0.62 \\
			\hline
			\multirow{2}{*}{\begin{tabular}{c} \WM \\ $[\text{ms}^{-1}]$ \end{tabular}} & NA\Tstrut & -2.8 & -0.01 & 0.94 & 0.36 & 0.44 & 0.57 \\
			& CS\Tstrut & 2.83 & 2.7 & 1.67 & 0.64 & 0.53 & 0.43 \\
			\hline
			\multirow{2}{*}{\begin{tabular}{c} \TA \\ $[\text{K}]$ \end{tabular}} & NA\Tstrut & 2.14 & 2.18 & 4.06 & 0.68 & 0.67 & 0.96 \\
			& CS\Tstrut & 1.92 & 4.43 & 5.95 & 0.68 & 0.89 & 0.92 \\
			\hline
		\end{tabular}%
	\end{center}
	
	\caption{Estimated expected change $\mathbb{E}(\Delta^Q)$ and probability of increase $\mathbb{P}(\Delta^Q>0)$ of 100-year return values over the period (2025,2125) for annual maxima of four climate variables at the central locations of the North Atlantic (NA) and Celtic Sea (CS) neighbourhoods. For further description, see Table~\ref{Tbl:ExpPrb:GATTTA}.}
	
	\label{Tbl:ExpPrb:NACS}
\end{table}
Relative to the probability estimates in Table~\ref{Tbl:ExpPrb:GATTTA}, probabilities in Table~\ref{Tbl:ExpPrb:NACS} for \WS, \WM and \RS are nearer 0.5 than zero or unity. Moreover, the expected values $\Delta^Q$ are small in magnitude. For comparison, there is evidence under \SH that climate models agree on an increasing \TA for both North Atlantic and Celtic Sea regions. Clear trends are present with climate scenario for both \RS and \TA, but not for \WS and \WM.

Table~SM1 of the SM estimates $\mathbb{E}(\Delta^Q)$ and $\mathbb{P}(\Delta^Q>0)$ for change $\Delta^Q$ in annual \TA minima from global and climate zones, quantifying the information already visualised in Figure~\ref{Fgr-ClmZon-BW-MnmMxm}(b). Similarly, Table~SM3 quantifies $\mathbb{E}(\Delta^M)$ and $\mathbb{P}(\Delta^M>0)$ for change $\Delta^M = M_{2125}-M_{2025}$ in the annual means for global and climate zones, showing clear trends with climate zone and climate scenario, already visualised in Figure~\ref{Fgr-ClmZon-BW-Men}.

\subsection{Quantifying the effect of climate scenario, and uncertainty due to climate model and ensemble member on $\Delta^Q$ and $\Delta^M$} \label{Sct:DscCnc:LnrMxdEff}
We can summarise sources of variability in our data for changes in 100-year return values $\Delta^Q$ and annual means $\Delta^M$ more precisely by estimating another statistical model. Here we estimate a linear mixed effects model (LMM, e.g. \citealt{WstEA22}) using $\Delta^Q$ (or $\Delta^M$) as a response, with climate scenarios as so-called \emph{fixed effects}, and climate models and nested climate ensemble members (for a given model) as \emph{random effects}. The advantage of this approach over linear regression (for all of scenario, model and ensemble) is that we assume that the climate model and ensemble output available to us are drawn randomly from large families of models and their ensemble members. We can therefore estimate the variability in $\Delta^Q$ (or $\Delta^M$) which can be attributed to our random choices of climate model and their ensemble members; that is, we can estimate how much of the variability in $\Delta^Q$ (or $\Delta^M$) is due to (apparently) random effects from different climate models, and from different ensemble members for a given climate model. 

Using estimates of $\Delta \in (\Delta^Q,\Delta^M)$ for a given climate variable in a given climate zone, we can express the linear mixed effects model as 
\begin{eqnarray}
	\Delta_{ijk\ell} = \iota + \gamma_j + \delta_k + \zeta_{k(\ell)} + \epsilon_{ijk\ell}
\end{eqnarray}
where $\Delta_{ijk\ell} \in \mathbb{R}$ is the $i$th observation of $\Delta$ from the extreme value model output, for climate scenario $j$ ($j=1,2,3$, corresponding to scenarios \SL, \SM and \SH), for climate model $k$ ($k=1,2,...$) and ensemble member $\ell$ ($\ell=1,2,...$) of model $k$. Parameter $\iota \in \mathbb{R}$ is the intercept. Parameter $\gamma_j \in \mathbb{R}$ represents the fixed effect of scenario $j$ on response. Parameter $\delta_k \in \mathbb{R}$ represents the random effect of climate model $k$ on response, and $\zeta_{k(\ell)} \in \mathbb{R}$ the nested random effect of climate ensemble member $\ell$ for model $k$. Parameters $\epsilon_{ijk\ell} \in \mathbb{R}$ are model error terms. For fitting, we assume that each of $\delta_k$, $\zeta_{k(\ell)}$ and $\epsilon_{ijk\ell}$ are independently normally-distributed with zero means and variances $\tau_\delta^2 \ge 0$, $\tau_\zeta^2 \ge 0$ and $\tau_\epsilon^2 \ge 0$. The objective of the analysis is to estimate the set $\{\gamma_j\}_{j=1}^3$ of fixed effects, the variances $\tau_\delta^2$ and $\tau_\zeta^2$ of random effects, and the error variance $\tau_\epsilon^2$. Inference was performed using algorithms from the MATLAB Statistics toolbox and checked using equivalent functionality in Julia, with scenario \SL ($j=1$) as reference category for the fixed climate scenario effect. That is, to avoid issues of multicollinearity, we actually undertake estimation of the sum $\iota + \gamma_1$, together with that of scenario difference effects $\gamma_2-\gamma_1$ and $\gamma_3-\gamma_1$. 

Results for analysis of change $\Delta^Q$ in the 100-year return value annual maxima per climate variable and zone are given in Table~\ref{Tbl:LME:GATTTA}. 
\begin{table}[!ht]
	\begin{center}
		\begin{tabular}{|c|c|c|c|c|r|r|r|r|r|r|r|}
			\hline
			\multicolumn{1}{|c|}{\multirow{2}{*}{Variable\Tstrut}} &
			\multicolumn{1}{|c|}{\multirow{2}{*}{Zone\Tstrut}} &
			\multicolumn{3}{|c|}{SSP effect\Tstrut} &
			\multicolumn{5}{|c|}{Model standard deviations\Tstrut} &
			\multicolumn{2}{|c|}{$R^2$\Tstrut} \\ \cline{3-12}
			& & \(\iota + \gamma_1\) & \(\gamma_2 - \gamma_1\) & \(\gamma_3 - \gamma_1\) & \(\tau_R\) & \(\tau_{\text{FE}}\) & \(\tau_{\epsilon}\) & \(\tau_{\delta}\) & \(\tau_{\zeta}\) & \(R^2_{\text{FE}}\) & \(R^2_{\text{ME}}\) \\ \hline
			\multirow{6}{*}{\begin{tabular}{c} \RS \\ $[\text{Wm}^{-2}]$ \end{tabular}} & GL\Tstrut & -4.47 & -1.27 & -7.26 & 7.62 & 6.93 & 6.25 & 1.56 & 2.51 & 0.17 & 0.33 \\
			\cline{2-12}
			& AN\Tstrut & -4.62 & -1.38 & -7.29 & 7.6 & 6.85 & 6.12 & 1.48 & 2.76 & 0.19 & 0.35 \\
			& TS\Tstrut & 3.83 & -6.43 & -9.61 & 10.29 & 9.62 & 8.32 & 1.36 & 5.1 & 0.12 & 0.35 \\
			& TR\Tstrut & 1.6 & 1.25 & -7.02 & 10.92 & 10.3 & 9.16 & 0.0 & 6.43 & 0.11 & 0.3 \\
			& TN\Tstrut & 0.22 & -3.91 & -7.95 & 12.06 & 11.68 & 10.17 & 3.04 & 4.69 & 0.06 & 0.29 \\
			& AR\Tstrut & 2.44 & -6.83 & -11.48 & 10.23 & 9.34 & 8.22 & 2.48 & 4.19 & 0.17 & 0.35 \\
			\hline
			\multirow{6}{*}{\begin{tabular}{c} \WS \\ $[\text{ms}^{-1}]$ \end{tabular}} & GL\Tstrut & 0.25 & 1.16 & 0.02 & 4.73 & 4.7 & 4.31 & 0.0 & 1.84 & 0.01 & 0.17 \\
			\cline{2-12}
			& AN\Tstrut & 0.59 & 1.4 & 0.62 & 5.34 & 5.31 & 4.95 & 0.0 & 1.94 & 0.01 & 0.14 \\
			& TS\Tstrut & 0.73 & -0.58 & -1.81 & 4.8 & 4.74 & 4.38 & 0.45 & 1.76 & 0.02 & 0.17 \\
			& TR\Tstrut & -0.58 & 1.06 & 2.48 & 6.1 & 6.05 & 5.5 & 0.0 & 3.13 & 0.02 & 0.19 \\
			& TN\Tstrut & 0.38 & -0.57 & -1.43 & 4.74 & 4.69 & 4.46 & 0.0 & 1.65 & 0.02 & 0.11 \\
			& AR\Tstrut & -0.44 & -0.17 & -3.16 & 6.83 & 6.63 & 5.6 & 2.45 & 3.28 & 0.06 & 0.33 \\
			\hline
			\multirow{6}{*}{\begin{tabular}{c} \WM \\ $[\text{ms}^{-1}]$ \end{tabular}} & GL\Tstrut & -1.5 & 1.55 & 3.23 & 4.31 & 4.1 & 3.71 & 0.76 & 1.58 & 0.09 & 0.26 \\
			\cline{2-12}
			& AN\Tstrut & 0.68 & -0.53 & 1.51 & 4.94 & 4.87 & 4.54 & 0.0 & 1.74 & 0.03 & 0.15 \\
			& TS\Tstrut & -1.05 & 0.6 & 0.66 & 4.93 & 4.92 & 4.57 & 0.0 & 1.83 & 0.0 & 0.14 \\
			& TR\Tstrut & 0.96 & 1.53 & 5.69 & 7.11 & 6.69 & 6.05 & 2.91 & 1.07 & 0.11 & 0.28 \\
			& TN\Tstrut & -0.33 & 0.47 & 1.45 & 4.74 & 4.71 & 4.12 & 0.73 & 2.16 & 0.02 & 0.25 \\
			& AR\Tstrut & -1.68 & 1.47 & -2.11 & 6.72 & 6.56 & 5.65 & 0.0 & 3.35 & 0.05 & 0.29 \\
			\hline
			\multirow{6}{*}{\begin{tabular}{c} \TA \\ $[\text{K}]$ \end{tabular}} & GL\Tstrut & 1.78 & 2.12 & 7.33 & 3.93 & 2.46 & 2.17 & 0.86 & 0.82 & 0.61 & 0.69 \\
			\cline{2-12}
			& AN\Tstrut & 0.54 & 1.52 & 4.11 & 3.53 & 3.13 & 2.27 & 1.73 & 1.57 & 0.22 & 0.59 \\
			& TS\Tstrut & 1.06 & 3.22 & 6.74 & 3.97 & 2.94 & 2.46 & 1.17 & 1.22 & 0.45 & 0.62 \\
			& TR\Tstrut & 1.09 & 2.79 & 7.13 & 3.76 & 2.46 & 2.04 & 1.04 & 1.3 & 0.57 & 0.7 \\
			& TN\Tstrut & 1.8 & 2.19 & 7.4 & 3.91 & 2.52 & 2.23 & 0.68 & 0.95 & 0.59 & 0.68 \\
			& AR\Tstrut & 4.01 & 1.31 & 6.27 & 8.2 & 7.8 & 4.82 & 4.83 & 3.05 & 0.09 & 0.65 \\
			\hline
		\end{tabular}%
	\end{center}
	\caption{Summary of linear mixed effects modelling for the change $\Delta^Q$ in the 100-year return value of annual maxima for four climate variables over 6 climate zones (GL: global; AN: Antarctic; TS: Temperate South; TR: Tropical; TN: Temperate North; AR: Arctic). Columns under ``SSP effect'' show intercept $\iota$ and fixed effect parameter estimates $\gamma_j$ for the change in 100-year return value of the given variable over the period (2025,2125) under scenario $j$, in the units of the variable; thus we estimate a reduction in 3.16ms$^{-1}$ in the return value for \WS under scenario \SH \ed{in the Arctic}. Columns under ``Model standard deviations'' provide estimates of the various standard deviations of model fitting. $\tau_R$: the (full unconditional) standard deviation of the response $R$; $\tau_{\text{FE}}$: the model error standard deviation after fitting only the fixed effects (FE, of climate scenario); $\tau_{\epsilon}$: the model error standard deviation after fitting the full mixed effects model. $\tau_{\delta}$: standard deviation of climate model random effect; $\tau_{\zeta}$: standard deviation due to nested random effect of climate ensemble within climate model. $R^2$ statistics are also provided, from fitting only the fixed effects (FE), and from fitting the full mixed effects (ME) model. Note that the first (year 2015) observation for \RS in all \CE runs is spurious and has been omitted. See also notes about suspect values for \WM under model \MR in the caption of Table~\ref{Tbl:ExpPrb:GATTTA}. Further, since sample size for model fitting is huge, estimates of uncertainties and ``significance'' are of little practical value, and are omitted.}
	\label{Tbl:LME:GATTTA}
\end{table}
For \WS and \WM, $R^2_{\text{ME}}$ values for the full mixed effect model are $<0.3$ (except for \WS in the Arctic), and $R^2_{\text{ME}}$ values are lower than those for \RS and (in particular) \TA. This indicates that the variation in $\Delta^Q$ explainable by the effects of scenario, model and ensemble is relatively small for \WS and \WM. The same feature can be seen by comparing the reduction in standard deviation (from $\tau_R$ to $\tau_\epsilon$) from fitting the full mixed effects model for different variables; the reduction is generally rather modest for \WS and \WM. Of course this also results in relative large values of error standard deviation $\tau_\epsilon$ for wind variables. Values of $R^2_{FE}$ indicate the quality of model fit using fixed scenario effects only. For \WS and \WM, $R^2_{\text{FE}} \approx 0$, indicating that scenario has little skill in explaining variation of $\Delta^Q$ of wind variables; it is instructive again to compare these $R^2_{\text{FE}}$ values with those for \RS and \TA. Further, we see that values of $\tau_\zeta$ are relatively large for wind variables, indicating large variability between estimates of $\Delta^Q$ from ensemble members; values of $\tau_\zeta$ for wind variables tend to be larger than those of $\tau_\delta$, indicating that the effect of variability between ensembles is (much) larger than that of variability between climate models. This feature is illustrated in Figures~SM28-31, showing box-whisker plots of the posterior distribution of $\Delta^Q$ for individual ensemble members, per climate variable and spatial zone. For \WS in the Temperate North e.g., variation in $\Delta^Q$ is dominated by ensemble effects, and the variation between climate models is small (leading to estimated $\tau_\delta=0$); for \WS in the Arctic in contrast, variation between climate models is relatively large (leading to estimated $\tau_\delta \approx \tau_\zeta$).  Note in comparison that the values of $\tau_\delta$ and $\tau_\zeta$ are also of similar magnitudes for \RS and \TA; for the latter, these are also of similar magnitude to $\tau_\epsilon$. The Arctic, exhibiting an interestingly large climate model random effect for \TA, is an exception.

Corresponding results for single location analyses of annual maxima from the centre points of the \NA and \CS region are given in Table~\ref{Tbl:LME:NACS}. Unsurprisingly, we see that the estimates for fixed scenario effects $\gamma_j$ ($j=1,2,3$) in Tables~\ref{Tbl:LME:GATTTA} and \ref{Tbl:LME:NACS} are similar in sign and general magnitude to the expected changes $\mathbb{E}(\Delta^Q)$ estimated in Tables~\ref{Tbl:ExpPrb:GATTTA} and \ref{Tbl:ExpPrb:NACS}. 
\begin{table}[!ht]
	\begin{center}
		\begin{tabular}{|c|c|c|c|c|r|r|r|r|r|r|r|}
			\hline
			\multicolumn{1}{|c|}{\multirow{2}{*}{Variable\Tstrut}} &
			\multicolumn{1}{|c|}{\multirow{2}{*}{Zone\Tstrut}} &
			\multicolumn{3}{|c|}{SSP effect\Tstrut} &
			\multicolumn{5}{|c|}{Model standard deviations\Tstrut} &
			\multicolumn{2}{|c|}{$R^2$\Tstrut} \\ \cline{3-12}
			& & \(\iota + \gamma_1\) & \(\gamma_2 - \gamma_1\) & \(\gamma_3 - \gamma_1\) & \(\tau_R\) & \(\tau_{\text{FE}}\) & \(\tau_{\epsilon}\) & \(\tau_{\delta}\) & \(\tau_{\zeta}\) & \(R^2_{\text{FE}}\) & \(R^2_{\text{ME}}\) \\ \hline
			\multirow{2}{*}{\begin{tabular}{c} \RS \\ $[\text{Wm}^{-2}]$ \end{tabular}} & NA\Tstrut & 7.86 & -3.9 & -10.44 & 22.55 & 22.17 & 20.44 & 0.0 & 8.8 & 0.03 & 0.18 \\ 
			& CS\Tstrut & 1.35 & -3.13 & -7.23 & 10.88 & 10.49 & 9.1 & 2.46 & 4.81 & 0.07 & 0.3 \\ 
			\hline
			\multirow{2}{*}{\begin{tabular}{c} \WS \\ $[\text{ms}^{-1}]$ \end{tabular}} & NA\Tstrut & -0.61 & 2.25 & -0.08 & 5.08 & 4.98 & 4.48 & 1.26 & 1.95 & 0.04 & 0.22 \\
			& CS\Tstrut & 0.91 & -0.64 & 0.5 & 4.33 & 4.29 & 4.11 & 0.0 & 1.51 & 0.02 & 0.1 \\
			\hline
			\multirow{2}{*}{\begin{tabular}{c} \WM \\ $[\text{ms}^{-1}]$ \end{tabular}} & NA\Tstrut & -1.67 & 0.79 & 2.22 & 6.83 & 6.77 & 6.24 & 0.0 & 3.11 & 0.02 & 0.17 \\
			& CS\Tstrut & 2.76 & 0.64 & -4.23 & 8.09 & 7.85 & 6.97 & 2.67 & 2.87 & 0.06 & 0.26 \\
			\hline
			\multirow{2}{*}{\begin{tabular}{c} \TA \\ $[\text{K}]$ \end{tabular}} & NA\Tstrut & 0.46 & 0.89 & 3.29 & 3.61 & 3.33 & 2.21 & 2.05 & 1.18 & 0.15 & 0.62 \\
			& CS\Tstrut & 1.72 & 1.84 & 4.42 & 4.35 & 3.96 & 3.46 & 1.26 & 1.45 & 0.17 & 0.36 \\
			\hline
		\end{tabular}%
	\end{center}
		
	\caption{Summary of linear mixed effects modelling for the change $\Delta^Q$ in the 100-year return value of annual maxima for four climate variables at the central locations of the North Atlantic (NA) and Celtic Sea (CS) neighbourhoods. For further description, see Table~\ref{Tbl:LME:GATTTA}.}
	
	\label{Tbl:LME:NACS}
	
\end{table}
$R^2_{\text{FE}}$ values in Table~\ref{Tbl:LME:NACS} indicate that fixed scenario effects are poorer in explaining variation in return value differences $\Delta^Q$ for the central locations of the North Atlantic and Celtic Sea regions, than for the climate zones considered in Table~\ref{Tbl:LME:GATTTA}; again, \TA shows the most obvious fixed effect.

Table~SM2 of the SM estimates a LMM for change $\Delta^Q$ of annual minima from global and climate zones, and similarly Table~SM4 estimates a LMM for change $\Delta^M$ in the annual means for global and climate zones. Table~SM4 illustrates clearly that changes $\Delta^M$ in annual mean are more predictable than changes $\Delta^Q$ of annual spatial maxima in Table~\ref{Tbl:LME:GATTTA}, even for \WS and \WM in the Temperate North and Arctic zones. However, the magnitudes of the changes $\Delta^M$ in wind variables (see Table~SM3) are very small.

\subsection{Concluding remarks}

As noted in Section~\ref{Sct:Int}, climate model output requires calibration to spatio-temporally local conditions to reduce its bias and potentially its variance. Given the obvious absence of measurement data for the future to 2100, any calibration is problematic, requiring the assumption that the characteristics of the estimated calibration model persist into the future. In this work we choose not to calibrate, preferring to estimate \textit{changes} in key distributional characteristics of the climate variables directly from the climate model output. \ed{By focussing on differences in return values over a period of 100 years, we are sure that caibration offsets will not affect our inferences. However, (unknown) linear and higher order ``gain'' terms in a calibration will typically result in systematic departures from the results given here.} \ed{We note further that current analysis is conditional on the quality of current GCM models; we might anticipate that findings would change, were the analysis repeated in future using improved (e.g. higher resolution) climate models.} Diagnostic plots for GEVR (Figures~\ref{Fgr-Glb-Cdf-UK-WM-Mxm} and \ref{Fgr-Glb-Cdf-UK-TA-Mxm}, and those of Section~SM5) suggest that when present, climate-related changes in 100-year return value are generally explainable by a change in GEVR location parameter (i.e. $\mu^1$) rather than changes in tail characteristics ($\sigma^1$ and $\xi^1$). \ed{We note that thorough visual checking of data and model fitting diagnostics is essential to reliable inferences for changes in the 100-year return value, and note the possibility to perform further quantitative analysis on the distributions of model parameter estimates from the work.}

It is important to consider why obvious climate effects in both annual mean differences $\Delta^M$ and return value differences $\Delta^Q$ are visible in global and climate zone annual means and extremes for \RS and in particular \TA, but not for wind variable \WS and \WM. Does this reflect inadequacy of the physical climate models for wind characterisation, or a genuine lack of climate effect on spatio-temporal extremes of wind variables \ed{due to prevailing atmospheric dynamics, at least in some regions}? Perhaps the more probable explanation is the limited spatio-temporal resolution of current climate models, which are therefore unable to capture local changes in atmospheric pressure fields driving winds (but see e.g. \citealt{EwnJnt23a} regarding tropical cyclones around Madagascar). Presumably the spatio-temporal scales of climate-related changes to \RS and \TA are considerably larger. It is also interesting to consider why there is most consistency in estimates for the distributions of both $\Delta^M$ and $\Delta^Q$ for \TA. Does this indicate that \TA is inherently easier to forecast in a climate model than other variables, or that more effort has been devoted to modelling \TA well, or perhaps that different modelling institutions make similar physical assumptions in their climate models regarding \TA?

\ed{The current work is based on analysis of annual extremes and annual means derived from daily average data for \RS, \TA and \WS, potentially over a grid of locations on the Earth's surface. The choice of daily average data for \RS and \TA was dictated by the fact that only daily average data for these variables are typically available from public repositories like CEDA. For wind however, \WM is also typically provided in CMIP6 output. Extreme value analysis using data based ultimately on daily averages will provide different inferences to analysis using data based ultimately on daily extremes. The lack of material differences in behaviour in extremes of \WS and \WM would suggest that this is not an obvious concern for wind. In marked contrast, the lack of climate effect on both \WS and \WM relative to \RS and \TA is more surprising and perhaps concerning. Moreover, we might expect that numerical models such as GCMs, operating at a given resolution, provide better estimates of daily mean climate characteristics rather than daily extremes (e.g. \citealt{JntEwnFrr08a}).}

\ed{Model choice for the current work assumes linear growth of all model parameters (for GEVR and NHGR). Possibly at the expense of some bias and variance from model misspecification, for the sake of clarity of exposition, a common linear functional form for variation of parameters was judged appropriate. For statistical modelling of annual extremes and means of \TA, particularly under scenario \SL and sometimes \SM, we observe differences in estimates for posterior distributions of the 2015 base year model parameters ($\mu^0$, $\sigma^0$, $\xi^0$ of GEVR, and $\alpha^0$, $\beta^0$ for NHGR) given GCM and climate ensemble member. These differences, although relatively small, point a level of inadequacy in the assumed linear form for the variation of GEVR and NHGR model parameters in time, and suggests it might be appropriate to consider a non-linear parameter variation (especially for GEVR level $\mu$). Elsewhere, we noted examples (such as for \WS) where a simpler non-time-varying model might be used. Ideally, a more sophisticated analysis might attempt model selection, using a criterion such as the divergence information criterion (DIC), across a hierarchy of Bayesian models of different temporal complexities. An additional step might be to estimate models for all scenarios (using data for a given combination of GCM, climate variable and ensemble member) simultaneously, imposing a common starting tail for the distribution of climate variable in 2015. We also note the possibility of including covariates other than time within the analysis.}

Previous studies (e.g. \citealt{MccEA20}, \citealt{EwnJnt23a}) of climate effects on the return value for significant wave height $H_S$ derived from CMIP5 and CMIP6 output, have claimed more confidence in inferences. For example, in broad terms, \citealt{MccEA20} claims a 10\% increase in return values for $H_S$ in the Southern Ocean. \cite{EwnJnt23a} claims weak effects with climate scenario and location offshore Southern Australia and Madagascar, the trends of which are consistent with \citealt{MccEA20}. However, other studies (e.g. \citealt{KmrEA15}) have found little evidence for trends in extreme wind speeds from climate model output. Nevertheless is important to consider why the current study does not find compelling evidence for climate change effects on \WS and \WM which drive $H_S$. One possible explanation is the analysis of annual maxima in the current work, compared with a peaks-over-threshold analysis in the two articles mentioned; although \cite{EwnJnt23a} also consider some analysis of temporal block maxima and comment on its inefficiency relative to peaks-over-threshold. In the current work, annual maximum analysis was adopted since it provided a straightforward pre-processing route to compile temporal and spatio-temporal block maxima, possibly at the expense of statistical efficiency. Another possibility is that taking maxima over large spatial domains (e.g. over climate zone or the whole globe) has the effect of mixing different extreme value tails together, thereby complicating estimation of tail characteristics. In this situation, it would be advisable to restrict spatial domains so that the locations they contain are likely to exhibit the same tail behaviour. However, results for single location analysis for the North Atlantic and Celtic Sea region also do not show obvious return value trends for wind variables. Another source of increased uncertainty in the current work is our consideration of not only multiple climate models, but also of multiple ensemble runs for those climate models. As shown in Section~\ref{Sct:DscCnc:LnrMxdEff}, uncertainty in estimating changes in return value is related at least to some degree to variability between estimates from different climate models and their constituent ensembles.

Characterising changes in large-scale spatio-temporal mean characteristics of wind variables (\WS and \WM), irradiance (\RS) and near-surface temperature (\TA) in output of CMIP6 GCMs is relatively straightforward. Estimating changes in extreme quantiles of the distributions of large-scale spatio-temporal maxima is considerably more challenging. The current analysis shows that CMIP6 output is informative for changes in extremes of annual spatial maxima of \TA and \RS (over large spatial domains and at specific locations in the North Sea and Celtic Sea regions). Further, the effect of climate scenario, climate model and constituent ensemble member variability on the change in 100-year value over the next 100 years can be quantified to a reasonable degree. For \WS and \WM however, there is less evidence in the CMIP6 output in general supporting changes in return value. Moreover, the observed variability in return value change is dominated by sources other than climate scenario.

\FloatBarrier
\section*{Acknowledgements}
%
The authors would like to acknowledge discussions with colleagues at Shell. Data used in the current study are available at \cite{Lch24}. Software of GEVR and NHGR model fitting is available at \cite{LchJnt24}.

\appendix
\section{Bayesian inference} \label{Sct:App:BysInf}
%
Inference for the GEV regression and NHGR models described in Section~\ref{Sct:Mth} is performed using Markov chain Monte Carlo (MCMC, see e.g. \citealt{GmrLps06}) following the method of \cite{RbrRsn09}. This procedure was reported previously in \cite{EwnJnt23a}, but is summarised here for completeness and ease of reference. All the parameters $\boldsymbol{\theta}$ of the model are jointly updated for a sequence of $n_B+n_I$ MCMC iterations. At each iteration, a new set of parameter values is proposed, and accepted according to the Metropolis-Hastings acceptance criterion based on (a) the sample likelihood evaluated at the current and candidate states, and (b) the values of the prior densities for parameters at the current and candidate states. Following $n_B$ burn-in iterations, the Markov chain is judged to have converged, so that the subsequent $n_I$ iterations provide a valid sample from the joint posterior distribution of parameters. \ed{Prior distributions were specified as follows for GEV: $\xi \sim U(-1,0.2)$; $\sigma \sim U(0,\infty)$; $\mu \sim U(-\infty, \infty)$ and for NHGR: $\alpha$, $\beta \sim U(0,\infty)$}. Likelihoods for the models are available from the distributions given in the main text. An appropriate starting solution $\boldsymbol{\theta}_1$ for the MCMC inference was obtained by random sampling from the prior distributions of parameters, ensuring a valid likelihood. 

For the first $n_S<n_B$ iterations, candidate parameter values $\boldsymbol{\theta}_k^c$ are proposed (independently) from $\boldsymbol{\theta}_k^c \sim N(\boldsymbol{0},0.1^2 \boldsymbol{I})$ following \cite{RbrRsn09}. Thereafter $\boldsymbol{\theta}_k^c \sim (1-\beta) N\left(\boldsymbol{\theta}_{k-1}, 2.38^{2} \Sigma_{k}\right)+\beta N\left(\boldsymbol{\theta}_{k-1}, 0.1^{2} / 4\right)$, where $\beta=0.05$, $\Sigma_{k}$ is the empirical variance-covariance matrix of parameters from the past $k$ iterations, and $\boldsymbol{\theta}_{k-1}$ is the current value of parameters. Throughout, a candidate state is accepted using the standard Metropolis-Hasting acceptance criterion. Since prior distributions for parameters are uniform, and proposals symmetric, this criterion is effectively just a likelihood ratio. That is, we accept the candidate state with probability $\min (1, L(\boldsymbol{\theta}^c)/L(\boldsymbol{\theta}))$, where $L(\boldsymbol{\theta})$ and $L(\boldsymbol{\theta}^c)$ are the likelihoods evaluated at the current and candidate states respectively, with candidates lying outside their prior domains rejected.

\FloatBarrier
\clearpage
\bibliographystyle{elsarticle-harv}
\bibliography{C:/Philip/Git/Cod/LaTeX/phil}

\end{document}


\maketitle

\renewcommand{\thesection}{SM\arabic{section}} 
\renewcommand{\thefigure}{SM\arabic{figure}} 
\renewcommand{\thetable}{SM\arabic{table}} 

\section*{Overview} \label{Ovr}

This supplementary material provides supporting information for the article referenced above.\\

Section~\ref{SM:TS:Glb} provides illustrations of global annual maximum and mean data for four climate variables (\RS, \WS, \WM, \TA) where available, and global minimum \TA, from each of seven GCMs (\AC, \CA, \CE, \EC, \MR, \No, \UK). For each combination of GCM and climate variable, figure panels give time series for multiple climate ensemble runs under three climate scenarios (\SL, green; \SM, orange; \SH, grey). Note that the data presented are not smoothed. Plots in this section complement discussion around Figure~1 of the main text.\\

Section~\ref{SM:TS:ClmZon} provides illustrations of climate zone annual maximum and mean data for the four climate variables where available, and climate zone minimum \TA, from each of the seven GCMs. For each GCM, two figures are presented: one for annual maximum and \TA minimum (in Section~\ref{SM:TS:ClmZon:MxmMnm}), the other for annual mean (in Section~\ref{SM:TS:ClmZon:Men}). Note that the data presented are not smoothed. Plots in this section complement discussion around Figure~2 of the main text.\\

Section~\ref{SM:TS:ClmZon:LnrRgrSlp} provides illustrations of estimated slopes of linear regression lines fitted directly to annual mimumum data for \TA, and also for annual means, per given climate zone, GCM, climate scenario and ensemble member. Plots in this section complement discussion around Figure~3 of the main text.\\

Section~\ref{SM:TS:NACS} illustrates time-series of annual data for the centre locations of the North Atlantic and Celtic Sea region neighbourhoods considered, for the four climate variables from each of the seven GCMs. For each combination of GCM and climate variable, figure panels give time series for multiple climate ensemble runs under the three climate scenarios. Note that the data presented are not smoothed. Plots in this section complement discussion in Section 2.2 of the main text.\\

Section~\ref{SM:Rsl:Glb} provides estimated posterior cumulative distribution functions for GEVR model parameters for global annual maxima and global annual minima for \TA (panels (a)), and NHGR model parameter for global means (panels (b)) from the \UK GCM, with separate figures for each of the climate variables. These plots support the discussion of Figures~5-7 of the main text.\\

Section~\ref{SM:DQDM} provides supporting information for linear mixed effects modelling of $\Delta^Q = Q_{2125}-Q_{2025}$ and $\Delta^M = M_{2125}-M_{2025}$. Section~\ref{SM:DQDM:DQMxm} shows box-whisker plots for the posterior distribution of $\Delta^Q$ estimated per climate variable, spatial zone, scenario, model and ensemble. Sections~\ref{SM:DQDM:TblDQMnm} and \ref{SM:DQDM:TblDM} give tables characterising changes $\Delta^Q$ in return values of the annual minimum, and $\Delta^M$ in the annual mean, to be compared with Tables~2 and 4 of the main text for the change $\Delta^Q$ in the 100-year value of the annual maximum for global and climate zone maxima.

\clearpage
\FloatBarrier
\section{Time-series of global annual maxima, \TA minima and means} \label{SM:TS:Glb}

\begin{figure}[!ht]
	\centering
	\includegraphics[width=1\textwidth]{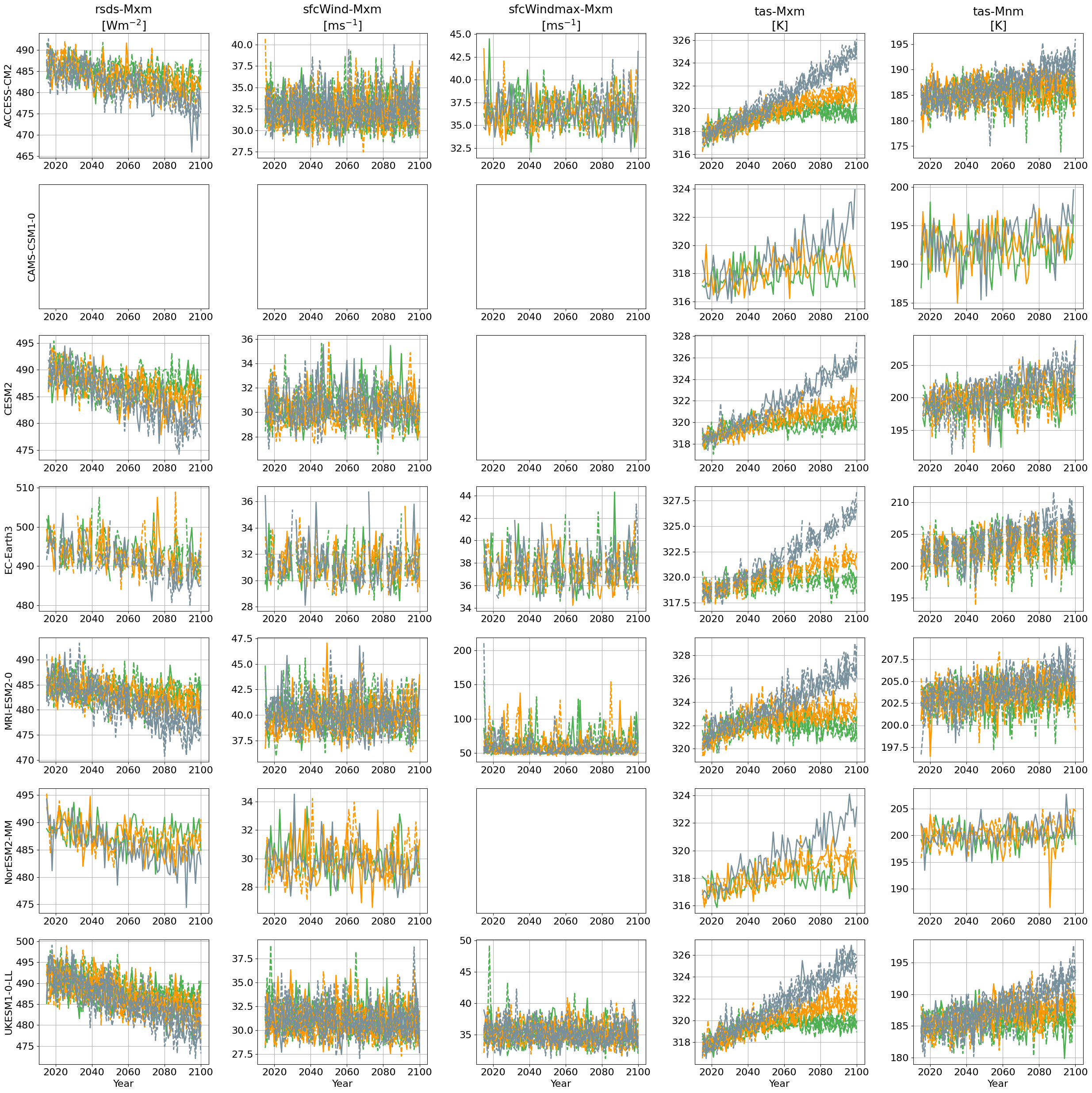}
	\caption{Time series of global annual maximum data for four climate variables (\RS, \WS, \WM, \TA) where available, and global minimum \TA, from each of seven GCMs (\AC, \CA, \CE, \EC, \MR, \No, \UK). Rows indicate GCM, columns indicate climate variable and type of extreme. For each combination of GCM and climate variable, panels give time series for multiple climate ensemble runs (distinguished by line style) under three climate scenarios (\SL, green; \SM, orange; \SH, grey). Empty panels indicate that data for the specific combination of GCM and climate variable were not available for analysis. Note a suspect value in ensemble \texttt{r1i1p1f1} for annual minimum \TA for \No under scenario \SM, discussed in Section~2.1 of the main text.} 
	\label{Fgr-TimSrsDgn-Glb-MxmMnm-Raw}
\end{figure}

\begin{figure}[!ht]
	\centering
	\includegraphics[width=1\textwidth]{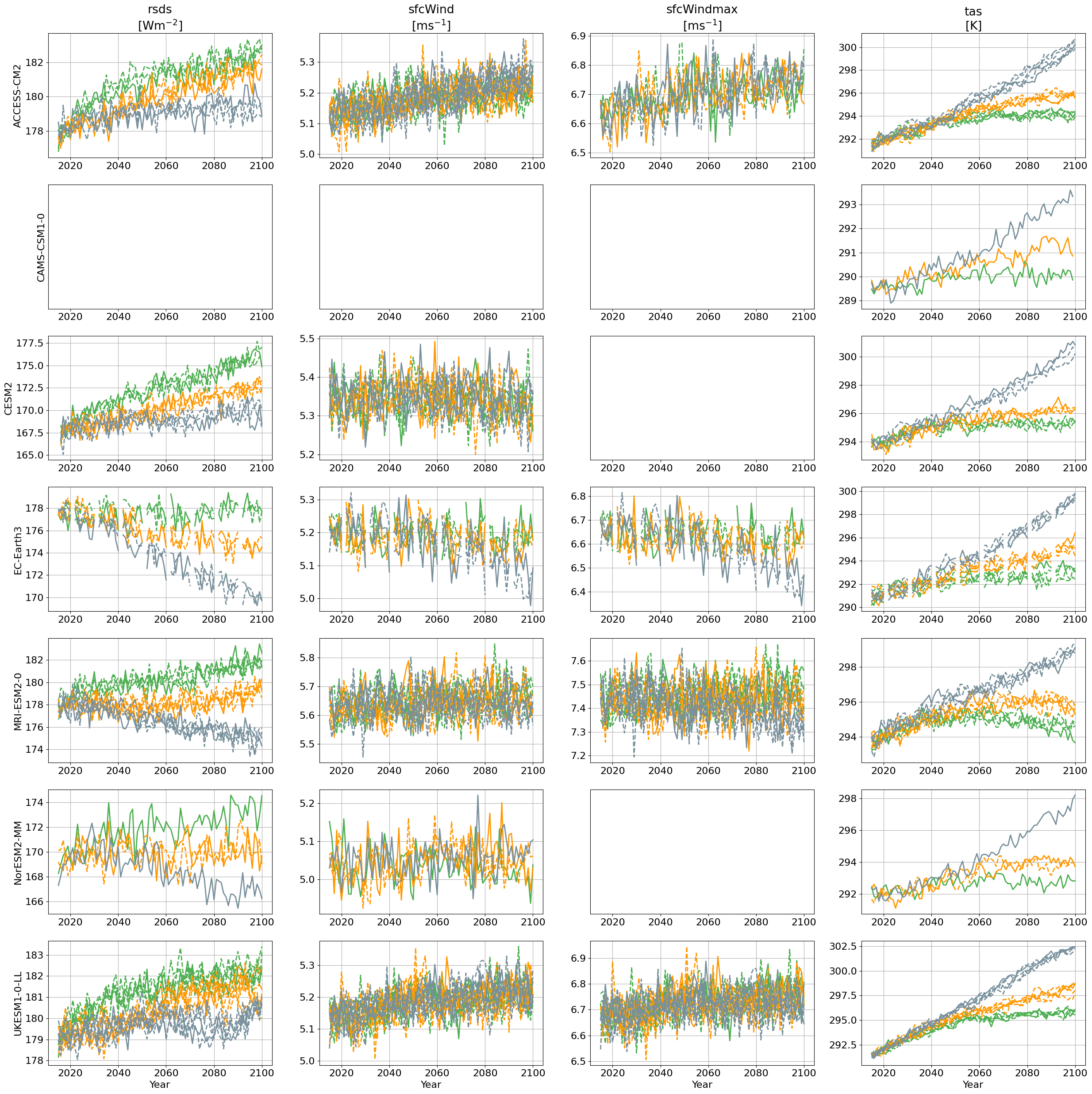}
	\caption{Time series of global annual mean data for four climate variables (\RS, \WS, \WM, \TA) where available, from each of seven GCMs (\AC, \CA, \CE, \EC, \MR, \No, \UK). Rows indicate GCM, columns indicate climate variable. For each combination of GCM and climate variable, panels give time series for multiple climate ensemble runs (distinguished by line style) under three climate scenarios (\SL, green; \SM, orange; \SH, grey). Empty panels indicate that data for the specific combination of GCM and climate variable were not available for analysis.}
	\label{Fgr-TimSrsDgn-Glb-Men-Raw}
\end{figure}

\clearpage
\FloatBarrier
\section{Time-series of annual maxima, \TA minima and means for climate zones} \label{SM:TS:ClmZon}

\FloatBarrier
\subsection{Maxima/Minima} \label{SM:TS:ClmZon:MxmMnm}

\foreach \gcm/\gcmv in {AC/ACV, CA/CAV, CE/CEV, EC/ECV, MR/MRV, No/NoV, UK/UKV} {
	\begin{figure}[!ht]
		\centering
		\includegraphics[width=1\textwidth]{Fgr-ClmZon-TS-\gcm-MxmMnm-Raw.png}
		\ifthenelse{\equal{\gcm}{MR}}
			{\caption{Time-series plots of annual maximum \csname \gcmv \endcsname (rows 1-4) and annual minimum \TA (row 5), for each of 5 climate zones (columns), for \csname \gcm \endcsname. Colour indicates climate scenario (\SL, green; \SM, orange; \SH, grey) with different line style for each ensemble member. No smoothing of time-series is performed. Note suspect values for \WM in the Temperate North.}}
			{\ifthenelse{\equal{\gcm}{No}}
				{\caption{Time-series plots of annual maximum \csname \gcmv \endcsname (rows 1-4) and annual minimum \TA (row 5), for each of 5 climate zones (columns), for \csname \gcm \endcsname. Colour indicates climate scenario (\SL, green; \SM, orange; \SH, grey) with different line style for each ensemble member. No smoothing of time-series is performed. Note suspect low value of \TA minimum in the Antarctic under scenario \SM.}}
				{\caption{Time-series plots of annual maximum \csname \gcmv \endcsname (rows 1-4) and annual minimum \TA (row 5), for each of 5 climate zones (columns), for \csname \gcm \endcsname. Colour indicates climate scenario (\SL, green; \SM, orange; \SH, grey) with different line style for each ensemble member. No smoothing of time-series is performed.}}}
				
		\label{Fgr-TimSrsDgn-\gcm-ClmZon-MxmMnm-Raw}
	\end{figure}
}

\FloatBarrier
\subsection{Means} \label{SM:TS:ClmZon:Men}

\foreach \gcm/\gcmv in {AC/ACV, CA/CAV, CE/CEV, EC/ECV, MR/MRV, No/NoV, UK/UKV} {
	\begin{figure}[!ht]
		\centering
		\includegraphics[width=1\textwidth]{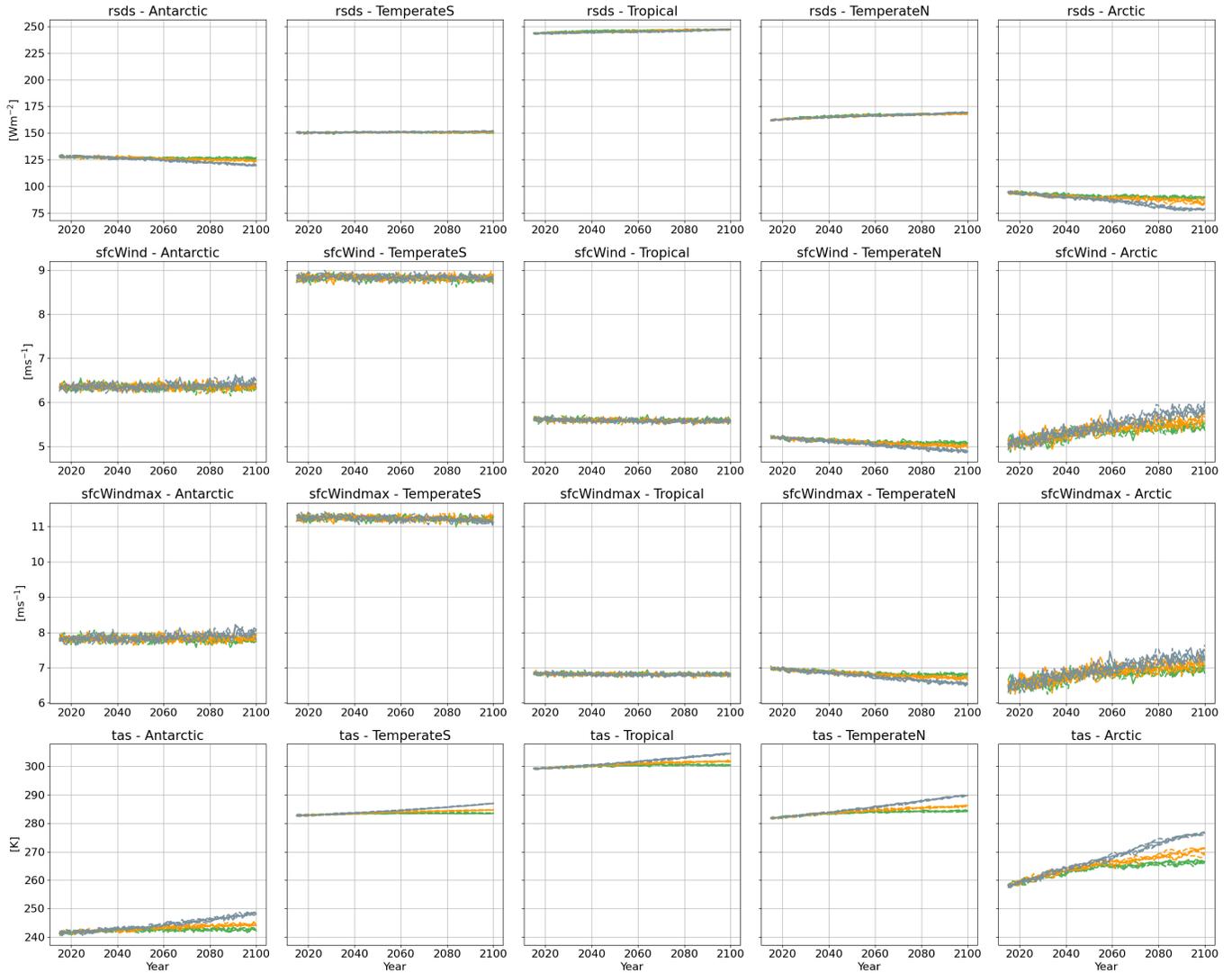}
		\caption{Time-series plots of annual mean \csname \gcmv \endcsname (rows), for each of 5 climate zones (columns), for \csname \gcm \endcsname. Colour indicates climate scenario (\SL, green; \SM, orange; \SH, grey) with different line style for each ensemble member. No smoothing of time-series is performed. Note unusually low value for \TA minimum in the Antarctic.}
		\label{Fgr-TimSrsDgn-\gcm-ClmZon-Men-Raw}
	\end{figure}
}

\clearpage
\FloatBarrier
\section{Slopes of linear regression lines through annualised data} \label{SM:TS:ClmZon:LnrRgrSlp}

\FloatBarrier
\subsection{Minima} \label{SM:TS:ClmZon:LnrRgrSlp:MxmMnm}

\begin{figure}[!ht]
	\centering
	\includegraphics[width=1\textwidth]{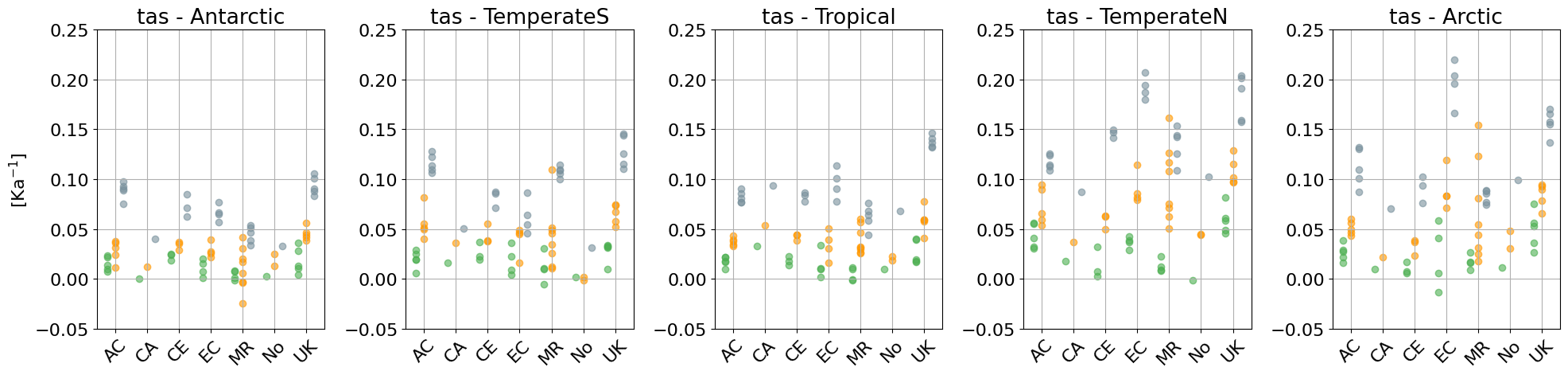}
	\caption{Estimated slopes of linear regression lines fitted directly to the annual mimumum data for \TA per given climate zone, GCM, climate scenario and ensemble member. Columns indicate climate zone. Discs indicate slopes for ensemble members, and disc colour indicates climate scenario (\SL, green; \SM, orange; \SH, grey). \ed{Two letter acronyms are used to refer to GCMs, as detailed in Table 1 of the main text.}}
	\label{Fgr-ClmZon-Mnm-LnrRgrTrn}
\end{figure}
 
\FloatBarrier
\subsection{Means} \label{SM:TS:ClmZon:LnrRgrSlp:Men}
 
\begin{figure}[!ht]
	\centering
	\includegraphics[width=1\textwidth]{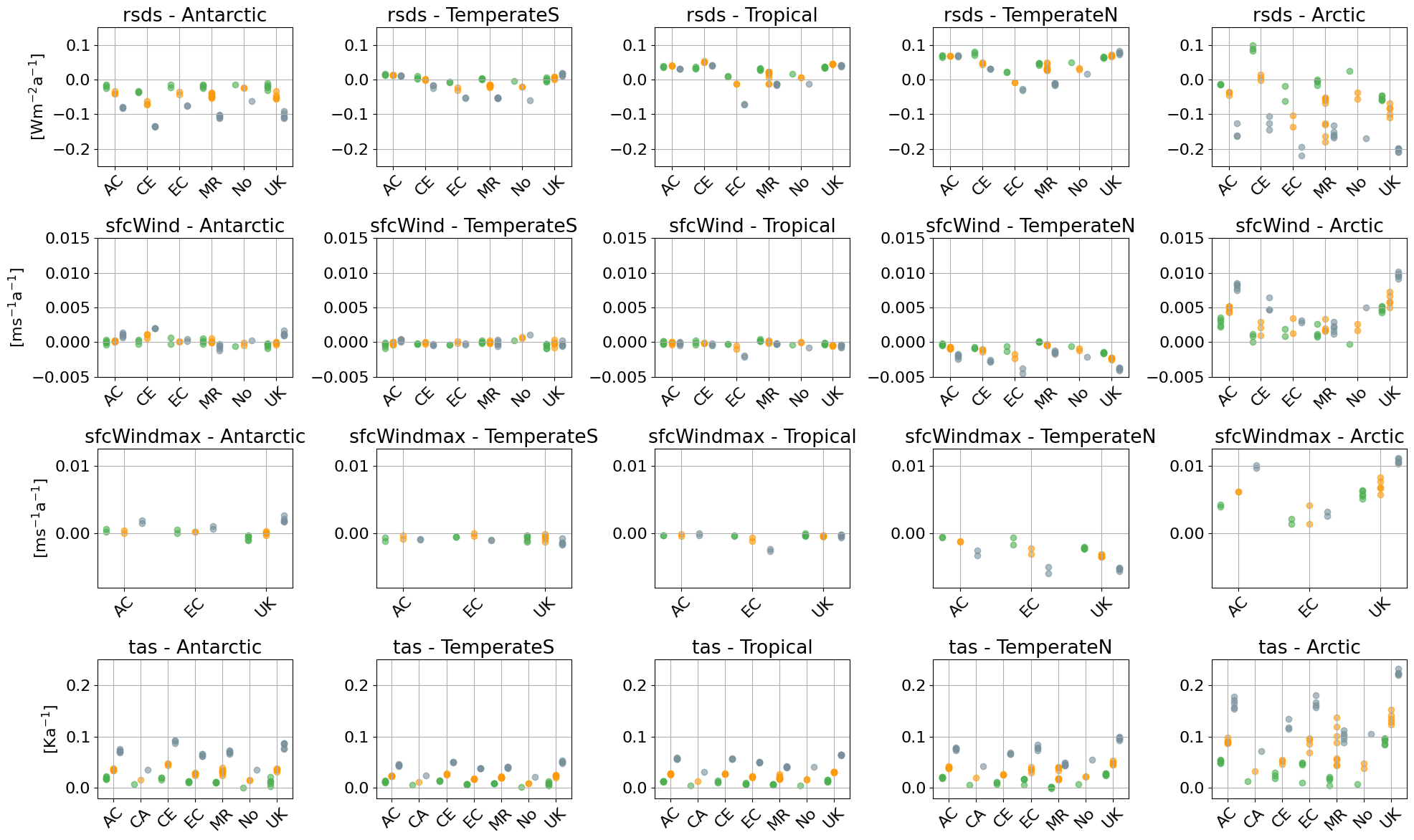}
	\caption{Slopes of linear regression lines fitted directly to the annual mean data for each climate variable, given climate zone, GCM, climate scenario and ensemble member. Rows indicate different climate variables. Columns indicate climate zone. Discs indicate slopes for ensemble members, and disc colour indicates climate scenario (\SL, green; \SM, orange; \SH, grey). }
	\label{Fgr-ClmZon-Men-LnrRgrTrn}
\end{figure}

\clearpage
\FloatBarrier
\section{Examples of time-series for annual maxima from the North Atlantic and Celtic Sea region} \label{SM:TS:NACS}

\begin{figure}[!ht]
	\centering
	\includegraphics[width=1\textwidth]{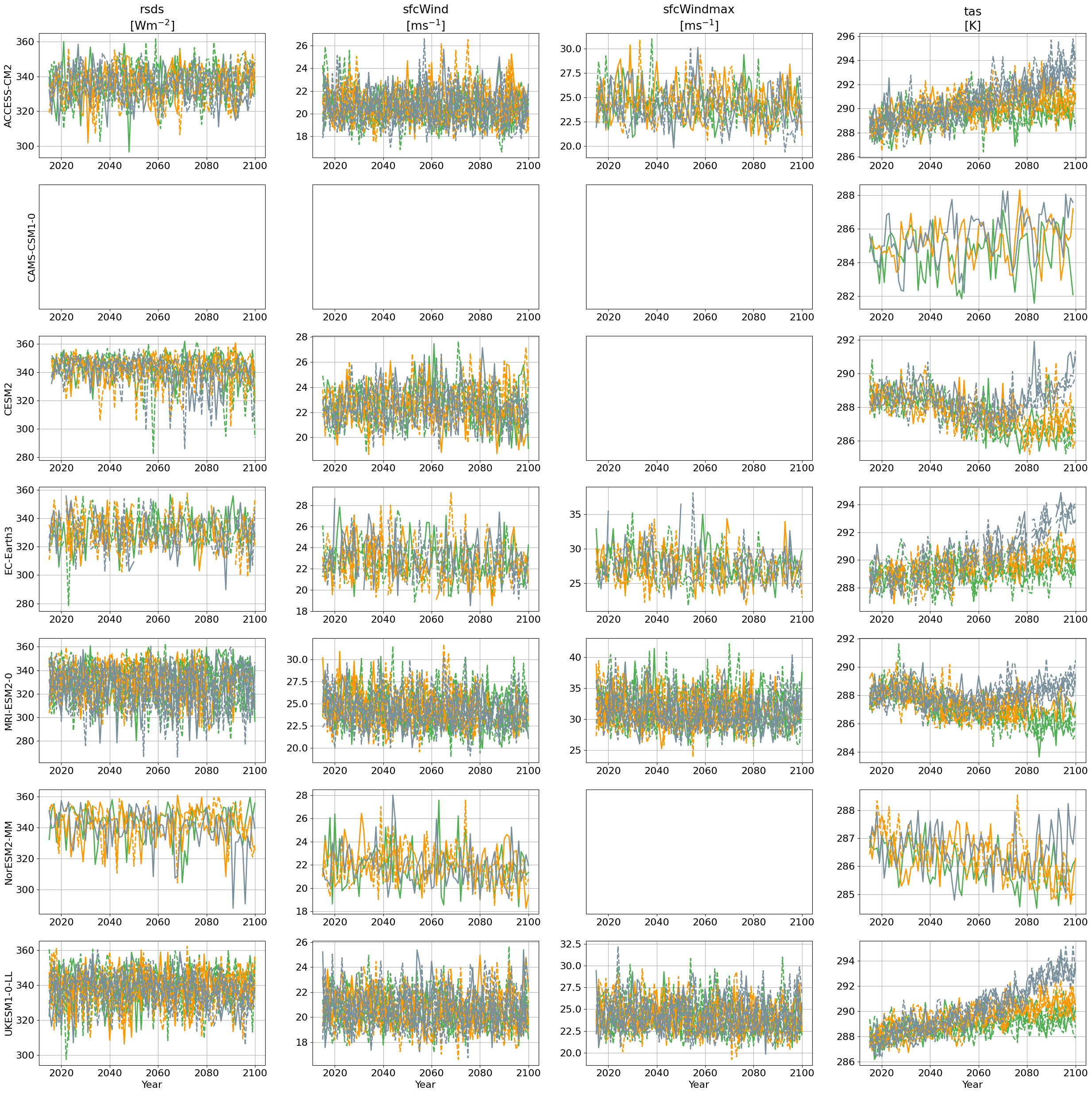}
	\caption{Time-series of annual maximum data for the centre location of the North Atlantic neighbourhood, for the four climate variables from each of the seven GCMs. For each combination of GCM and climate variable, panels give time series for multiple climate ensemble runs (distinguished by line style) under the three climate scenarios (\SL, green; \SM, orange; \SH, grey). Note that the data presented are not smoothed. \ed{Empty panels indicate that data for the specific combination of GCM and climate variable were not available for analysis.}}
	\label{Fgr-TimSrsDgn-NA-Cnt}
\end{figure}

\begin{figure}[!ht]
	\centering
	\includegraphics[width=1\textwidth]{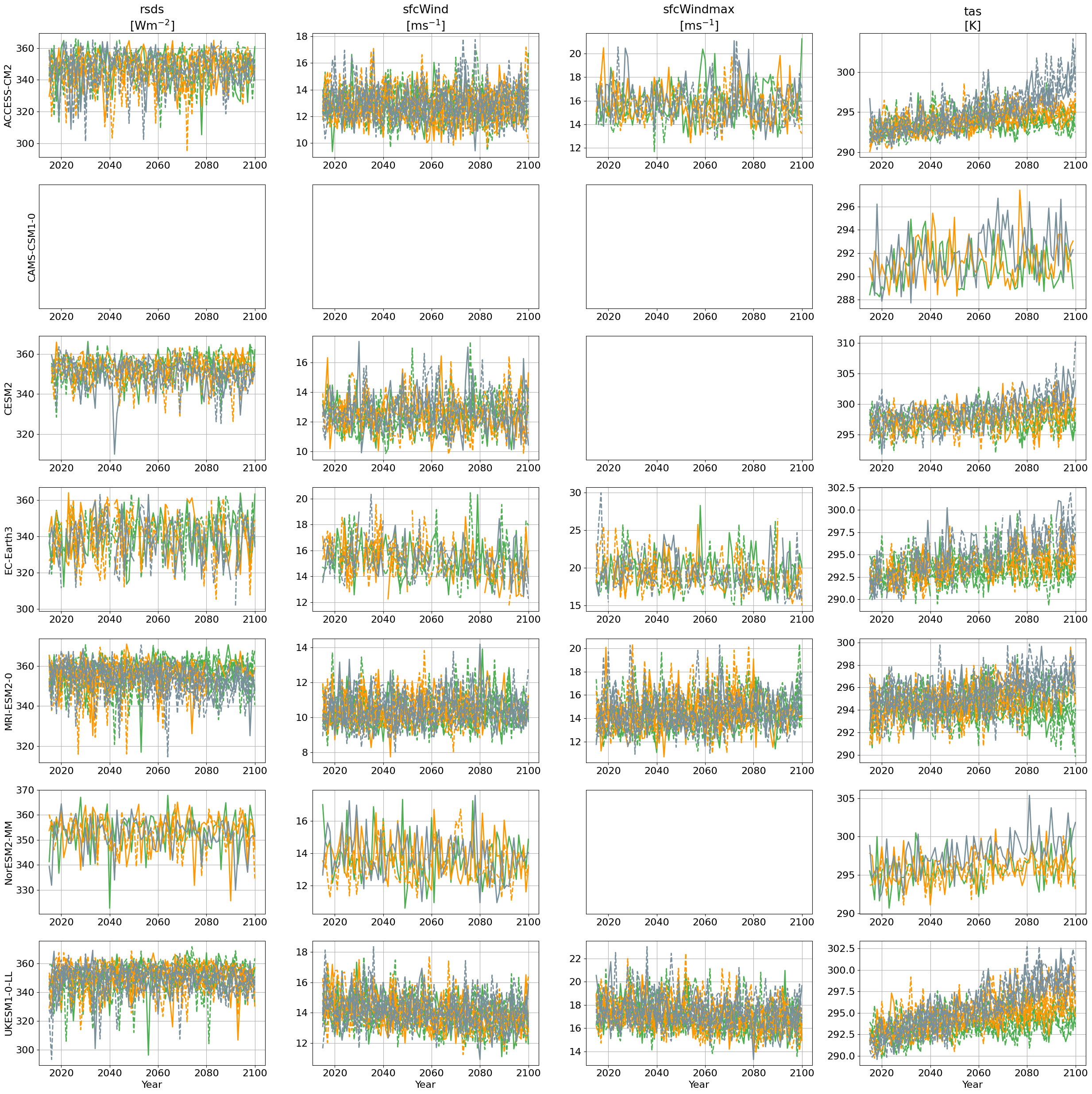}	\caption{Time-series of annual maximum data for the centre location of the Celtic Sea neighbourhood, for the four climate variables from each of the seven GCMs. For each combination of GCM and climate variable, panels give time series for multiple climate ensemble runs (distinguished by line style) under the three climate scenarios (\SL, green; \SM, orange; \SH, grey). Note that the data presented are not smoothed. \ed{Empty panels indicate that data for the specific combination of GCM and climate variable were not available for analysis.}}
	\label{Fgr-TimSrsDgn-CS-Cnt}
\end{figure}

\clearpage
\FloatBarrier
\section{Diagnostic plots for GEVR model fitting to global annual data, and NHGR modelling fitting to global means, for \UK} \label{SM:Rsl:Glb}

\foreach \var in {RS, WS, WM, TA} {
	\begin{figure}[!ht]
		\centering
		\begin{subfigure}{0.7\textwidth}
			\includegraphics[width=1\textwidth]{Fgr-Glb-Cdf-UK-\var-Mxm.png}
			\caption{Estimated posterior cumulative distribution functions for GEVR model parameters and 100-year return values ($Q_{2025}$, $Q_{2125}$, for years 2025 and 2125), for global annual maxima from the specified GCM and climate variable. Colours indicate climate scenario (\SL, green; \SM, orange; \SH, grey). Dashed line styles indicate inferences for different ensemble members; a solid line corresponds to the ensemble mean cumulative distribution function. Model parameters are the GEV location $\mu^0$, scale $\sigma^0$ and shape $\xi^0$ in 2015, and the changes $\mu^1$, $\sigma^1$ and $\xi^1$ in those parameters over the period (2015,2100). Return value for year $t$ is denoted by $Q_t$.}
		\end{subfigure}
		\begin{subfigure}{0.7\textwidth}
			\includegraphics[width=1\textwidth]{Fgr-Glb-DgnCdf-UK-\var-Men.png}
			\caption{Estimated posterior cumulative distribution functions for NHGR  model parameters and predicted values ($M_{2025}$, $M_{2125}$, for years 2025 and 2125), for the global annual mean from the specified GCM and climate variable. Colours indicate climate scenario: SSP126 (green), SSP245 (orange), SSP585 (grey). Dashed line styles indicate inferences for different ensemble members; solid line corresponds to the ensemble mean. Model parameters are the NHGR mean $\alpha^0$, scale $\beta^0$ in 2015, and the changes $\alpha^1$, $\beta^1$ in those parameters over the period (2015,2100). Predicted value for year $t$ is denoted by $M_t$.}
		\end{subfigure}
		\caption{\UK, \csname \var \endcsname}
	\end{figure}
}

\begin{figure}[!ht]
	\centering
	\begin{subfigure}{0.7\textwidth}
		\includegraphics[width=1\textwidth]{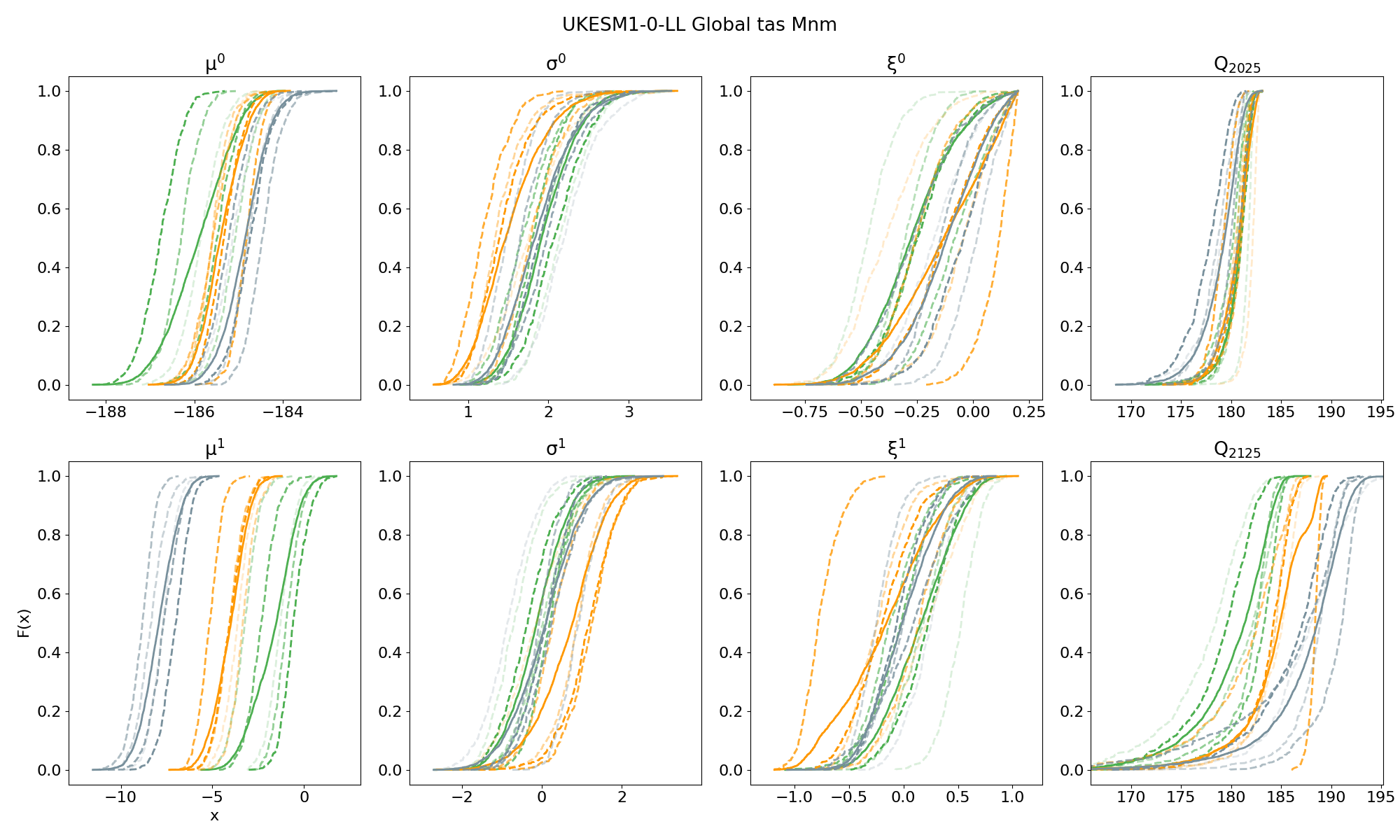}
		\caption*{Estimated posterior cumulative distribution functions for GEVR model parameters and 100-year return values ($Q_{2025}$, $Q_{2125}$, for years 2025 and 2125), for global annual minima from the \TA from \UK. Colours indicate climate scenario (\SL, green; \SM, orange; \SH, grey). Dashed line styles indicate inferences for different ensemble members; a solid line corresponds to the ensemble mean cumulative distribution function. Note that extreme value analysis is performed on negated annual minimum \TA. Model parameters are the GEV location $\mu^0$, scale $\sigma^0$ and shape $\xi^0$ in 2015, and the changes $\mu^1$, $\sigma^1$ and $\xi^1$ in those parameters over the period (2015,2100). 100-year return values for annual minimum \TA in year $t$ is denoted by $Q_t$. Results indicate that extremely low temperatures become less frequent in future with increasing climate forcing.}
	\end{subfigure}
	\caption{\TA for \UK}
\end{figure}

\clearpage
\FloatBarrier
\section{Supporting information for synoptic modelling of $\Delta^Q$ and $\Delta^M$} \label{SM:DQDM}

\subsection{Box-whisker plots for $\Delta^Q$ per climate ensemble for global climate zones} \label{SM:DQDM:DQMxm}

\begin{figure}[!ht]
	\centering
	\includegraphics[width=1\textwidth]{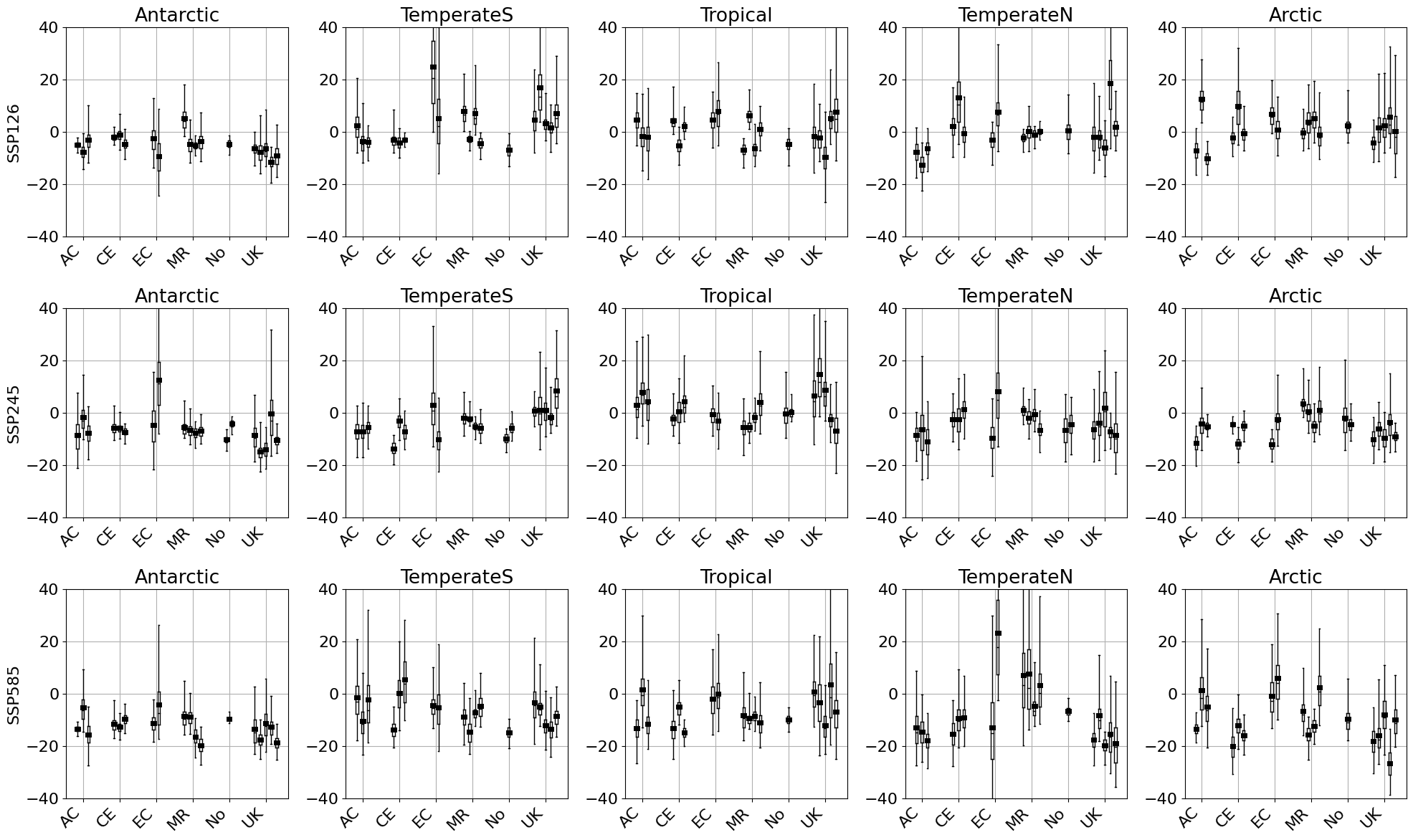}
	\caption{Box-whisker plots for the posterior distribution of the change $\Delta^Q$ in 100-year return value $Q_t$ of spatial annual maximum over the next 100-years per climate ensemble, for \RS. Rows represent climate scenarios (\SL, \SM, \SH), and columns spatial domain (Antarctic, Temperate South, Tropical, Temperate North and Arctic). Each panel gives box-whiskers per climate ensemble grouped by GCM (\AC, \CE, \EC, \MR, \No and \UK). The first two characters of the GCM name are used for concise labelling. For each box-whisker, the box represents the central (25\%, 75\%) interval for the posterior distribution, and the whisker the central (2.5\%, 97.5\%) interval. The posterior median is shown as a thin horizontal line (of the same width as the box), and the posterior mean as a thicker horizontal line. Vertical extents of panels have been restricted so that all central (25\%, 75\%) intervals is clear. \ed{See Section 4.2 and Figure 9 of the main text for discussion}.}
	\label{Fgr-ClmZon-BW-Mxm-Ens-RS}
\end{figure}

\begin{figure}[!ht]
	\centering
	\includegraphics[width=1\textwidth]{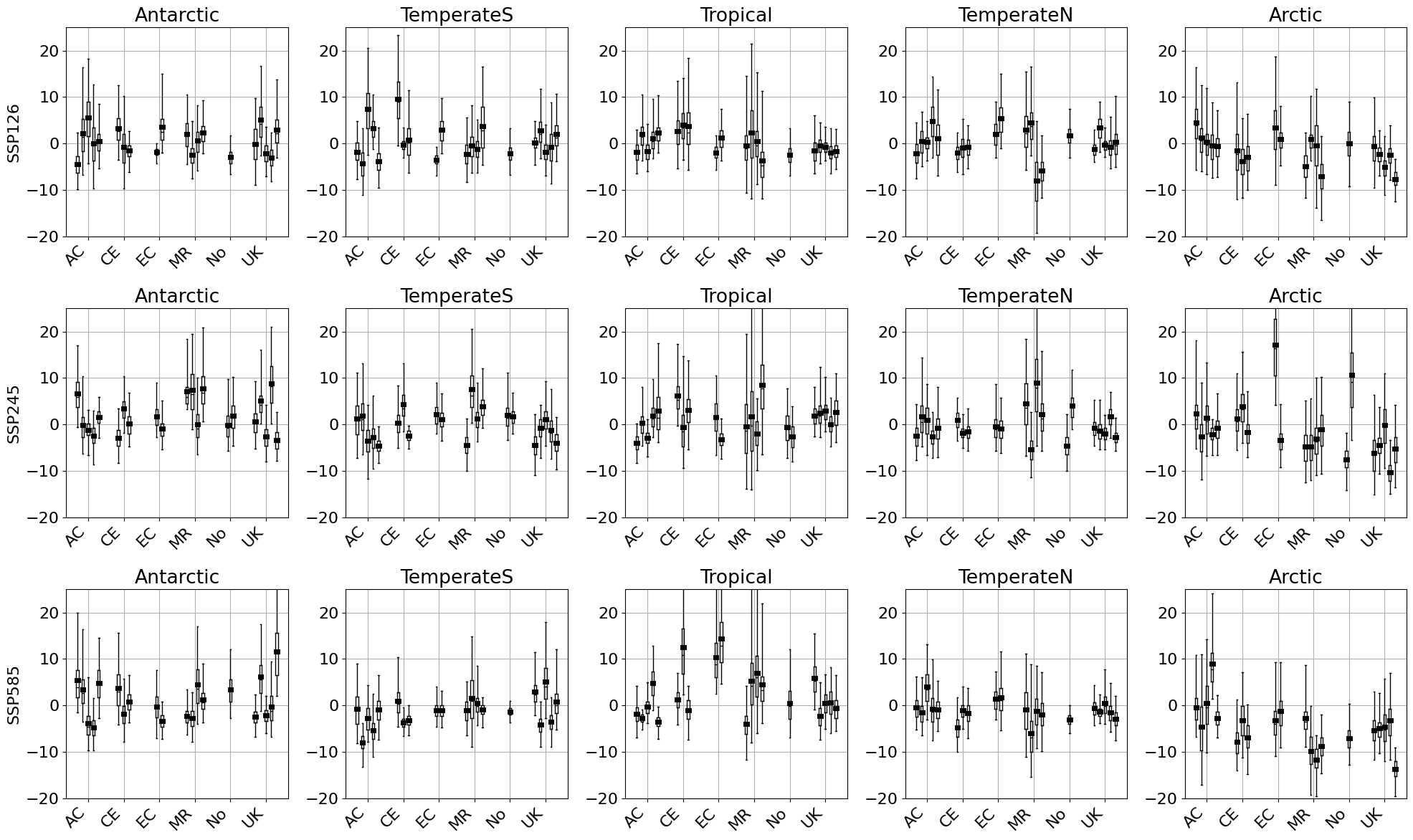}
	\caption{Box-whisker plots for the posterior distribution of the change $\Delta^Q$ in 100-year return value $Q_t$ of spatial annual maximum over the next 100-years per climate ensemble, for \WS. For details, see Figure~\ref{Fgr-ClmZon-BW-Mxm-Ens-RS}.}
	\label{Fgr-ClmZon-BW-Mxm-Ens-WS}
\end{figure}

\begin{figure}[!ht]
	\centering
	\includegraphics[width=1\textwidth]{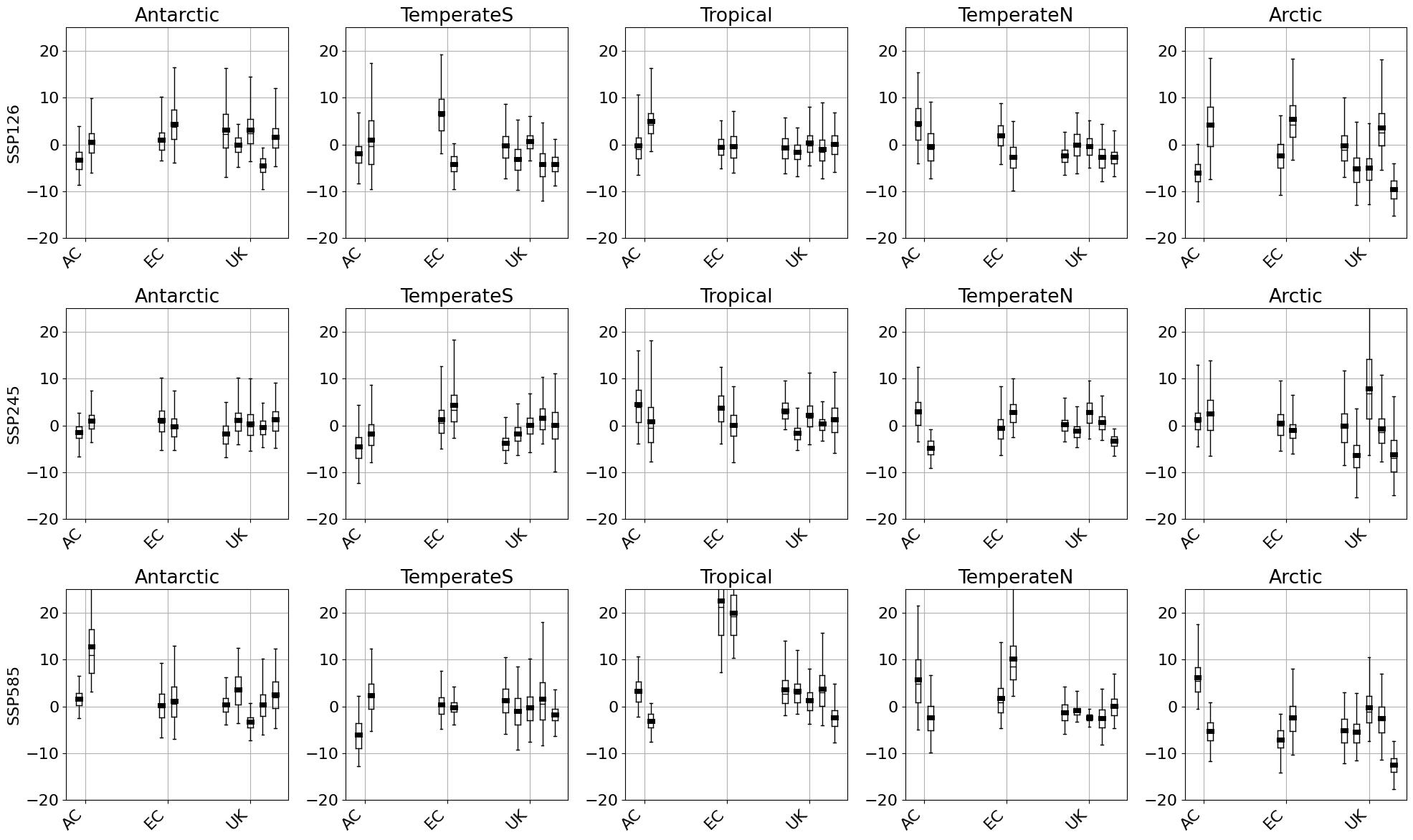}
	\caption{Box-whisker plots for the posterior distribution of the change $\Delta^Q$ in 100-year return value $Q_t$ of spatial annual maximum over the next 100-years per climate ensemble, for \WM. For details, see Figure~\ref{Fgr-ClmZon-BW-Mxm-Ens-RS}.}
	\label{Fgr-ClmZon-BW-Mxm-Ens-WM}
\end{figure}

\begin{figure}[!ht]
	\centering
	\includegraphics[width=1\textwidth]{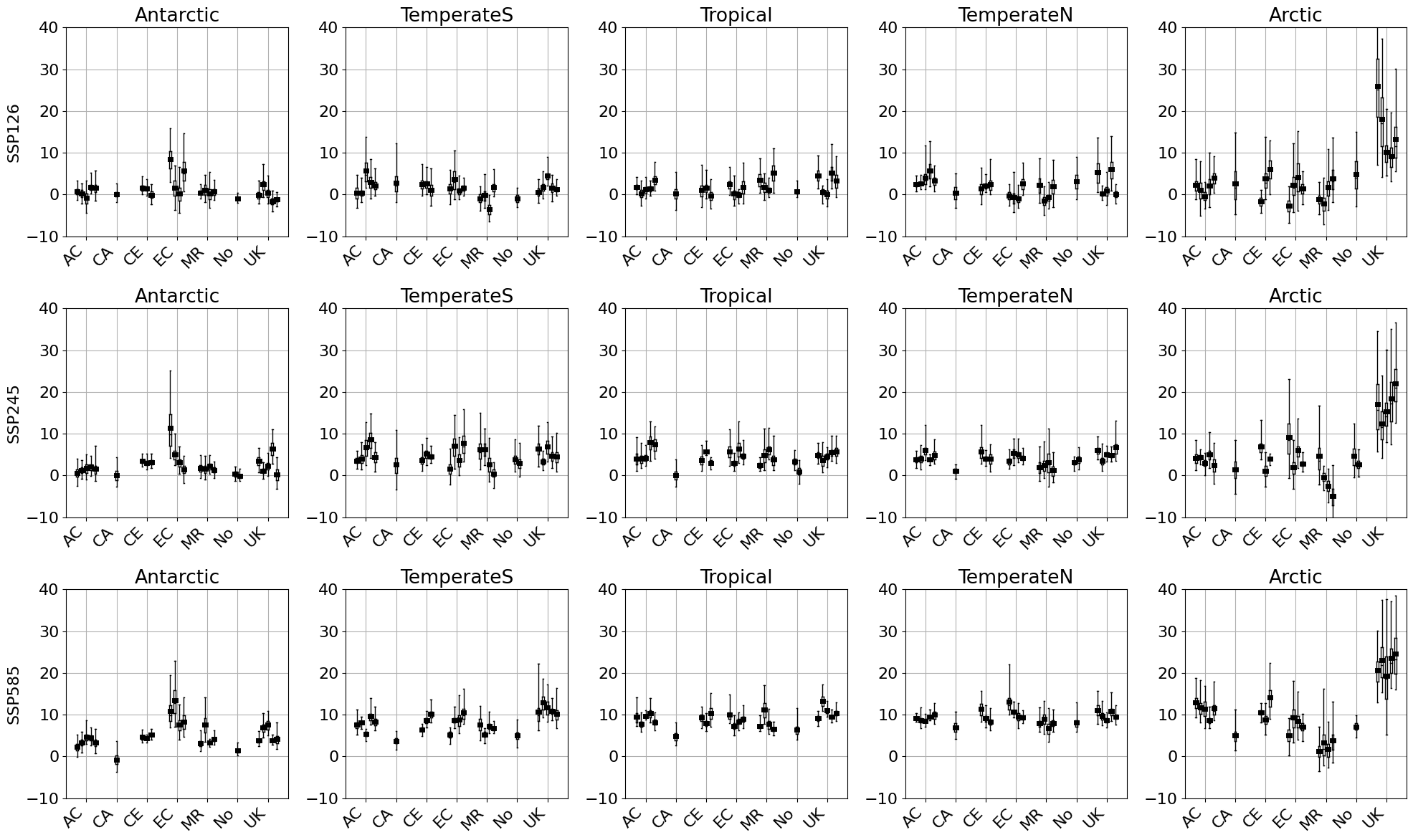}
	\caption{Box-whisker plots for the posterior distribution of the change $\Delta^Q$ in 100-year return value $Q_t$ of spatial annual maximum over the next 100-years per climate ensemble, for \TA. For details, see Figure~\ref{Fgr-ClmZon-BW-Mxm-Ens-RS}.}
	\label{Fgr-ClmZon-BW-Mxm-Ens-TA}
\end{figure}

\clearpage
\FloatBarrier
\subsection{Summary tables for the change $\Delta^Q$ in the annual minimum} \label{SM:DQDM:TblDQMnm}

\begin{table}[!ht]
	\begin{center}
		\begin{tabular}{|c|c|c|c|c|r|r|r|}
			\hline
			\multicolumn{1}{|c|}{\multirow{2}{*}{Variable\Tstrut}} &
			\multicolumn{1}{|c|}{\multirow{2}{*}{Zone\Tstrut}} &
			\multicolumn{3}{|c|}{\(\mathbb{E}(\Delta^Q)\)\Tstrut} &
			\multicolumn{3}{|c|}{\(\mathbb{P}(\Delta^Q > 0)\)\Tstrut} \\ \cline{3-8}
			& & \(126\) & \(245\) & \(585\) & \(126\) & \(245\) & \(585\) \\ \hline
			\multirow{6}{*}{\begin{tabular}{c} \TA \\ $[\text{K}]$ \end{tabular}} 
			& GL\Tstrut & -0.61 & 2.2 & 5.84 & 0.56 & 0.74 & 0.87 \\
			\cline{2-8}
			& AN\Tstrut & -0.28 & 2.25 & 5.88 & 0.58 & 0.74 & 0.87 \\
			& TS\Tstrut & 1.65 & 5.24 & 9.18 & 0.67 & 0.85 & 0.89 \\
			& TR\Tstrut & 1.33 & 2.44 & 8.86 & 0.66 & 0.74 & 0.93 \\
			& TN\Tstrut & 2.94 & 7.02 & 17.42 & 0.74 & 0.87 & 0.98 \\
			& AR\Tstrut & 1.64 & 3.16 & 11.38 & 0.65 & 0.75 & 0.92 \\
			\hline
		\end{tabular}%
	\end{center}

	\caption{Estimated expected change $\mathbb{E}(\Delta^Q)$ and probability of increase $\mathbb{P}(\Delta^Q>0)$ of the 100-year return values over the period (2025,2125) for annual \TA minima over 6 climate zones (GL: global; AN: Antarctic; TS: Temperate South; TR: Tropical; TN: Temperate North; AR: Arctic). Columns under ``$\mathbb{E}(\Delta^Q)$'' show expected changes per scenario, in the units of the variable; thus we estimate an increase of 7.02K in the 100-year minimum \TA in the Temperate North under scenario \ed{\SM} over the next 100 years. Columns under ``$\mathbb{P}(\Delta^Q>0)$'' show corresponding probabilities of increasing return value. Estimates are calculated assuming equal weighting for each climate model. See notes about suspect values for \WM under model \MR, and suspect run \texttt{r1i1p1f1} for annual minimum \TA for \No under scenario \SM, discussed in Section~2.1 of the main text.}

	\label{Tbl:ExpPrb:GATTTA:Mnm}

\end{table}

\begin{table}[!ht]
	\begin{center}
		\begin{tabular}{|c|c|c|c|c|r|r|r|r|r|r|r|}
			\hline
			\multicolumn{1}{|c|}{\multirow{2}{*}{Variable\Tstrut}} &
			\multicolumn{1}{|c|}{\multirow{2}{*}{Zone\Tstrut}} &
			\multicolumn{3}{|c|}{SSP effect\Tstrut} &
			\multicolumn{5}{|c|}{Model standard deviations\Tstrut} &
			\multicolumn{2}{|c|}{$R^2$\Tstrut} \\ \cline{3-12}
			& & \(\iota + \gamma_1\) & \(\gamma_2 - \gamma_1\) & \(\gamma_3 - \gamma_1\) & \(\tau_R\) & \(\tau_{\text{FE}}\) & \(\tau_{\epsilon}\) & \(\tau_{\delta}\) & \(\tau_{\zeta}\) & \(R^2_{\text{FE}}\) & \(R^2_{\text{ME}}\) \\ \hline
			\multirow{6}{*}{\begin{tabular}{c} \TA \\ $[\text{K}]$ \end{tabular}} 
			& GL\Tstrut & -0.69 & 3.0 & 6.45 & 6.51 & 5.95 & 5.49 & 0.32 & 2.26 & 0.16 & 0.29 \\
			\cline{2-12}
			& AN\Tstrut & -0.33 & 2.7 & 6.16 & 6.38 & 5.86 & 5.45 & 0.52 & 2.09 & 0.16 & 0.27 \\
			& TS\Tstrut & 0.97 & 3.65 & 7.53 & 7.77 & 7.13 & 6.32 & 2.26 & 2.43 & 0.16 & 0.34 \\
			& TR\Tstrut & 1.48 & 0.85 & 7.51 & 6.61 & 5.72 & 5.04 & 1.03 & 2.47 & 0.25 & 0.42 \\
			& TN\Tstrut & 1.85 & 3.93 & 14.49 & 9.38 & 7.11 & 6.16 & 3.53 & 2.34 & 0.43 & 0.57 \\
			& AR\Tstrut & 1.67 & 1.46 & 9.74 & 9.97 & 9.0 & 8.46 & 0.0 & 3.02 & 0.18 & 0.28 \\
			\hline
		\end{tabular}%
	\end{center}

	\caption{Summary of linear mixed effects modelling for the change $\Delta^Q$ in the 100-year return value of annual minima for \TA over 6 climate zones (GL: global; AN: Antarctic; TS: Temperate South; TR: Tropical; TN: Temperate North; AR: Arctic). Columns under ``SSP effect'' show intercept $\iota$ and fixed effect parameter estimates $\gamma_j$ for the change in 100-year return value of the given variable over the period (2025,2125) under scenario $j$, in the units of the variable. Columns under ``Model standard deviations'' provide estimates of the various standard deviations of model fitting. $\tau_R$: the (full unconditional) standard deviation of the response $R$; $\tau_{\text{FE}}$: the model error standard deviation after fitting only the fixed effects (FE, of climate scenario); $\tau_{\epsilon}$: the model error standard deviation after fitting the full mixed effects model. $\tau_{\delta}$: standard deviation of climate model random effect; $\tau_{\zeta}$: standard deviation due to nested random effect of climate ensemble within climate model. $R^2$ statistics are also provided, from fitting only the fixed effects (FE), and from fitting the full mixed effects (ME) model. Further, since sample size for model fitting is huge, estimates of uncertainties and ``significance'' are of little practical value, and are omitted. See notes about suspect values for \WM under model \MR, and suspect run \texttt{r1i1p1f1} for annual minimum \TA for \No under scenario \SM, discussed in Section~2.1 of the main text.}
	
	\label{Tbl:LME:GATTTA:Mnm}

\end{table}

\clearpage
\FloatBarrier
\subsection{Summary tables for the change $\Delta^M$ in the annual mean} \label{SM:DQDM:TblDM}

\begin{table}[!ht]
	\begin{center}
		\begin{tabular}{|c|c|c|c|c|r|r|r|}
			\hline
			\multicolumn{1}{|c|}{\multirow{2}{*}{Variable\Tstrut}} &
			\multicolumn{1}{|c|}{\multirow{2}{*}{Zone\Tstrut}} &
			\multicolumn{3}{|c|}{\(\mathbb{E}(\Delta^M)\)\Tstrut} &
			\multicolumn{3}{|c|}{\(\mathbb{P}(\Delta^M > 0)\)\Tstrut} \\ \cline{3-8}
			& & \(126\) & \(245\) & \(585\) & \(126\) & \(245\) & \(585\) \\ \hline
			\multirow{6}{*}{\begin{tabular}{c} \RS \\ $[\text{Wm}^{-2}]$ \end{tabular}} & GL\Tstrut & 3.5 & 2.1 & -1.19 & 0.95 & 0.83 & 0.53 \\
			\cline{2-8}
			& AN\Tstrut & -1.91 & -3.56 & -8.56 & 0.08 & 0.03 & 0.0 \\
			& TS\Tstrut & 0.21 & -0.5 & -1.71 & 0.57 & 0.4 & 0.41 \\
			& TR\Tstrut & 2.39 & 2.08 & 0.82 & 0.96 & 0.8 & 0.62 \\
			& TN\Tstrut & 4.3 & 3.34 & 2.53 & 1.0 & 0.91 & 0.65 \\
			& AR\Tstrut & -0.66 & -5.08 & -14.12 & 0.35 & 0.09 & 0.0 \\
			\hline
			\multirow{6}{*}{\begin{tabular}{c} \WS \\ $[\text{ms}^{-1}]$ \end{tabular}} & GL\Tstrut & 0.04 & 0.05 & 0.04 & 0.7 & 0.71 & 0.68 \\
			\cline{2-8}
			& AN\Tstrut & -0.01 & 0.02 & 0.06 & 0.48 & 0.55 & 0.68 \\
			& TS\Tstrut & -0.02 & -0.0 & -0.01 & 0.39 & 0.48 & 0.47 \\
			& TR\Tstrut & -0.01 & -0.02 & -0.04 & 0.45 & 0.38 & 0.3 \\
			& TN\Tstrut & -0.06 & -0.11 & -0.22 & 0.19 & 0.05 & 0.0 \\
			& AR\Tstrut & 0.2 & 0.31 & 0.51 & 0.82 & 0.9 & 0.94 \\
			\hline
			\multirow{6}{*}{\begin{tabular}{c} \WM \\ $[\text{ms}^{-1}]$ \end{tabular}} & GL\Tstrut & 0.07 & 0.05 & 0.0 & 0.77 & 0.69 & 0.6 \\
			\cline{2-8}
			& AN\Tstrut & -0.01 & 0.01 & 0.14 & 0.46 & 0.53 & 0.84 \\
			& TS\Tstrut & -0.06 & -0.04 & -0.1 & 0.27 & 0.32 & 0.16 \\
			& TR\Tstrut & -0.02 & -0.04 & -0.06 & 0.36 & 0.26 & 0.28 \\
			& TN\Tstrut & -0.13 & -0.23 & -0.41 & 0.05 & 0.01 & 0.0 \\
			& AR\Tstrut & 0.38 & 0.49 & 0.74 & 0.93 & 0.95 & 0.96 \\
			\hline
			\multirow{6}{*}{\begin{tabular}{c} \TA \\ $[\text{K}]$ \end{tabular}} & GL\Tstrut & 1.76 & 3.97 & 7.99 & 0.9 & 1.0 & 1.0 \\
			\cline{2-8}
			& AN\Tstrut & 1.16 & 2.52 & 6.06 & 0.91 & 0.99 & 1.0 \\
			& TS\Tstrut & 0.77 & 1.6 & 3.77 & 0.99 & 1.0 & 1.0 \\
			& TR\Tstrut & 0.83 & 1.91 & 4.48 & 0.97 & 1.0 & 1.0 \\
			& TN\Tstrut & 1.02 & 2.63 & 6.08 & 0.85 & 1.0 & 1.0 \\
			& AR\Tstrut & 3.45 & 6.51 & 13.06 & 0.93 & 1.0 & 1.0 \\
			\hline
		\end{tabular}%
	\end{center}
	
	\caption{Estimated expected change $\mathbb{E}(\Delta^M)$ and probability of increase $\mathbb{P}(\Delta^M>0)$ of the annual mean over the period (2025,2125), for four climate variables over 6 climate zones (GL: global; AN: Antarctic; TS: Temperate South; TR: Tropical; TN: Temperate North; AR: Arctic). Columns under ``$\mathbb{E}(\Delta^M)$'' show expected changes per scenario, in the units of the variable. Columns under ``$\mathbb{P}(\Delta^M>0)$'' show corresponding probabilities of increasing change. Estimates are calculated assuming equal weighting for each climate model. Note that the first (year 2015) observation for \RS in all \CE runs is spurious and has been omitted. Multiple values for \WM from \MR are also suspect (with some values $>100$ ms$^{-1}$); for this reason \MR output is ignored for variable \WM.}

\end{table}

\begin{table}[!ht]
	\begin{center}
		\begin{tabular}{|c|c|c|c|c|r|r|r|r|r|r|r|}
			\hline
			\multicolumn{1}{|c|}{\multirow{2}{*}{Variable\Tstrut}} &
			\multicolumn{1}{|c|}{\multirow{2}{*}{Zone\Tstrut}} &
			\multicolumn{3}{|c|}{SSP effect\Tstrut} &
			\multicolumn{5}{|c|}{Model standard deviations\Tstrut} &
			\multicolumn{2}{|c|}{$R^2$\Tstrut} \\ \cline{3-12}
			& & \(\iota + \gamma_1\) & \(\gamma_2 - \gamma_1\) & \(\gamma_3 - \gamma_1\) & \(\tau_R\) & \(\tau_{\text{FE}}\) & \(\tau_{\epsilon}\) & \(\tau_{\delta}\) & \(\tau_{\zeta}\) & \(R^2_{\text{FE}}\) & \(R^2_{\text{ME}}\) \\ \hline
			\multirow{6}{*}{\begin{tabular}{c} \RS \\ $[\text{Wm}^{-2}]$ \end{tabular}} & GL\Tstrut & 3.05 & -1.32 & -4.69 & 3.49 & 2.89 & 1.5 & 2.76 & 0.32 & 0.31 & 0.82 \\
			\cline{2-12}
			& AN\Tstrut & -1.42 & -2.2 & -6.65 & 3.3 & 1.83 & 1.33 & 1.09 & 1.03 & 0.69 & 0.84 \\
			& TS\Tstrut & -0.03 & -0.6 & -1.76 & 1.99 & 1.84 & 1.24 & 1.28 & 0.65 & 0.15 & 0.61 \\
			& TR\Tstrut & 1.71 & 0.25 & -1.45 & 2.47 & 2.38 & 1.47 & 2.05 & 0.62 & 0.07 & 0.65 \\
			& TN\Tstrut & 3.79 & -0.48 & -1.57 & 2.76 & 2.67 & 1.52 & 2.24 & 0.41 & 0.06 & 0.7 \\
			& AR\Tstrut & -0.38 & -4.51 & -13.49 & 6.7 & 3.94 & 2.42 & 3.14 & 0.81 & 0.65 & 0.87 \\
			\hline
			\multirow{6}{*}{\begin{tabular}{c} \WS \\ $[\text{ms}^{-1}]$ \end{tabular}} & GL\Tstrut & 0.02 & 0.01 & -0.0 & 0.1 & 0.1 & 0.08 & 0.06 & 0.01 & 0.0 & 0.34 \\
			\cline{2-12}
			& AN\Tstrut & -0.01 & 0.02 & 0.07 & 0.12 & 0.12 & 0.11 & 0.03 & 0.03 & 0.05 & 0.14 \\
			& TS\Tstrut & -0.02 & 0.02 & 0.02 & 0.09 & 0.09 & 0.09 & 0.02 & 0.01 & 0.01 & 0.08 \\
			& TR\Tstrut & -0.01 & -0.01 & -0.04 & 0.07 & 0.07 & 0.06 & 0.03 & 0.01 & 0.04 & 0.18 \\
			& TN\Tstrut & -0.06 & -0.05 & -0.17 & 0.11 & 0.08 & 0.04 & 0.06 & 0.02 & 0.42 & 0.82 \\
			& AR\Tstrut & 0.15 & 0.12 & 0.32 & 0.29 & 0.26 & 0.19 & 0.16 & 0.05 & 0.2 & 0.58 \\
			\hline
			\multirow{6}{*}{\begin{tabular}{c} \WM \\ $[\text{ms}^{-1}]$ \end{tabular}} & GL\Tstrut & 0.06 & -0.02 & -0.07 & 0.13 & 0.13 & 0.1 & 0.1 & 0.01 & 0.05 & 0.43 \\
			\cline{2-12}
			& AN\Tstrut & -0.01 & 0.02 & 0.15 & 0.15 & 0.13 & 0.13 & 0.0 & 0.02 & 0.2 & 0.22 \\
			& TS\Tstrut & -0.06 & 0.02 & -0.04 & 0.1 & 0.1 & 0.09 & 0.0 & 0.02 & 0.05 & 0.1 \\
			& TR\Tstrut & -0.03 & -0.02 & -0.04 & 0.08 & 0.08 & 0.07 & 0.04 & 0.01 & 0.05 & 0.26 \\
			& TN\Tstrut & -0.11 & -0.1 & -0.27 & 0.15 & 0.09 & 0.06 & 0.07 & 0.03 & 0.59 & 0.84 \\
			& AR\Tstrut & 0.32 & 0.11 & 0.37 & 0.33 & 0.29 & 0.23 & 0.19 & 0.04 & 0.21 & 0.52 \\
			\hline
			\multirow{6}{*}{\begin{tabular}{c} \TA \\ $[\text{K}]$ \end{tabular}} & GL\Tstrut & 1.21 & 2.27 & 6.23 & 3.17 & 1.87 & 0.86 & 1.57 & 0.35 & 0.65 & 0.93 \\
			\cline{2-12}
			& AN\Tstrut & 0.75 & 1.64 & 4.93 & 2.33 & 1.16 & 0.84 & 0.89 & 0.49 & 0.75 & 0.87 \\
			& TS\Tstrut & 0.51 & 1.01 & 3.02 & 1.36 & 0.57 & 0.32 & 0.52 & 0.3 & 0.83 & 0.94 \\
			& TR\Tstrut & 0.58 & 1.29 & 3.69 & 1.66 & 0.73 & 0.43 & 0.61 & 0.23 & 0.81 & 0.93 \\
			& TN\Tstrut & 0.67 & 1.87 & 5.16 & 2.42 & 1.27 & 0.58 & 1.03 & 0.3 & 0.72 & 0.94 \\
			& AR\Tstrut & 2.33 & 3.64 & 9.81 & 5.16 & 3.37 & 1.56 & 2.92 & 0.7 & 0.57 & 0.91 \\
			\hline
		\end{tabular}%
	\end{center}

	\caption{Summary of linear mixed effects modelling for change $\Delta^M$ in the annual mean of four climate variables over 6 climate zones (GL: global; AN: Antarctic; TS: Temperate South; TR: Tropical; TN: Temperate North; AR: Arctic). Columns under ``SSP effect'' show fixed effect parameter estimates $\gamma_j$ for the change in 100-year return value of the given variable over the period (2025,2125) under scenario $j$, in the units of the variable; \ed{thus we estimate an increase in $6.23K$ in the return value for \TA globally under scenario \SH.}. Columns under ``Model standard deviations'' provide estimates of the various standard deviations of model fitting. $\tau_R$: the (full unconditional) standard deviation of the response $R$; $\tau_{\text{FE}}$: the model error standard deviation after fitting only the fixed effects (FE, of climate scenario); $\tau_{\epsilon}$: the model error standard deviation after fitting the full mixed effects model. $\tau_{\delta}$: standard deviation of climate model random effect; $\tau_{\zeta}$: standard deviation due to nested random effect of climate ensemble within climate model. $R^2$ statistics are also provided, from fitting only the fixed effects (FE), and from fitting the full mixed effects (ME) model. Note that the first (year 2015) observation for \RS in all \CE runs is spurious and has been omitted. Multiple values for \WM from \MR are also suspect (with some values $>100$ ms$^{-1}$); for this reason \MR output is ignored for variable \WM. Further, since sample size for model fitting is huge, estimates of uncertainties and ``significance'' are of little practical value, and are omitted.}

\end{table}